\documentclass[12pt]{article}
\pdfoutput=1
\usepackage{pstricks}
\usepackage{color}
\usepackage{amssymb,amsmath,bm,bbm}
\usepackage{epsf}
\usepackage{epsfig}
\usepackage{afterpage}
\usepackage{longtable}
\usepackage{cite}
\usepackage{latexsym,mathrsfs,dsfont}
\usepackage{graphics}
\usepackage{url}
\usepackage{paralist}
\usepackage{bbold}

\newcommand{\svs}{\vbox{\vskip 5mm}}

\newcommand{\ord}{\mathcal{O}}

\newcommand{\IM}{\rm{Im}}
\newcommand{\RE}{\rm{Re}}

\newcommand{\tev}{\, {\rm TeV}}
\newcommand{\gev}{\, {\rm GeV}}
\newcommand{\mev}{\, {\rm MeV}}

\newcommand{\vcb}{|V_{cb}|}

\newcommand{\vub}{|V_{ub}|}
\newcommand{\vts}{|V_{ts}|}
\newcommand{\vus}{|V_{us}|}

\newcommand{\bsi}{B_6^{(1/2)}}
\newcommand{\bei}{B_8^{(3/2)}}

\def\epe{\varepsilon'/\varepsilon}
\newcommand{\beq}{\begin{equation}}
\newcommand{\eeq}{\end{equation}}
\newcommand{\be}{\begin{equation}}
\newcommand{\ee}{\end{equation}}
\newcommand{\bi}{\begin{itemize}}
\newcommand{\ei}{\end{itemize}}
\newcommand{\ba}{\begin{array}}
\newcommand{\ea}{\end{array}}
\newcommand{\beqa}{\begin{eqnarray}}
\newcommand{\eeqa}{\end{eqnarray}}
\newcommand{\bea}{\begin{eqnarray}}
\newcommand{\eea}{\end{eqnarray}}
\newcommand{\beqn}{\begin{eqnarray}}
\newcommand{\eeqn}{\end{eqnarray}}

\newcommand{\D}{\Delta}

\newcommand{\eps}{\epsilon}

\newcommand{\re}{{\rm Re}}
\newcommand{\im}{{\rm Im}}

\definecolor{red}{cmyk}{0,1,1,0.4}

\def\kpn{K^+\rightarrow\pi^+\nu\bar\nu}
\def\klpn{K_{L}\rightarrow\pi^0\nu\bar\nu}

\usepackage{fancyhdr}
\advance \headheight by 3.0truept       % for 12pt mandatory...
\begin{document}

\begin{flushright}
    {FLAVOUR(267104)-ERC-65}\\
    {BARI-TH/14-688}
\end{flushright}

\medskip

\begin{center}
{\LARGE\bf
\boldmath{$\Delta I=1/2$ Rule, $\epe$ and $K\to\pi\nu\bar\nu$ in $Z^\prime(Z)$\\
and $G^\prime$ Models with FCNC Quark Couplings}}
\\[0.8 cm]
{\bf Andrzej~J.~Buras$^{a,b}$, Fulvia~De~Fazio$^{c}$ and
Jennifer Girrbach$^{a,b}$
 \\[0.5 cm]}
{\small
$^a$TUM Institute for Advanced Study, Lichtenbergstr. 2a, D-85747 Garching, Germany\\
$^b$Physik Department, Technische Universit\"at M\"unchen,
James-Franck-Stra{\ss}e, \\D-85747 Garching, Germany\\
$^c$Istituto Nazionale di Fisica Nucleare, Sezione di Bari, Via Orabona 4,
I-70126 Bari, Italy}
\end{center}

\vskip0.41cm

%{\em Version of \today}

\abstract{%
\noindent
The experimental value for the isospin amplitude ${\rm Re}A_2$ in  $K\to\pi\pi$ decays 
has been successfully explained within the Standard Model (SM), both within 
large $N$ approach to QCD and 
by QCD lattice calculations. On the other hand within large $N$ approach  
 the value of ${\rm Re}A_0$ is by at least $30\%$ below the data. 
While this deficit could be the result of theoretical uncertainties in this 
approach and could be removed by future precise QCD lattice calculations, it cannot be excluded that the missing piece in  ${\rm Re}A_0$
comes from New Physics (NP).  We demonstrate that this deficit can be significantly softened by 
tree-level FCNC transitions  mediated by a heavy colourless $Z^\prime$ gauge boson with flavour violating {\it left-handed} coupling  $\Delta^{sd}_L(Z^\prime)$ and approximately 
universal flavour diagonal {\it right-handed} coupling  $\Delta^{qq}_R(Z^\prime)$ to quarks. The approximate flavour universality of the latter coupling assures negligible NP contributions to  ${\rm Re}A_2$.
 This property together with the breakdown of GIM mechanisms at
tree-level allows to enhance significantly the contribution of the leading 
QCD penguin operator $Q_6$ to ${\rm Re}A_0$. A large fraction of the missing piece in the $\Delta I=1/2$ rule can be
explained in this manner for $M_{Z^\prime}$ in the reach of the LHC, while satisfying constraints from $\varepsilon_K$, $\epe$, $\Delta M_K$, LEP-II and the LHC. 
The presence of a small right-handed flavour violating coupling $\Delta^{sd}_R(Z^\prime)\ll\Delta^{sd}_L(Z^\prime)$ and of enhanced matrix elements 
of $\Delta S=2$ left-right operators allows to satisfy simultaneously the constraints from   ${\rm Re}A_0$ and $\Delta M_K$, although this requires some 
fine-tuning. We identify {\it quartic} correlation 
between  $Z^\prime$ contributions to ${\rm Re}A_0$, $\epe$, $\varepsilon_K$ and 
$\Delta M_K$.
The tests of this proposal will require 
much improved evaluations of  ${\rm Re}A_0$ and $\Delta M_K$ within the SM, of $\langle Q_6 \rangle_0$ as well as precise tree level determinations of $\vub$ and $\vcb$. We present correlations between $\epe$, $\kpn$ and $\klpn$ 
with and without the $\Delta I=1/2$ rule constraint and generalize the 
whole analysis to $Z^\prime$ with colour ($G^\prime$) and $Z$ with FCNC couplings. In the 
latter case no improvement on  ${\rm Re}A_0$  can be achieved without 
destroying the agreement of the SM with the data on ${\rm Re}A_2$. Moreover, this scenario is 
very tightly constrained by $\epe$.  On the other hand in the context of the 
$\Delta I=1/2$ rule  $G^\prime$ is even more effective than 
$Z^\prime$: it provides the missing piece in  ${\rm Re}A_0$ for $M_{G^\prime}=(3.5-4.0)\tev$. 
}

\thispagestyle{empty}
\newpage
\setcounter{page}{1}

\tableofcontents

\section{Introduction}
The non-leptonic $K_L\to\pi\pi$ decays have played already for almost sixty 
years an important role in particle physics and were instrumental in the 
construction of the Standard Model (SM) and in the selection of allowed 
extensions of this model. The three pillars in these decays are:
\begin{itemize}
\item
The real parts of the amplitudes $A_I$ for a kaon to decay into two pions 
with isospin $I$ which are measured to be \cite{Beringer:1900zz}
\be\label{N1}
{\rm Re}A_0= 27.04(1)\times 10^{-8}~\gev, 
\quad {\rm Re}A_2= 1.210(2)   \times 10^{-8}~\gev,
\ee
and express the so-called $\Delta I=1/2$ rule \cite{GellMann:1955jx,GellMann:1957wh}
\be\label{N1a}
R=\frac{{\rm Re}A_0}{{\rm Re}A_2}=22.35.
\ee
\item
The parameter $\varepsilon_K$, a measure of indirect CP-violation in 
$K_L\to\pi\pi$ decays, which is found to be
\be\label{N2}
\varepsilon_K=2.228(11)\times 10^{-3}e^{i\phi_\varepsilon},
\ee
where $\phi_\varepsilon=43.51(5)^\circ$. 
\item
The ratio of the direct CP-violation and indirect CP-violation in $K_L\to\pi\pi$ decays  measured 
to be \cite{Beringer:1900zz,Batley:2002gn,AlaviHarati:2002ye,Worcester:2009qt} 
\be\label{eprime}
\RE(\epe)=(16.5\pm 2.6)\times 10^{-4}.
\ee
\end{itemize}
Also the strongly suppressed branching ratio for the  rare decay $K_L\to \mu^+\mu^-$
 and the tiny experimental value for $K_L-K_S$ mass difference
\be\label{DMK}
(\Delta M_{K})_{\rm exp} = 3.484(6) 10^{-15} \ \textrm{GeV}=5.293(9)\text{ps}^{-1}
\ee
were strong motivations for the GIM mechanism \cite{Glashow:1970gm} and in turn allowed to predict 
not only the existence of the charm quark but also approximately its mass 
\cite{Gaillard:1974hs}.

While due to the GIM mechanism $\varepsilon_K$, $\epe$ and $\Delta M_K$ receive 
contributions from the SM dynamics first at one-loop level and as such are 
sensitive to NP contributions, the $\Delta I=1/2$ rule involving tree-level 
decays has been expected already for a long time to be governed by SM dynamics.
Unfortunately due to non-perturbative nature of non-leptonic decays  precise 
calculation of the amplitudes ${\rm Re}A_0$ and ${\rm Re}A_2$ do not exist 
even today. However, a significant progress in reaching this goal over last 
forty years has been made. 

Indeed, after pioneering calculations of short distance QCD effects in 
the amplitudes ${\rm Re}A_0$ and ${\rm Re}A_2$ \cite{Gaillard:1974nj,Altarelli:1974exa}, termed in the past as {\it octet enhancement}, and the discovery of QCD
penguin operators \cite{Shifman:1975tn} which in the isospin limit 
contribute only to ${\rm Re}A_0$, the dominant dynamics behind the 
$\Delta I=1/2$ has been identified in \cite{Bardeen:1986vz}. To this 
end  an {\it analytic}  approximate approach based on the dual representation of QCD as a theory of weakly interacting mesons for large $N$, advocated 
previously in \cite{'tHooft:1973jz,'tHooft:1974hx,Witten:1979kh,Treiman:1986ep}, has been used. In this approach $\Delta I=1/2$ rule for $K\to\pi\pi$ decays 
has a simple origin. The octet enhancement through the long but slow quark-gluon  renormalization group evolution down to the scales $\ord(1\gev)$, analyzed first in   \cite{Gaillard:1974nj,Altarelli:1974exa},   is continued as a short but 
fast meson evolution down to zero momentum scales at which the factorization of 
hadronic matrix elements is at work. The recent inclusion of lowest-lying vector meson contributions in addition to the pseudoscalar ones and of NLO QCD 
corrections to Wilson coefficients in a momentum scheme improved significantly the matching between quark-gluon and meson evolutions \cite{Buras:2014maa}. In this approach QCD penguin operators 
play a subdominant role but one can uniquely predict an enhancement of ${\rm Re}A_0$ through QCD penguin contributions.  Working at scales 
$\ord(1\gev)$ 
this enhancement amounts to roughly $15\%$ of the experimental value  of ${\rm Re}A_0$ 
subject to uncertainties to which we will return below.

In the present era of the dominance of non-perturbative QCD calculations by lattice simulations 
 with dynamical fermions, that have a higher control over uncertainties than the approach in \cite{Bardeen:1986vz,Buras:2014maa}, it is
very encouraging that the structure of the enhancement of ${\rm Re}A_0$ and suppression of ${\rm Re}A_2$, identified already in \cite{Bardeen:1986vz}, has 
also been found  by RBC-UKQCD collaboration  \cite{Boyle:2012ys,Blum:2011pu,Blum:2011ng,Blum:2012uk}. The comparison between the results of both approaches in 
 \cite{Buras:2014maa} indicates that the experimental value of the 
amplitude  ${\rm Re}A_2$ can be well described within the SM, in particular, as the calculations in these papers have been performed at 
rather different scales and using a different technology.

On the other hand  both approaches cannot 
presently obtain sufficiently large value of 
${\rm Re}A_0$. Within the dual QCD approach one finds then $R=16.0\pm1.5$, while 
the first lattice results for ${\rm Re}A_0$ imply $R\approx 11$. However, 
the latter result has been obtained with non-physical kinematics and it is 
to be expected that larger values of $R$, even as high as its experimental 
value in (\ref{N1a}), could be obtained in lattice QCD in the future.

Presently  theoretical value of ${\rm Re}A_0$  within dual QCD approach is by  $30\%$ below the data and even more in the case of lattice QCD. 
While this deficit could be the result of theoretical uncertainties in both approaches, it cannot be excluded that the missing piece in  ${\rm Re}A_0$
comes from New Physics (NP).  In this context we would like to emphasize, that 
although the explanation of the dynamics behind the $\Delta I=1/2$ rule is 
not any longer at the frontiers of particle physics, it is important to 
determine precisely the room for NP contribution left  not only in  ${\rm Re}A_0$ but also  ${\rm Re}A_2$. From the present perspective 
only lattice simulations with dynamical fermions 
can provide  precise values of  ${\rm Re}A_{0,2}$ one day, but this may still 
take several years of intensive efforts by the lattice community \cite{Tarantino:2012mq,Sachrajda:2013fxa,Christ:2013lxa}.  Having precise 
SM values for 
${\rm Re}A_{0,2}$ would give us two observables which could be used to 
constrain NP. Our paper demonstrates explicitly the impact of such 
constraints.

In this context we would like to strongly emphasize that while the dominant 
part of the $\Delta I=1/2$ rule originates in the SM dynamics it is legitimate 
to ask whether some subleading part of it comes from much shorter distance 
scales and either exclude this possibility or demonstrate that 
this indeed could be the case under certain assumptions.

In what follows our working assumption will be that roughly $30\%$ of 
 ${\rm Re}A_0$ comes from some kind of NP which does not affect  ${\rm Re}A_2$
in order not to spoil the agreement of the SM with the data. As the missing 
piece in  ${\rm Re}A_0$ is by about eight times larger than the measured 
value of ${\rm Re}A_2$, the required NP must have a particular structure: tiny 
or absent contributions to  ${\rm Re}A_2$ and at the same time large contributions to  ${\rm Re}A_0$. Moreover it should satisfy other constraints coming 
from $\varepsilon_K$, $\Delta M_K$, $\epe$ and rare kaon decays.

 As $K\to\pi\pi$ decays originate already at tree-level, we expect that NP contributing 
to these decays at one-loop level will not help us in reaching our goal. 
 Consequently we have to look for NP that contributes to $K\to\pi\pi$ decays
 already at tree-level  as well. Moreover in order 
not to spoil the agreement of the SM with the data for ${\rm Re}A_2$ only 
Wilson coefficients of QCD penguin operators should be modified. In this 
context we recall that in  \cite{Bardeen:1986uz}
an additional (with respect to previous estimates) enhancement of the QCD
penguin contributions to ${\rm Re}A_0$ has been identified. It comes from 
an incomplete GIM cancellation above the charm quark mass. But as the analyses 
in  \cite{Bardeen:1986vz,Buras:2014maa} show, this enhancement is insufficient 
to reproduce fully the experimental value of  ${\rm Re}A_0$.

However, the observation that the breakdown of GIM mechanism and the enhanced 
contributions of QCD penguin operators could in principle provide the 
missing part of the  $\Delta I=1/2$ rule,  gives us a hint what kind of NP could do the job here. We have to break GIM mechanism at a much higher 
scale than scales $\ord(m_c)$ and allow the QCD renormalization group evolution 
to enhance the Wilson coefficient of the leading QCD penguin operator $Q_6$ by 
a larger amount than it is possible within the SM.

It turns then out that a tree-level exchange 
of heavy neutral gauge boson, colourless ($Z^\prime$) or carrying colour ($G^\prime$)  can provide a significant part of the missing piece of ${\rm Re}A_0$  but 
the  couplings of these heavy gauge bosons  to SM 
fermions must have a very special structure in order to satisfy existing 
constraints from other observables. Assuming $M_{Z^\prime} (M_{G^\prime})$ to be in the 
ballpark of a few $\tev$ and denoting  left-handed (LH) and 
right-handed (RH) couplings of $Z^\prime(G^\prime)$ to two SM fermions  with 
flavours $i$ and $j$,  as in \cite{Buras:2012jb}, by $\Delta_{L,R}^{ij}(Z^\prime)$,  we find that in the mass eigenstate basis for all particles involved, a 
$Z^\prime$ or $G^\prime$ with the following general structure of its couplings is required:
\begin{itemize}
\item
${\rm Re}\Delta_L^{s d}(Z^\prime)=\ord(1)$ and ${\rm Re}\Delta_R^{qq}(Z^\prime)=\ord(1)$ in order to generate $Q_6$ penguin operator with sizable Wilson 
coefficient in the presence of a heavy $Z^\prime$.
\item
The diagonal couplings $\Delta_R^{qq}(Z^\prime)$ must be 
flavour universal in order not to affect the amplitude ${\rm Re}A_2$. But this universality cannot be exact as this would not 
allow to generate a small ${\rm Re}\Delta_R^{s d}(Z^\prime)=\ord(10^{-3})$ 
coupling which is required in order to satisfy the constraint on 
$\Delta M_K$ in the presence of ${\rm Re}\Delta_L^{s d}(Z^\prime)=\ord(1)$.
\item
${\rm Im}\Delta_L^{s d}(Z^\prime)$ and ${\rm Im}\Delta_R^{qq}(Z^\prime)$ 
must be typically $\ord(10^{-3}-10^{-4})$ in order to be consistent with the data 
on $\varepsilon_K$ and $\epe$.
\item
The couplings to leptons must be sufficiently small in order not to 
violate the existing bounds on rare kaon decays. This is automatically 
satisfied for  $G^\prime$. 
\item
Finally, $\Delta_L^{uu}(Z^\prime)$ must be small in order not to generate 
large contributions to the current-current operators $Q_1$ and $Q_2$ that 
could affect the amplitude ${\rm Re}A_2$.
\end{itemize}

We observe, that indeed the structure of the $Z^\prime$ or $G^\prime$ 
couplings must be 
rather special. But in the context of $\epe$ it is interesting to note 
that in this NP scenario, as opposed to many NP scenarios, 
there is no modification of Wilson coefficients of electroweak penguin operators up to tiny renormalization group effects that can be neglected for all practical purposes. NP part of $\epe$ involves only QCD penguin operators, in particular $Q_6$, and the size of this effect, as we will demonstrate below, is 
correlated with NP contribution to  ${\rm Re}A_0$, $\varepsilon_K$ and 
$\Delta M_K$.

Now comes an important point. While  SM contribution to ${\rm Re}A_0$ practically does not involve any CKM uncertainties, this is not the case of $\varepsilon_K$, $\epe$ and branching ratios on rare kaon decays which all involve potential 
uncertainties due to present inaccurate knowledge of the elements of the 
CKM matrix $\vub$ and $\vcb$. Therefore there are uncertainties in the 
room left for NP in these observables and these uncertainties in turn 
affect indirectly the allowed size of NP contribution to ${\rm Re}A_0$.
 Therefore it will be of interest to consider several scenarios for the pair 
$\vub$ and $\vcb$ and investigate in each case whether $Z^\prime$ couplings required to improve the situation with the $\Delta I=1/2$ rule could also help in 
explaining the data on $\varepsilon_K$, $\epe$, $\Delta M_K$ and rare kaon 
 decays in
case the SM would fail to do it one day. Of course presently one cannot 
reach clear cut conclusions on these matters due to hadronic uncertainties affecting 
 $\varepsilon_K$, $\epe$ and $\Delta M_K$ but it is expected that the situation 
will improve in this decade.

In order to be able to discuss implications for $\kpn$ and 
$\klpn$ we will assume in the first  part of our paper that $Z^\prime$ is 
colourless. This is also the case analyzed in all our previous $Z^\prime$ 
papers \cite{Buras:2012sd,Buras:2012jb,Buras:2012dp,Buras:2013uqa,Buras:2013rqa,Buras:2013raa,Buras:2013qja,Buras:2013dea}. Subsequently,  we will discuss how our analysis changes in the case of $G^\prime$.
The fact that in this case 
$G^\prime$ does not contribute to $\kpn$ and $\klpn$ allows already 
to distinguish this case from the colourless $Z^\prime$ but also the LHC 
bounds on the couplings of such bosons  and the 
 NP contributions to ${\rm Re}A_0$, $\epe$, $\varepsilon_K$ and $\Delta M_K$ 
 are different in these two cases. In our presentation we will also first 
assume exact flavour universality for $\Delta_R^{qq}(Z^\prime)$  and 
 $\Delta_R^{qq}(G^\prime)$  couplings in 
order to demonstrate that in this case the experimental constraints from ${\rm Re}A_0$ and $\Delta M_K$ cannot be simultaneously satisfied. Fortunately 
already a very small violation of flavour universality in  $\Delta_R^{qq}(Z^\prime)$  or $\Delta_R^{qq}(G^\prime)$  allows to cure  this 
problem because of the enhanced matrix elements of left-right operators contributing in this case to $\Delta M_K$.

Our paper is organized as follows. In Section~\ref{sec:2a} we briefly describe 
some general aspects of $Z^\prime$  and $G^\prime$ models considered by us. In Section~\ref{sec:2} we present general formulae for the 
effective 
Hamiltonian for $K\to\pi\pi$ decays including all operators, list the initial conditions for Wilson coefficients at $\mu=M_{Z^\prime}$ for the case of a colourless $Z^\prime$ and find the 
expressions for ${\rm Re}A_0$ and $\epe$ that include SM and $Z^\prime$ 
contributions. In Section~\ref{sec:3} we discuss briefly $\varepsilon_K$, 
$\Delta M_K$, $\kpn$ and $\klpn$, again for a colourless $Z^\prime$, 
referring for details to our previous papers. In Section~\ref{sec:4} we 
present numerical analysis of  ${\rm Re}A_0$, $\epe$ and $\kpn$ and $\klpn$
taking into account the constraints from $\varepsilon_K$ and 
$\Delta M_K$. We consider two scenarios. One in which we impose the 
$\Delta I=1/2$ constraint (Scenario A) and one in which we ignore this 
constraint (Scenario B). These two scenarios can be clearly distinguished through the rare decays $\kpn$ and $\klpn$ and their correlation with $\epe$.
 In Section~\ref{sec:5}  we repeat the full analysis for $G^\prime$ 
and in Section~\ref{sec:5b} for the $Z$ boson 
with flavour violating couplings.  We conclude in Section~\ref{sec:6}.

\boldmath
\section{General Aspects of $Z^\prime$ and $G^\prime$ Models}\label{sec:2a}
\unboldmath
The present paper is the continuation of our extensive  study of NP 
represented by a new neutral heavy gauge boson ($Z^\prime$) in the context 
of a general parametrization of its couplings to SM fermions and within
 specific models like 331 models \cite{Buras:2012sd,Buras:2012jb,Buras:2012dp,Buras:2013uqa,Buras:2013rqa,Buras:2013raa,Buras:2013qja,Buras:2013dea}. 
The new aspect of the present paper is the generalization of these studies 
to $K\to\pi\pi$ decays with the goal to answer three questions:
\begin{itemize}
\item
Whether the existence of a $Z^\prime$ or $G^\prime$ with a mass in the reach of the LHC 
could have an impact on the $\Delta I=1/2$ rule, in particular on the 
amplitude ${\rm Re}A_0$.
\item
Whether such gauge bosons could have sizable impact on the ratio $\epe$.
\item
What is the impact of $\epe$ constraint on FCNC couplings of the SM $Z$ 
boson.
\end{itemize}

To our knowledge the first question has not been addressed in the literature, 
while selected analyses of $\epe$ within models with tree-level flavour changing neutral currents can be found in \cite{Gedalia:2009ws,Bauer:2009cf}. However, in these papers 
NP entered $\epe$ through electroweak penguin operators while in the case of 
$Z^\prime$ scenarios considered here
 only QCD penguin operators are relevant. Concerning the last point we 
refer to earlier analyses in \cite{Buras:1998ed,Buras:1999da}. The present paper provides a modern look at this scenario and in particular investigates the 
sensitivity to CKM parameters. A review of $Z^\prime$ models can be found 
in \cite{Langacker:2008yv} and a collection of papers related mainly to 
$B_{s,d}$ decays can be found in \cite{Buras:2012jb}.

Our paper will deal with NP in $K^0-\bar K^0$ mixing, $K\to\pi\pi$ and rare 
$K$ decays dominated either by a heavy $Z^\prime$, heavy $G^\prime$ or 
FCNC processes mediated by $Z$. We will not provide a complete model in 
which other fields like heavy vector-like fermions, heavy Higgs scalars and 
charged gauge bosons are generally present and gauge anomalies are properly 
 canceled. Examples of such models can be found 
in \cite{Langacker:2008yv} and the 331 models analyzed by us can be mentioned 
 here \cite{Buras:2012sd,Buras:2013dea}. A general discussion can also 
be found in \cite{Fox:2011qd} and among more recent papers we 
refer to \cite{Dobrescu:2013cmh} and \cite{Altmannshofer:2014cfa}.
 But none of these 
papers discusses the hierarchy of the couplings of $Z^\prime$ and $G^\prime$ 
couplings which is required to make these gauge bosons to be relevant for 
the $\Delta I=1/2$ rule. Our goal then is  to find this hierarchy 
first and postpone the construction of a concrete model to a future analysis.

$Z^\prime$ contributions to  ${{\RE}A_0}$, ${{\RE}A_2}$ and $\epe$ involve generally in 
addition to $M_{Z^\prime}$ the following couplings:
\be\label{couplings}
\Delta_L^{s d}(Z^\prime),\qquad \Delta_R^{s d}(Z^\prime), \qquad \Delta_L^{q q}(Z^\prime),\qquad \Delta_R^{qq}(Z^\prime),
\ee
where $q=u,d,c,s,b,t$. The same applies to $G^\prime$. The diagonal couplings can be generally flavour dependent but 
as we already stated above in order to protect the small amplitude  ${{\RE}A_2}$ from 
significant NP contributions in the process of modification of the large amplitude 
 ${{\RE}A_0}$ either the coupling $\Delta_L^{q q}(Z^\prime)$ or the 
coupling $\Delta_R^{qq}(Z^\prime)$ 
must be approximately flavour universal. They cannot be both flavour universal as then 
it would not be possible to generate large flavour violating couplings in the mass eigenstate basis. In what follows we will assume that $\Delta_R^{qq}(Z^\prime)$ are either exactly 
flavour universal or flavour universal to a high degree still allowing for a strongly 
suppressed but non-vanishing coupling $\Delta_R^{s d}(Z^\prime)$.

For the left-handed couplings it will turn out that $\Delta_L^{s d}(Z^\prime)=\ord(1)$ in order to reach the first goal on our list. Such a coupling 
could be in principle generated in the presence of heavy vectorial fermions or other 
dynamics at scales above $M_{Z^\prime}$. In order to simplify our analysis 
and reduce the number of free parameters, we will finally assume that 
$\Delta_L^{q q}(Z^\prime)$ are very small. Thus in summary the hierarchy of 
couplings in the present paper will be assumed to be as follows:
\be\label{couplingsh}
\Delta_L^{s d}(Z^\prime)\gg \Delta_L^{qq}(Z^\prime), \qquad \Delta_R^{sd}(Z^\prime)\ll \Delta_R^{qq}(Z^\prime), \qquad \Delta_L^{s d}(Z^\prime)\gg \Delta_R^{sd}(Z^\prime)
\ee
with the same hierarchy assumed for $G^\prime$.

Only the coupling $\Delta_{L,R}^{s d}(Z^\prime)$ will be assumed to be complex while as we will see 
in the context of our analysis the remaining two can be assumed to be 
real without particular loss of generality. We should note that the hierarchy 
in (\ref{couplingsh}) will suppress in the case of $K\to\pi\pi$ decays 
the primed operators that are absent in the SM anyway.

In our previous papers we have considered a number of scenarios for 
flavour violating $Z^\prime$ couplings to quarks. These are defined
as follows:
\begin{enumerate}
\item
Left-handed Scenario (LHS) with complex $\Delta_L^{sd}\not=0$  and $\Delta_R^{sd}=0$,
\item
Right-handed Scenario (RHS) with complex $\Delta_R^{sd}\not=0$  and $\Delta_L^{sd}=0$,
\item
Left-Right symmetric Scenario (LRS) with complex  
$\Delta_L^{sd}=\Delta_R^{sd}\not=0$,
\item
Left-Right asymmetric Scenario (ALRS) with complex
$\Delta_L^{sd}=-\Delta_R^{sd}\not=0$.
\end{enumerate}

Among them only LHS scenario is consistent with  (\ref{couplingsh}) if 
$\Delta_R^{sd}$ is assumed to vanish. But as we will demonstrate in this 
case it is not possible to satisfy simultaneously the constraints from 
 ${{\RE}A_0}$ and $\Delta M_K$. Consequently $\Delta_R^{sd}$ has to be non-vanishing, although very small, in order to satisfy these two constraints simultaneously.  Thus in the scenarios considered in our previous papers the status of the $\Delta I=1/2$ rule cannot be improved with respect to the SM.

\boldmath
\section{General Formulae for $K\to\pi\pi$ Decays}\label{sec:2}
\unboldmath
\subsection{General Structure}
Let us begin our presentation with the general formula for the effective 
Hamiltonian relevant for $K\to\pi\pi$ 
decays in the model in question
\be\label{general}
\mathcal{H}_{\rm eff}(K\to\pi\pi)= \mathcal{H}_{\rm eff}(K\to\pi\pi)({\rm SM})+ 
\mathcal{H}_{\rm eff}(K\to\pi\pi)(Z^\prime)
\ee
where  the SM part is given by \cite{Buras:1993dy}
\be\label{basic}
 \mathcal{H}_{\rm eff}(K\to\pi\pi)({\rm SM})=\frac{G_F}{\sqrt{2}}V_{ud}V_{us}^*\sum_{i=1}^{10}(z^{\rm SM}_i(\mu)+\tau y^{\rm SM}_i(\mu))Q_i, \qquad
\tau=-\frac{V_{td}V_{ts}^*}{V_{ud}V_{us}^*},
\ee
and
the operators $Q_i$  as follows:

{\bf Current--Current:}
\begin{equation}\label{O1} 
Q_1 = (\bar s_{\alpha} u_{\beta})_{V-A}\;(\bar u_{\beta} d_{\alpha})_{V-A}
~~~~~~Q_2 = (\bar su)_{V-A}\;(\bar ud)_{V-A} 
\end{equation}

{\bf QCD--Penguins:}
\begin{equation}\label{O2}
Q_3 = (\bar s d)_{V-A}\sum_{q=u,d,s,c,b,t}(\bar qq)_{V-A}~~~~~~   
 Q_4 = (\bar s_{\alpha} d_{\beta})_{V-A}\sum_{q=u,d,s,c,b,t}(\bar q_{\beta} 
       q_{\alpha})_{V-A} 
\end{equation}
\begin{equation}\label{O3}
 Q_5 = (\bar s d)_{V-A} \sum_{q=u,d,s,c,b,t}(\bar qq)_{V+A}~~~~~  
 Q_6 = (\bar s_{\alpha} d_{\beta})_{V-A}\sum_{q=u,d,s,c,b,t}
       (\bar q_{\beta} q_{\alpha})_{V+A} 
\end{equation}

{\bf Electroweak Penguins:}
\begin{equation}\label{O4} 
Q_7 = \frac{3}{2}\;(\bar s d)_{V-A}\sum_{q=u,d,s,c,b,t}e_q\;(\bar qq)_{V+A} 
~~~~~ Q_8 = \frac{3}{2}\;(\bar s_{\alpha} d_{\beta})_{V-A}\sum_{q=u,d,s,c,b,t}e_q
        (\bar q_{\beta} q_{\alpha})_{V+A}
\end{equation}
\begin{equation}\label{O5} 
 Q_9 = \frac{3}{2}\;(\bar s d)_{V-A}\sum_{q=u,d,s,c,b,t}e_q(\bar q q)_{V-A}
~~~~~Q_{10} =\frac{3}{2}\;
(\bar s_{\alpha} d_{\beta})_{V-A}\sum_{q=u,d,s,c,b,t}e_q\;
       (\bar q_{\beta}q_{\alpha})_{V-A} 
\end{equation}
Here, $\alpha,\beta$ denote colours and $e_q$ denotes the electric quark charges reflecting the
electroweak origin of $Q_7,\ldots,Q_{10}$. Finally,
$(\bar sd)_{V-A}\equiv \bar s_\alpha\gamma_\mu(1-\gamma_5) d_\alpha$. 

The coefficients $z^{\rm SM}_i(\mu)$ and $y^{\rm SM}_i(\mu)$ are the Wilson coefficients  of these operators within the SM. They
are known at the NLO level in the renormalization group improved perturbation theory including both QCD and QED corrections  \cite{Buras:1993dy,Ciuchini:1993vr}. Also some elements of NNLO corrections can be found in the literature \cite{Buras:1999st,Gorbahn:2004my}.

As discussed in the previous section $Z^\prime$ contributions to $K\to\pi\pi$ in  the class of $Z^\prime$ models discussed by us can be well approximated by the following  effective Hamiltonian 
\be\label{ZprimeA}
\mathcal{H}_{\rm eff}(K\to\pi\pi)(Z^\prime)=\sum_{i=3}^{6}(C_i(\mu)Q_i+C_i^\prime(\mu) Q_i^\prime),
\ee
where the primed operators $Q_i^\prime$ are obtained from $Q_i$ by interchanging 
$V-A$ and $V+A$. For the sake of completeness we keep still 
 $ Q_i^\prime$ operators even if at the end due to the hierarchy of 
couplings in (\ref{couplingsh}), $Z^\prime$ contributions will be 
well approximated by $Q_i$ and the contributions from  $ Q_i^\prime$ operators 
can be neglected.

Due to the fact that $M_{Z^\prime}\gg m_t$ the summation over 
flavours in (\ref{O2})-(\ref{O5}) includes now also the top quark.
This structure is valid for both $Z^\prime$ and $G^\prime$.
 As the hadronic matrix elements of $Q_i$ do not 
depend on the properties of $Z^\prime$ or $G^\prime$, these two cases can only be distinguished by the values of the coefficients $C_i(\mu)$ and $C_i^\prime(\mu)$. In this 
and  two following sections we analyze the case of $Z^\prime$. 
But in Section~\ref{sec:5} we will also discuss $G^\prime$.

The important feature of the effective Hamiltonian in (\ref{ZprimeA}) is the absence of $Q_{1,2}$ operators 
dominating the $A_2$ amplitude and the absence of electroweak penguin operators 
which in some of the extensions of the SM are problematic for $\epe$. In 
our model NP effects in ${\rm Re}A_0$, relevant for the $\Delta I=1/2$ rule 
and ${\rm Im}A_0$, relevant for $\epe$, will enter only through QCD penguin  contributions. This is a novel feature when 
compared with other scenarios, like LHT \cite{Blanke:2007wr} and Randall-Sundrum scenarios \cite{Gedalia:2009ws,Bauer:2009cf}, where NP contributions to 
$\epe$ are dominated by electroweak penguin operators. In particular, in 
the latter case, where FCNCs are mediated by new heavy Kaluza-Klein gauge bosons, 
the flavour universality of their diagonal couplings to quarks is absent 
due to different positions of light and heavy quarks in the bulk. Consequently 
the pattern of NP contributions to $\epe$ differs from the one in the models 
discussed here.

Denoting by $\Delta_{L,R}^{ij}$, as in \cite{Buras:2012jb}, the couplings of $Z^\prime$ to two quarks with 
flavours $i$ and $j$, a tree level $Z^\prime$  exchange generates in 
our model
only the operators $Q_3$, $Q_5$, $Q^\prime_3$ and $Q^\prime_5$ at 
$\mu=M_{Z^\prime}$. The inclusion of QCD effects, in particular the renormalization group evolution down to low energy scales, generates the remaining QCD penguin operators.
In principle using the two-loop anomalous dimensions of \cite{Buras:1993dy,Ciuchini:1993vr} and the $\ord(\alpha_s)$ corrections to the coefficients $C_i$ and $C_i^\prime$ at $\mu_{Z^\prime}=\ord(M_{Z^\prime})$ in the NDR-${\rm \overline{MS}}$ scheme in \cite{Buras:2012gm} the full NLO analysis of $Z^\prime$ contributions 
could be performed. However, due to the fact that the mass of $Z^\prime$ is free and 
other parametric and hadronic uncertainties, a leading order analysis of NP 
contributions is sufficient for our purposes. 
In this manner it will also be possible to see certain properties analytically.

The non-vanishing Wilson coefficients at $\mu=M_{Z^\prime}$ are then given at the 
LO
as follows
\begin{align}
\begin{split}
C_3(M_{Z^\prime})
& = \frac{\Delta_L^{s d}(Z^\prime)\Delta_L^{q q}(Z^\prime)}{4 M^2_{Z^\prime}}, \qquad 
C_3^\prime(M_{Z^\prime})
 = \frac{\Delta_R^{s d}(Z^\prime)\Delta_R^{q q}(Z^\prime)}{4 M^2_{Z^\prime}}
 \,,\end{split}\label{C3}\\
\begin{split}
C_5(M_{Z^\prime})
& = \frac{\Delta_L^{s d}(Z^\prime)\Delta_R^{q q}(Z^\prime)}{4 M^2_{Z^\prime}}, \qquad
C_5^\prime(M_{Z^\prime})
  = \frac{\Delta_R^{s d}(Z^\prime)\Delta_L^{q q}(Z^\prime)}{4 M^2_{Z^\prime}}
\,.\end{split}\label{C6}
\end{align}

\subsection{Renormalization Group Analysis (RG)}
With these results at hand we will perform RG analysis of NP contributions at the LO level\footnote{SM contributions are evaluated including NLO QCD corrections.}. We will then  see that the only operator that matters at scales $\ord(1 \gev)$ in our 
$Z^\prime$ 
models is either
$Q_6$ or $Q_6^\prime$. This is to be expected if 
we recall that at $\mu=M_W$ the Wilson coefficient of the electroweak penguin 
operator $Q_8$, the electroweak analog of $Q_6$, also vanishes. But due  
to its large anomalous dimension and enhanced hadronic $K\to \pi \pi$ 
matrix elements $Q_8$ is by far the dominant electroweak penguin operator 
in $\epe$ within the SM, leaving behind the $Q_7$ operator whose Wilson coefficient  does not vanish at $\mu=M_W$. Even if the structure of  the present RG analysis 
differs from the SM one, due to the absence of the remaining operators in the NP part, in particular the absence of $Q_2$,  much longer RG 
evolution from 
$M_{Z^\prime}$ and not $M_W$ down to low energies makes $Q_6$ or $Q_6^\prime$ the winner at the 
end. This fact as we will see simplifies significantly the phenomenological 
analysis of NP contributions to ${\rm Re}A_0$ and $\epe$.

The relevant $4\times4$  one-loop anomalous dimension matrix
\begin{equation}
\hat\gamma_s(\alpha_s)=\hat\gamma_s^{(0)}\frac{\alpha_s}{4\pi}
\label{eq:gsexpKpp}
\end{equation}
 can  be extracted from the known  $6\times6$ matrix \cite{Gilman:1979bc}. 
The evolution of the operators in the NP part is then governed in the 
$(Q_3,Q_4,Q_5,Q_6)$ basis by 
\begin{equation}
\hat \gamma^{(0)}_s = 
\left(
\begin{array}{cccc}
\frac{-22}{9} & \frac{22}{3} & -\frac{4}{9} & \frac{4}{3}
\\ \svs
6 - f\frac{2}{9} & -2 + f \frac{2}{3} & -f\frac{2}{9} & f\frac{2}{3} \\ \svs
0 & 0 & {2} & -6  \\ \svs
-f \frac{2}{9} & f \frac{2}{3} & -f\frac{2}{9} & -16
   + f\frac{2}{3}
\end{array}
\right),
\label{eq:gs0Kpp}
\end{equation}
where $f$ is the number of effective flavours: $f=6$ for $\mu\ge m_t$ and 
$f=3$ for $\mu\le m_c$. The same matrix governs the evolution of primed operators.

In order to see what happens analytically we then assume first that in 
the mass eigenstate basis only the couplings  $\Delta_L^{s d}$ and $\Delta_R^{qq}$ are non-vanishing with $\Delta_R^{qq}$ being exactly flavour universal. While, the 
coefficients of the operators $Q_3$ and $Q_4$ can still be generated through 
RG evolution, these effects are very small and can be 
neglected. Then 
to an excellent approximation only the operators $Q_5$ and $Q_6$ matter and 
RG evolution is governed by the reduced $2\times 2$ anomalous dimension 
 matrix given in the $(Q_5,Q_6)$ basis as follows
\begin{equation}
\hat \gamma^{(0)}_s = 
\left(
\begin{array}{cc}
2 & -6  \\ \svs
-f\frac{2}{9} & -16
   + f\frac{2}{3}
\end{array}
\right).
\label{reduced}
\end{equation}

Denoting then by $\vec C(M_{Z^\prime})$ the column vector with components given 
by the  Wilson coefficients $C_5$ and $C_6$ at $\mu=M_{Z^\prime}$ we 
find their values at $\mu=m_c$ by means of\footnote{The reason for choosing 
$\mu=m_c$ will be explained below.}
\begin{equation}\label{cmub1}
 \vec C(m_c)=\hat U(m_c,M_{Z^\prime}) \vec C(M_{Z^\prime})
\end{equation}
where
\begin{equation}\label{cmub1a}
 \hat U(m_c,M_{Z^\prime})=\hat U^{(f=4)}(m_c,m_b)\hat U^{(f=5)}(m_b, m_t) \hat U^{(f=6)}(m_t, M_{Z^\prime})
\end{equation}
and \cite{Buras:1998raa}
\begin{equation}\label{u0vd0} 
\hat U^{(f)}(\mu_1,\mu_2)= \hat V
\left({\left[\frac{\alpha_s(\mu_2)}{\alpha_s(\mu_1)}
\right]}^{\frac{\vec\gamma^{(0)}}{2\beta_0}}
   \right)_D \hat V^{-1}.   \end{equation}
Here $\hat V$ diagonalizes ${\hat\gamma^{(0)T}}$
\begin{equation}\label{ga0d} 
\hat\gamma^{(0)}_D=\hat V^{-1} {\hat\gamma^{(0)T}} \hat V
  \end{equation}
and $\vec\gamma^{(0)}$ is the vector containing the diagonal elements of
the diagonal matrix :
\begin{equation}\label{g120d} \hat\gamma^{(0)}_D=
 \left(\begin{array}{cc} \gamma^{(0)}_+ & 0 \\
                          0 & \gamma^{(0)}_-
    \end{array}\right).  \end{equation}
with 
\be
\beta_0= \frac{33-2f}{3}\, .
\ee

For $\alpha_s(M_Z)=0.1185$,  $m_c=1.3\gev$ and $M_{Z^\prime}=3\tev$ we have
\begin{equation}\label{C5C6} 
\left[\begin{array}{c} C_5(m_c) \\
                         C_6(m_c)
    \end{array}\right]
 = \left[\begin{array}{cc} 0.86 & 0.19 \\
                          1.13 & 3.60
    \end{array}\right]  \left[\begin{array}{c} 1 \\
                         0
    \end{array}\right] \frac{\Delta_L^{s d}(Z^\prime)\Delta_R^{q q}(Z^\prime)}{4 M^2_{Z^\prime}}.
  \end{equation}
Consequently 
\be\label{LOC5C6}
 C_5(m_c)= 0.86  \frac{\Delta_L^{s d}(Z^\prime)\Delta_R^{q q}(Z^\prime)}{4 M^2_{Z^\prime}}\qquad   C_6(m_c)= 1.13\frac{\Delta_L^{s d}(Z^\prime)\Delta_R^{q q}(Z^\prime)}{4 M^2_{Z^\prime}}.
\ee

Due to the large element $(1,2)$ in the matrix (\ref{reduced}) and 
the large anomalous dimension of the $Q_6$ operator represented by the $(2,2)$ 
element of this matrix, $C_6(m_c)$ is by a factor of $1.3$ larger than 
$C_5(m_c)$ even if $C_6(M_{Z^\prime})$ vanishes at LO. Moreover the matrix element $\langle Q_5\rangle_0$ is colour suppressed 
which is not the case of  $\langle Q_6\rangle_0$ and within a good 
approximation we can neglect the contribution of $Q_5$. In summary, it 
is sufficient to keep only $Q_6$ contribution in the decay amplitude in this scenario for $Z^\prime$ couplings.

\boldmath
\subsection{The Total $A_0$ Amplitude}
\unboldmath

Adding NP contributions to the SM contribution we find 
\be
A_0= A_0^{\rm SM}+A_0^{\rm NP}, \qquad 
\ee
with the SM contribution given by 
\be
{\RE} A_0^{\rm SM}=\frac{G_F}{\sqrt{2}}\lambda_u\sum_{i=1}^{10} z^{\rm SM}_i(\mu)\langle Q_i(\mu)\rangle_0,
\ee
\be
{\IM} A_0^{\rm SM}=-\frac{G_F}{\sqrt{2}}{\IM} \lambda_t\sum_{i=3}^{10} y^{\rm SM}_i(\mu)\langle Q_i(\mu)\rangle_0\, .
\ee
Here 
\be
 \lambda_i=V_{id}V_{is}^*
\ee
is the usual CKM factor.
As NP enters only Wilson coefficients and 
\be
\langle Q^\prime_i(\mu)\rangle_0=-\langle Q_i(\mu)\rangle_0,
\ee
NP contributions can be included by modifying $z_i$ and $y_i$ with $i=3-6$ as follows
\be\label{zidelta}
\Delta z_i(\mu)=\frac{\sqrt{2}}{\lambda_u G_F}\left({\RE} C_i(\mu)-{\RE} C^\prime_i(\mu)\right)
\ee
and
\be\label{yidelta}
\Delta y_i(\mu)=-\frac{\sqrt{2}}{{\IM}\lambda_t G_F}\left({\IM} C_i(\mu)-{\IM} C^\prime_i(\mu)\right).
\ee

In the scenario just discussed only  $Q_6$ operator is relevant and 
we have 
\be\label{RENP}
{\RE} A_0^{\rm NP}=\frac{G_F}{\sqrt{2}}\lambda_u \Delta z_6(\mu)\langle Q_6(\mu)\rangle_0= {\RE} C_6(\mu)\langle Q_6(\mu)\rangle_0
\ee
\be\label{IMNP}
{\IM} A_0^{\rm NP}=-\frac{G_F}{\sqrt{2}}{\IM}\lambda_t \Delta y_6(\mu)\langle Q_6(\mu)\rangle_0={\IM} C_6(\mu)\langle Q_6(\mu)\rangle_0,
\ee
where we have written two equivalent expressions so that one can either work 
with $z_6$ and $y_6$ as in the SM or directly with the NP coefficient
$C_6$. The latter expressions exhibit better the fact that 
NP contributions do not depend explicitly  on CKM parameters. For the matrix element $\langle Q_6(\mu)\rangle_0$ we will use the large $N$  result
 \cite{Bardeen:1986vz,Buras:2014maa} 
\be\label{eq:Q60}
\langle Q_6(\mu) \rangle_0=-\,4 
\left[ \frac{m_{\rm K}^2}{m_s(\mu) + m_d(\mu)}\right]^2 (F_K-F_\pi)
\,B_6^{(1/2)}\,,
\ee
except that we will allow for variation of $\bsi$ around its strict large $N$ 
limit $\bsi=1$.
In writing this formula we have removed the factor $\sqrt{2}$ from formula (97) 
in \cite{Buras:2014maa} in order 
to compensate 
for the fact that our $F_K$ and $F_\pi$ are larger by this factor relative 
their definition in \cite{Buras:2014maa}. Their numerical values are given 
 in Table~\ref{tab:input}.

In our numerical analysis 
we will use for the quark masses the values from FLAG 2013  
\cite{Aoki:2013ldr}
\be
m_s(2\gev)=(93.8\pm2.4) \mev, \qquad
m_d(2\gev)=(4.68\pm0.16)\mev.
\ee
Then at the nominal value $\mu=m_c=1.3\gev$ we have
\be
m_s(m_c)=(108.6\pm2.8) \mev, \qquad
m_d(m_c)=(5.42\pm 0.18)\mev.
\ee
Consequently for $\mu=\ord(m_c)$ a useful formula is the following one: 
\be\label{eq:Q600}
\langle Q_6(\mu) \rangle_0=-0.50 \, \left[ \frac{114\mev}{m_s(\mu) + m_d(\mu)}\right]^2 \,B_6^{(1/2)}\,\gev^3\, .
\ee

The final expressions for $Z^\prime$ contributions to $A_0$ are 
\be\label{RENPFINAL}
{\RE} A_0^{\rm NP}={{\RE} \Delta_L^{sd}}(Z^\prime) K_6(M_{Z^\prime}) \left[1.4\times 10^{-8}\gev\right], 
\ee
\be\label{IMNPFINAL}
{\IM} A_0^{\rm NP}={{\IM}\Delta_L^{sd}}(Z^\prime) K_6(M_{Z^\prime}) \left[1.4\times 10^{-8}\gev\right],
\ee
where we have defined $\mu$-independent factor
\be
K_6(M_{Z^\prime})=-r_6(\mu) \Delta_R^{qq}(Z^\prime)\, \left[ \frac{3\tev}{M_{Z^\prime}}\right]^2 \, \left[ \frac{114\mev}{m_s(\mu) + m_d(\mu)}\right]^2 \,B_6^{(1/2)}\, 
\ee
with the renormalization group factor $r_6(\mu)$ defined by
\be
C_6(\mu) =
\frac{\Delta_L^{s d}(Z^\prime)\Delta_R^{q q}(Z^\prime)}{4 M^2_{Z^\prime}} r_6(\mu).
\ee
For $\mu=1.3\gev$, as seen in (\ref{LOC5C6}), we find $r_6=1.13$.

Demanding now that $P\%$ of the experimental value of ${\rm Re}A_0$ in 
(\ref{N1}) comes from $Z^\prime$ contribution, we arrive at the condition: 
\be\label{condition1}
{\rm Re} \Delta_L^{sd}(Z^\prime) K_6(Z^\prime) = 3.9\, \left[\frac{P\%}{20\%}\right].
\ee
Evidently the couplings ${\rm Re} \Delta_L^{sd}$ and $\Delta_R^{q q}(Z^\prime)$ 
must have opposite signs and must satisfy 
\be\label{condition2}
{\rm Re} \Delta_L^{sd}(Z^\prime)\Delta_R^{q q}(Z^\prime)\left[ \frac{3\tev}{M_{Z^\prime}}\right]^2\bsi= -3.4\, \left[\frac{P\%}{20\%}\right].
\ee
We also find
\be
{\IM} A_0^{\rm NP}=\frac{\IM \Delta_L^{sd}}{\RE \Delta_L^{sd}} \left[\frac{P\%}{20\%}\right]\left[5.4\times 10^{-8}\gev\right]\,
\ee
with implications for $\epe$ which we will discuss below.

From (\ref{condition2}) we observe that for $M_{Z^\prime}\approx 3\tev$ and 
$\bsi=1.0\pm 0.25$ as expected from the large-$N$ approach,
the product $|{\rm Re} \Delta_L^{sd}(Z^\prime){\rm Re} \Delta_R^{qq}(Z^\prime)|$ must be larger than unity unless $P$ is smaller than $7$.
The strongest bounds on  ${\rm Re} \Delta_L^{sd}(Z^\prime)$ come from 
$\Delta M_K$ while the ones on ${\rm Re} \Delta_R^{qq}(Z^\prime)$ from the 
LHC. 

In what follows we will discuss first $\epe$, subsequently $\varepsilon_K$ 
and $\Delta M_K$ and finally in Section~\ref{sec:4} the constraints from the LHC.

\boldmath
\subsection{The Ratio $\epe$}
\unboldmath
\subsubsection{Preliminaries}
The ratio $\epe$  measures the size of the direct CP
violation in $K_L\to\pi\pi$ 
relative to the indirect CP violation described by $\varepsilon_K$. 
In the SM $\varepsilon^\prime$ is governed by QCD penguins but 
receives also an important destructively interfering 
 contribution from electroweak penguins that is generally much more sensitive to NP than the QCD penguin contribution. The interesting feature of NP presented 
here is that the electroweak penguin part of $\epe$ remains as in the SM and only the QCD penguin part gets modified.

The big challenge in  making predictions for $\epe$ within the SM and 
its extensions is the strong cancellation of 
QCD penguin contributions and electroweak penguin contributions to
 this ratio. In the SM QCD penguins give positive contribution, while the electroweak 
penguins  negative one.  In order to obtain useful prediction for $\epe$ in the SM the corresponding hadronic parameters
$\bsi$ and $\bei$ have to  be known with the accuracy of at least $10\%$. 
Recently significant progress has been made by RBC-UKQCD collaboration in the case of $\bei$ that is 
relevant for electroweak penguin contribution \cite{Blum:2011ng} but the calculation of $\bsi$, which will enter our analysis is even more important. There are some hopes that also this parameter could be known from lattice QCD with satisfactory precision in this decade \cite{Christ:2009ev,Christ:2013lxa}.
 
On the other hand the calculations of  short distance contributions to this ratio (Wilson coefficients of QCD and electroweak penguin operators)  within the SM have been  known already for twenty years at the NLO level \cite{Buras:1993dy,Ciuchini:1993vr} and present technology could extend them to the NNLO level if necessary. First steps in this direction have been done in \cite{Buras:1999st,Gorbahn:2004my}. As we have seen above due to the NLO calculations in \cite{Buras:2012gm} a complete NLO analysis of $\epe$ can also be performed in the NP models 
considered here.

Selected analyses of $\epe$ in various extension of the SM and its correlation 
with $\varepsilon_K$, $\kpn$ and $\klpn$ can be found in 
\cite{Buras:1998ed,Buras:1999da,Blanke:2007wr,Bauer:2009cf}.
 Useful information can also be found in 
\cite{Bertolini:1998vd,Buras:2003zz,Pich:2004ee,Cirigliano:2011ny,Bertolini:2012pu}.
\boldmath
\subsubsection{$\epe$ in the Standard Model}
\unboldmath
In the SM all QCD penguin and electroweak penguin operators in (\ref{O2})-(\ref{O5}) contribute to $\epe$. The NLO renormalization group analysis of these operators is rather involved 
\cite{Buras:1993dy,Ciuchini:1993vr} but eventually one can derive an analytic 
formula for $\epe$  \cite{Buras:2003zz} in terms of the basic one-loop functions
\begin{equation}\label{X0}
X_0(x_t)={\frac{x_t}{8}}\left[{\frac{x_t+2}{x_t-1}} 
+ {\frac{3 x_t-6}{(x_t -1)^2}}\; \ln x_t\right],
\end{equation}
\begin{equation}\label{Y0}
Y_0(x_t)={\frac{x_t}{8}}\left[{\frac{x_t -4}{x_t-1}} 
+ {\frac{3 x_t}{(x_t -1)^2}} \ln x_t\right],
\end{equation}
\bea\label{Z0}
Z_0(x_t)~\!\!\!\!&=&\!\!\!\!~-\,{\frac{1}{9}}\ln x_t + 
{\frac{18x_t^4-163x_t^3 + 259x_t^2-108x_t}{144 (x_t-1)^3}}+
\nonumber\\ 
&&\!\!\!\!+\,{\frac{32x_t^4-38x_t^3-15x_t^2+18x_t}{72(x_t-1)^4}}\ln x_t
\eea
\begin{equation}\label{E0}
E_0(x_t)=-\,{\frac{2}{3}}\ln x_t+{\frac{x_t^2(15-16x_t+4x_t^2)}{6(1-x_t)^4}}
\ln x_t+{\frac{x_t(18-11x_t-x_t^2)}{12(1-x_t)^3}} ~,
\end{equation}
where $x_t=m^2_t/M_W^2$.

The updated 
version of this formula used in the present paper is given as follows
\be 
\left(\frac{\varepsilon'}{\varepsilon}\right)_{\rm SM}= a\,\IM\lambda_t
\cdot F_{\varepsilon'}(x_t)
\label{epeth}
\ee
where $a=0.92\pm0.03$ represents the correction coming from $\Delta I=5/2$ 
transitions \cite{Cirigliano:2003nn} that has not been included in 
\cite{Buras:2003zz}. Next
\be
F_{\varepsilon'}(x_t) =P_0 + P_X \, X_0(x_t) + 
P_Y \, Y_0(x_t) + P_Z \, Z_0(x_t)+ P_E \, E_0(x_t)~,
\label{FE}
\ee
with the first term dominated by QCD-penguin contributions, the next three 
terms by electroweak penguin contributions and the last term being 
totally negligible. 
The coefficients $P_i$ are given in terms of the non-perturbative parameters
$R_6$ and $R_8$ defined in (\ref{RS}) as follows:
\begin{equation}
P_i = r_i^{(0)} + 
r_i^{(6)} R_6 + r_i^{(8)} R_8 \,.
\label{eq:pbePi}
\end{equation}
The coefficients $r_i^{(0)}$, $r_i^{(6)}$ and $r_i^{(8)}$ comprise
information on the Wilson-coefficient functions of the $\Delta S=1$ weak
effective Hamiltonian at the NLO. Their numerical values extracted from  \cite{Buras:2003zz} are given in the NDR renormalization scheme for $\mu=m_c$ and three  values of $\alpha_s(M_Z)$
in Table~\ref{tab:pbendr}\footnote{We thank Matthias Jamin for providing this table for the most recent values of $\alpha_s(M_Z)$.}. While other values of 
$\mu$ could be considered the procedure for finding the coefficients 
$r_i^{(0)}$, $r_i^{(6)}$ and $r_i^{(8)}$ is most straight forward at $\mu=m_c$.

The details on the procedure in question can be found in  \cite{Buras:1993dy,Buras:2003zz}.
In particular in obtaining the numerical values in  Table~\ref{tab:pbendr}
 the experimental value for 
${\RE} A_2$ has been imposed to determine  hadronic 
matrix elements of subleading electroweak penguin operators ($Q_9$ and $Q_{10}$). The matrix elements of $(V-A)\otimes(V-A)$ penguin operators have been 
bounded by relating them to the matrix elements $\langle Q_{1,2}\rangle_0$ 
that govern the octet enhancement of ${\RE} A_0$.
Moreover, as $\epe$ involves ${\RE} A_0$ also 
this amplitude has been taken from experiment. This procedure can also 
be used in $Z^\prime$ models as here experimental value of  ${\RE} A_0$ will 
constitute an important constraint and the contributions of operators 
$Q_9$ and $Q_{10}$ are unaffected by new $Z^\prime$ contributions up to 
tiny $\ord(\alpha)$  effects from mixing with the operator $Q_6$. 

The dominant  dependence on the hadronic matrix elements in $\epe$ resides in 
 the QCD-penguin operator $Q_6$ and the  electroweak penguin operator $Q_8$. 
Indeed from Table~\ref{tab:pbendr} we find that
the largest are the coefficients $r_0^{(6)}$ and $r_Z^{(8)}$ representing QCD-penguin and electroweak penguin contributions, respectively.
The fact that these coefficients are of  similar size but having opposite 
signs has been a problem since the end of 1980s when the electroweak penguin contribution  increased  in importance due to the large 
top-quark mass \cite{Flynn:1989iu,Buchalla:1989we}.

\begin{table}[!tb]
\begin{center}
\begin{tabular}{|c||c|c|c||c|c|c||c|c|c|}
\hline
&\multicolumn{3}{c||}{$\alpha_s(M_Z)= 0.1179$}&
 \multicolumn{3}{c||}{$\alpha_s(M_Z)= 0.1185$} &
  \multicolumn{3}{c|}{$\alpha_s(M_Z)= 0.1191$} \\
  \hline
$i$ & $r_i^{(0)}$ & $r_i^{(6)}$ & $r_i^{(8)}$ &
      $r_i^{(0)}$ & $r_i^{(6)}$ & $r_i^{(8)}$ &
       $r_i^{(0)}$ & $r_i^{(6)}$ & $r_i^{(8)}$ \\
      \hline
0 &--3.572 &  16.424 &   1.818 &
   --3.580 &  16.801 &   1.782 &
   --3.588 &  17.192 &   1.744 \\
$X_0$ &
      0.575 &   0.029 &       0 &
     0.572 &   0.030 &       0 &
     0.569 &   0.031 &       0 \\   
$Y_0$ &
       0.405 &   0.119 &       0 &
     0.401 &   0.121 &       0 &
     0.398 &   0.123 &       0 \\ 
$Z_0$ &
       0.709 &  --0.022 &  --12.447 &
     0.724 &  --0.023 &  --12.631 &
     0.739 &  --0.023 &  --12.822 \\
$E_0$ &
      0.215 &  --1.898 &   0.546 &
     0.211 &  --1.929 &   0.557 &
     0.208 &  --1.961 &   0.568 \\
\hline
\end{tabular}
\end{center}
\caption{\it The coefficients $r_i^{(0)}$, $r_i^{(6)}$ and $r_i^{(8)}$ of
formula (\ref{eq:pbePi}) in the NDR scheme for three values of $\alpha_s(M_Z)$.
\label{tab:pbendr}}~\\[-2mm]\hrule
\end{table}

The parameters
$R_6$ and $R_8$ are directly related to the parameters $\bsi$ and $\bei$ 
representing the hadronic matrix elements of $Q_6$ and $Q_8$, respectively. 
They are defined as
\be\label{RS}
R_6\equiv 1.13\,\bsi\left[ \frac{114\mev}{m_s(m_c)+m_d(m_c)} \right]^2,
\qquad
R_8\equiv 1.13\,\bei\left[ \frac{114\mev}{m_s(m_c)+m_d(m_c)} \right]^2,
\ee
where the factor $1.13$ signals the decrease of the value of $m_s$ since 
the analysis in  \cite{Buras:2003zz} has been done.

There is no reliable result on $\bsi$ from lattice QCD. On the other 
hand  one can  extract the lattice value for $\bei$ from 
\cite{Blum:2012uk}.
We find 
\be
\bei(3\gev)= 0.65\pm 0.05 \qquad {\rm (lattice)}.
\ee
As $\bei$ depends very weakly on the renormalization scale \cite{Buras:1993dy}, 
the same value can be used at $\mu=m_c$. In the absence of the value 
for $\bsi$ from lattice, we will investigate how the result on $\epe$ 
changes when $\bsi$ is varied within $25\%$ from its large $N$ value 
$\bsi=1$ \cite{Bardeen:1986uz}. Similar to $\bei$, the parameter $\bsi$ exhibits very weak $\mu$ dependence \cite{Buras:1993dy}.

\boldmath
\subsubsection{$Z^\prime$ Contribution to $\epe$}
\unboldmath
We will next present  $Z^\prime$ contributions to $\epe$. A straight forward 
calculation gives 
\be\label{eprimeZprime}
\left(\frac{\varepsilon'}{\varepsilon}\right)_{Z^\prime}=
-\frac{{\IM} A_0^{\rm NP}}{{\RE}A_0}\left[\frac{\omega_+}{|\varepsilon_K|\sqrt{2}}\right](1-\Omega_{\rm eff}),
\ee
where  \cite{Cirigliano:2003nn}
\be
\omega_+=a\frac{{\RE} A_2}{{\RE}A_0}=(4.1\pm0.1)\times 10^{-2}, \qquad \Omega_{\rm eff}=(6.0\pm7.7)\times 10^{-2}.
\ee
In order to obtain the first number we set $a=0.92\pm0.02$ and as in the case of the SM we use  the experimental values for ${\rm Re} A_0$ and  ${\rm Re} A_2$ in (\ref{N1}). 
Also the experimental values for $|\varepsilon_K|$  and $\RE A_0$ should be used in (\ref{eprimeZprime}).

The final expression for $\epe$ is given by 
\be\label{epsifinal}
\left(\frac{\varepsilon'}{\varepsilon}\right)_{\rm tot}=\left(\frac{\varepsilon'}{\varepsilon}\right)_{\rm SM}+\left(\frac{\varepsilon'}{\varepsilon}\right)_{Z^\prime}
\ee

\boldmath
\subsubsection{Correlation between $Z^\prime$ Contributions to $\epe$ and  ${\rm Re}A_0$}
\unboldmath
In our favourite scenarios only 
the couplings  $\Delta_L^{sd}(Z^\prime)$, $\Delta_R^{qq}(Z^\prime)$ and the operator $Q_6$ 
will be relevant in $K\to\pi\pi$ decays. In this case the expressions presented above allow
to derive the relation 
\be\label{Basicrelation}
\left(\frac{\varepsilon'}{\varepsilon}\right)_{Z^\prime}= -12.3
\left[\frac{{\rm Re}A^{\rm NP}_0}{{\rm Re}A_0}\right]
\left[\frac{{\IM \Delta_L^{sd}(Z^\prime)}}{{\RE \Delta_L^{sd}(Z^\prime)}}\right]=
-2.5\, \left[\frac{P\%}{20\%}\right] \left[\frac{{\IM \Delta_L^{sd}(Z^\prime)}}{{\RE \Delta_L^{sd}(Z^\prime)}}\right]
\ee
which is free from the uncertainties in the CKM matrix and  
$\langle Q_6\rangle_0$. But the most important message that follows from this relation is that
\be\label{epsicon}
\left[\frac{{\IM \Delta_L^{sd}(Z^\prime)}}{{\RE \Delta_L^{sd}(Z^\prime)}}\right]=\ord(10^{-4})
\ee
if we want to obtain $20\%$ shift in ${\rm Re}A_0$  and simultaneously be 
consistent with the data on $\epe$. This also implies that $Z^\prime$ contributions to $\varepsilon_K$ and $\klpn$ which require complex CP-violating phases 
will be easier to keep under control than it is the case of $\Delta M_K$ and $\kpn$ which are CP conserving. In order to put these expectations on a 
firm footing we have to discuss now $\varepsilon_K$, $\Delta M_K$ and $K\to\pi\nu\bar\nu$.

\boldmath
\section{Constraints from $\varepsilon_K$, $\Delta M_K$ and $K\to\pi\nu\bar\nu$}\label{sec:3}
\unboldmath
\boldmath
\subsection{$\varepsilon_K$ and $\Delta M_K$}
\unboldmath
In the models in question we have
\be
\Delta M_K=(\Delta M_K)_\text{SM}+\Delta M_K(Z^\prime),\qquad 
\varepsilon_K=(\varepsilon_K)_\text{SM}+\varepsilon_K(Z^\prime)
\ee
and  similar for $G^\prime$. 
A very detailed analysis of these observables in a general $Z^\prime$ 
model with $\Delta_L^{s d}(Z^\prime)$ and  $\Delta_R^{s d}(Z^\prime)$ 
couplings in  LHS, RHS, LRS and ALRS scenarios has been presented in \cite{Buras:2012jb}. We will not repeat the relevant formulae for $\varepsilon_K$ and $\Delta M_K$ which can be found there. Still it is useful to recall the 
operators contributing in the general case. These are:
\begin{align}\label{equ:operatorsZ}
\begin{split}
{Q}_1^\text{VLL}=\left(\bar s\gamma_\mu P_L d\right)\left(\bar s\gamma^\mu P_L d\right)\,, & \qquad
{Q}_1^\text{VRR}=\left(\bar s\gamma_\mu P_R d\right)\left(\bar s\gamma^\mu P_R d\right)\,,\end{split}\\
\begin{split}
{Q}_1^\text{LR}=\left(\bar s\gamma_\mu P_L d\right)\left(\bar s\gamma^\mu P_R d\right)\,,& \qquad
{Q}_2^\text{LR}=\left(\bar s P_L d\right)\left(\bar s P_R d\right)\,,
\end{split}
 \end{align}
where $P_{R,L}=(1\pm\gamma_5)/2$ and 
we suppressed colour indices as they are summed up in each factor. For instance $\bar s\gamma_\mu P_L d$ stands for $\bar s_\alpha\gamma_\mu P_L d_\alpha$ and similarly for other factors. In the SM only ${Q}_1^\text{VLL}$ is present. 
 This operator basis applies also to $G^\prime$ but the Wilson coefficients 
of these operators at $\mu=M_{G^\prime}$ will be different as we will see in 
Section~\ref{sec:5}.

If only the Wilson coefficient of the operator  ${Q}_1^\text{VLL}$ is 
affected by $Z^\prime$ contributions, as is the case of the LHS scenario, then 
 NP effects in $\varepsilon_K$ and $\Delta M_K$ can be summarized by the modification of the one-loop function $S$:
\be
S(K)=S_0(x_t)+\Delta S(K)
\ee
with the SM contribution represented by 
\begin{align}
S_0(x_t)  = \frac{4x_t - 11 x_t^2 + x_t^3}{4(1-x_t)^2}-\frac{3 x_t^2\log x_t}{2
(1-x_t)^3} = 2.31 \left[\frac{m_t(m_t)}{163\gev}\right]^{1.52} 
\end{align}
and
the one from $Z^\prime$ by
\be\label{Zprime1}
\Delta S(K)=
\left[\frac{\Delta_L^{sd}(Z^\prime)}{\lambda_t}\right]^2
\frac{4\tilde r}{M^2_{Z^\prime}g_{\text{SM}}^2},\qquad
g_{\text{SM}}^2=4\frac{G_F}{\sqrt 2}\frac{\alpha}{2\pi\sin^2\theta_W}=1.781\times10^{-7}\,\gev^{-2} .
\ee
Here $\tilde r$ is a QCD factor calculated in \cite{Buras:2012dp} at the NLO 
level. 
 One finds $\tilde r=0.965$, $\tilde r=0.953$ and $\tilde r = 0.925$ for $M_{Z^\prime} =2,~3, ~10\tev$, respectively.  Neglecting logarithmic scale dependence of $\tilde r$ we find then
\be\label{Zprime1VLL}
\Delta S(K)=2.4
\left[\frac{\Delta_L^{sd}(Z^\prime)}{\lambda_t}\right]^2
\left[\frac{3\tev}{M_{Z^\prime}}\right]^2.
\ee
For $\Delta^{sd}_L(Z^\prime)$ with a small phase, as in (\ref{epsicon}), one can still 
satisfy the $\varepsilon_K$ constraint but if we want to  explain $30\%$ 
of ${\rm Re} A_0$ the bound from $\Delta M_K$ is violated by several orders 
of magnitude. Indeed allowing conservatively that NP contribution is at most as large as the short distance SM contribution to $\Delta M_K$ we find the bound 
on a real $\Delta^{sd}_L(Z^\prime)$ 
\be\label{DeltaMKbound}
|\Delta^{sd}_L(Z^\prime)|\le 0.65\vus \sqrt{\frac{\eta_{cc}}{\eta_{tt}}}\frac{m_c}{M_W} \left[\frac{M_{Z^\prime}}{3\tev}\right]= 0.004\, \left[\frac{M_{Z^\prime}}{3\tev}\right].
\ee
This bound, as seen in (\ref{condition1}), does not allow any significant contribution to  ${\rm Re} A_0$ unless the coupling 
$\Delta_R^{qq}$ and or $\bsi$ are very large. We also note that the increase 
of $M_{Z^\prime}$ makes the situation even worse because the required value
of ${\rm Re} \Delta^{sd}_L(Z^\prime)$   by the condition (\ref{condition1}) grows quadratically 
with $M_{Z^\prime}$, whereas this mass enters only linearly in (\ref{DeltaMKbound}).  Evidently the LHS scenario does not provide any
relevant NP contribution to  ${\rm Re} A_0$ when the constraint from 
$\Delta M_K$ is imposed. On the other hand in this 
scenario still interesting results for $\epe$, $\kpn$ and $\klpn$ can be obtained. 

In order to remove
the incompatibility of ${\rm Re} A_0$ and $\Delta M_K$ constraints we have to
suppress somehow $Z^\prime$ contribution to  $\Delta M_K$ in the presence 
of a coupling $\Delta^{sd}_L(Z^\prime)$ that is sufficiently large so that 
the contribution of $Z^\prime$ to  ${\rm Re} A_0$ is relevant. To this end 
we introduce an effective $[\Delta^{sd}_L(Z^\prime)]_{\rm eff}$ to be used 
only in $\Delta S=2$ transitions and given by 
\be
[\Delta^{sd}_L(Z^\prime)]_{\rm eff}=\Delta^{sd}_L(Z^\prime)\delta
\ee
with $\Delta^{sd}_L(Z^\prime)$ still denoting the coupling used for the evaluation of ${\rm Re} A_0$ and  $\delta$ a suppression factor.
  We do not care about the sign of $\Delta^{sd}_L(Z^\prime)$  which
can be adjusted by the sign of $\Delta_R^{qq}(Z^\prime)$. Imposing then the 
constraint (\ref{condition1})  but demanding that simultaneously (\ref{DeltaMKbound}) is satisfied with $\Delta^{sd}_L(Z^\prime)$ replaced by $[\Delta^{sd}_L(Z^\prime)]_{\rm eff}$ 
 we find 
that the required $\delta$ is given as follows:
\be\label{delta}
\delta=\left[\frac{r_6(m_c)}{1.13}\right] \Delta_R^{qq}(Z^\prime)\, \left[ \frac{3\tev}{M_{Z^\prime}}\right]B_6^{(1/2)}\left[\frac{20\%}{P\%}\right]\, 10^{-3}\,.
\ee
Here we neglected the small uncertainty in the quark masses. Evidently, 
increasing simultaneously  $\Delta_R^{qq}(Z^\prime)$ and $\bsi$ above unity, 
decreasing $M_{Z^\prime}$ below $3\tev$ and $P$ below $20\%$ can increase $\delta$  but then one has to check other constraints, in particular from the LHC. 
We will study this issue below.

Such a small $\delta$ can be generated in the presence of flavour-violating right-handed couplings in addition to the 
left-handed ones. In this case at 
 NLO the values of the Wilson coefficients of $\Delta S=2$  operators at $\mu=M_{Z^\prime}$ generated through $Z^\prime$ tree level exchange 
are given in the NDR scheme as follows \cite{Buras:2012fs}
\begin{align}
C_1^\text{VLL}(M_{Z^\prime})&=\frac{(\Delta_L^{sd}(Z^\prime))^2}{2M_{Z^\prime}^2}\left(1+\frac{11}{3}\frac{\alpha_s(M_{Z^\prime})}{4\pi}\right),\\
C_1^\text{VRR}(M_{Z^\prime}) & =\frac{(\Delta_R^{sd}(Z^\prime))^2}{2M_{Z^\prime}^2}\left(1+\frac{11}{3}\frac{\alpha_s(M_{Z^\prime})}{4\pi}\right),\\
 C_1^\text{LR}(M_{Z^\prime}) & =\frac{\Delta_L^{sd}(Z^\prime)\Delta_R^{sd}(Z^\prime)}{
 M_{Z^\prime}^2}\left(1-\frac{1}{6}\frac{\alpha_s(M_{Z^\prime})}{4\pi}\right), \\
C_2^\text{LR}(M_{Z^\prime})&= -\frac{\Delta_L^{sd}(Z^\prime)\Delta_R^{sd}(Z^\prime)}{
 M_{Z^\prime}^2}\frac{\alpha_s(M_{Z^\prime})}{4\pi} \,.
\end{align}
The information about hadronic  matrix elements of these operators calculated by various lattice QCD collaborations is given in  the review \cite{Buras:2013ooa}.  

Now, it is known that similar to $Q_6$ and $Q_6^\prime$, the LR operators have 
 in the case of $K$ meson system chirally enhanced matrix 
elements over those of VLL and VRR operators and as LR operators have also large anomalous dimensions, their contributions to $\varepsilon_K$ and $\Delta M_K$ dominate NP contributions in LRS and ALRS scenarios, while they are absent in LHS and RHS scenarios.

In order to see how the problem with $\Delta M_K$ is solved in this case 
we calculate  $\Delta M_K$  in a general case assuming for simplicity that the 
couplings $\Delta_{L,R}(Z^\prime)$ are real. We find
\be\label{DMZnew}
 \Delta M_K(Z^\prime)   =
 \frac{(\Delta_L^{sd}(Z^\prime))^2}{M_{Z^\prime}^2} \langle \hat Q_1^\text{VLL}(M_{Z^\prime})\rangle \left[1+\left(\frac{\Delta_R^{sd}(Z^\prime)}{\Delta_L^{sd}(Z^\prime)}\right)^2+2\left(\frac{\Delta_R^{sd}(Z^\prime)}{\Delta_L^{sd}(Z^\prime)}\right) 
\frac{\langle \hat Q_1^\text{LR}(M_{Z^\prime})\rangle}{\langle \hat Q_1^\text{VLL}(M_{Z^\prime})\rangle}\right],
 \ee
where using the technology in \cite{Buras:2001ra,Buras:2012fs} we have expressed the final result in terms of the renormalization scheme independent 
matrix elements
\be
\langle\hat Q_1^\text{VLL}(M_{Z^\prime})\rangle= \langle Q_1^\text{VLL}(M_{Z^\prime})\rangle\left(1+\frac{11}{3}\frac{\alpha_s(M_{Z^\prime})}{4\pi}\right)
\ee
\be
\langle \hat Q_1^\text{LR}(M_{Z^\prime})\rangle=\langle  Q_1^\text{LR}(M_{Z^\prime})\rangle\left(1-\frac{1}{6}\frac{\alpha_s(M_{Z^\prime})}{4\pi}\right) -\frac{\alpha_s(M_{Z^\prime})}{4\pi}\langle Q_2^\text{LR}(M_{Z^\prime})\rangle\,.
\ee
Here   $\langle Q_1^\text{VLL}(M_{Z^\prime})\rangle$ and $\langle Q_{1,2}^\text{LR}(M_{Z^\prime})\rangle$  are the matrix elements evaluated at $\mu=M_{Z^\prime}$ in 
the NDR scheme and the presence of $\ord(\alpha_s)$ corrections removes 
the scheme dependence.

 But in the case of $K^0-\bar K^0$ matrix elements for $\mu=M_{Z^\prime}=3\tev$
\be
 \langle \hat Q^\text{VLL}(M_{Z^\prime})\rangle >0, \quad  
\langle \hat Q_1^\text{LR}(M_{Z^\prime})\rangle<0, \quad 
 |\langle \hat Q_1^\text{LR}(M_{Z^\prime})\rangle |\approx 97 \, |\langle \hat Q^\text{VLL}(M_{Z^\prime})\rangle|\,.
\ee
The signs are independent of the scale $\mu=M_{Z^\prime}$ but the numerical 
factor in the last relation increases logarithmically with this scale.
Consequently in LR and ALR scenarios the last term in (\ref{DMZnew}) 
dominates so that the problem with $\Delta M_K$ is even worse. We conclude 
therefore that in LHS, RHS, LRS and ALRS scenarios analyzed in our previous 
papers \cite{Buras:2012sd,Buras:2012jb,Buras:2012dp,Buras:2013uqa,Buras:2013rqa,Buras:2013raa,Buras:2013qja,Buras:2013dea}, the problem in question remains.

On the other hand we note that for a non-vanishing but small $\Delta_R^{sd}(Z^\prime)$ coupling 
\be\label{deltasupp}
 \delta=\left[1+\left(\frac{\Delta_R^{sd}(Z^\prime)}{\Delta_L^{sd}(Z^\prime)}\right)^2+2\left(\frac{\Delta_R^{sd}(Z^\prime)}{\Delta_L^{sd}(Z^\prime)}\right) 
\frac{\langle \hat Q_1^\text{LR}(M_{Z^\prime})\rangle}{\langle \hat Q_1^\text{VLL}(M_{Z^\prime})\rangle}\right]^{1/2} \, ,
 \ee
can be made very small and 
$Z^\prime$ contribution to $\Delta M_K$ and also $\varepsilon_K$ can be 
suppressed sufficiently and even totally eliminated.

In order to generate a non-vanishing $\Delta_R^{sd}(Z^\prime)$ in the mass eigenstate basis the exact flavour universality has to be violated generating 
a small contribution to  ${\rm Re} A_2$ but in view of the required size of 
$\Delta_R^{sd}(Z^\prime)=\ord(10^{-3})$ this effect can be neglected. Thus 
the presence of a small $\Delta_R^{sd}(Z^\prime)$ coupling has basically no impact on $K\to\pi\pi$ decays and serves only to avoid the problem with $\Delta M_K$  which we found in the LHS scenario. Even if this 
solution appears at first sight to be fine-tuned, its existence is interesting.
 Therefore we will analyze it numerically  below for a $Z^\prime$ in 
 a toy model for the coupling  $\Delta_R^{sd}(Z^\prime)$ which satisfies 
(\ref{deltasupp}) but allows for a non-vanishing $\delta$. The case of $G^\prime$ will be analyzed in Section~\ref{sec:5}.

\boldmath
\subsection{$\kpn$ and $\klpn$}
\unboldmath
A very detailed analysis of these decays in  a general $Z^\prime$ 
model with $\Delta_L^{s d}(Z^\prime)$ and  $\Delta_R^{s d}(Z^\prime)$ couplings
in various combinations has been presented in \cite{Buras:2012jb} and we 
will use the formulae of that paper. Still it is useful to recall the 
expression for the shift caused by $Z^\prime$ tree-level exchanges in the relevant function $X(K)$. One has now
\be
X(K)=X_0(x_t)+ \Delta X(K)
\ee
with $X_0(x_t)$ given in (\ref{X0}) and $Z^\prime$ contribution by
\be\label{XLK}
\Delta X(K)=\left[\frac{\Delta_L^{\nu\nu}(Z')}{g^2_{\rm SM}M_{Z'}^2}\right]
              \frac{\left[\Delta_L^{sd}(Z')+\Delta_R^{sd}(Z')\right]}{\lambda_t}.
\ee

We note that  in addition to the $\Delta_{L,R}^{s d}(Z^\prime)$ couplings 
that will be constrained by the $\Delta S=2$  observables as discussed above, 
also the unknown coupling $\Delta_L^{\nu\nu}(Z^\prime)$ will be involved 
and consequently it will not be possible to make definite predictions for 
the branching ratios for these decays.  However, it will be possible to learn something about the correlation between them. Evidently in the presence of a large 
$\Delta_L^{sd}(Z^\prime)$ coupling the present bounds on $K\to\pi\nu\bar\nu$ 
branching ratios can be avoided by choosing sufficiently low value of 
$\Delta_L^{\nu\bar\nu}(Z')$. In the case of Scenario B, in which 
we ignore the $\Delta I=1/2$ rule issue and work only with left-handed 
$Z^\prime$-couplings, $\Delta_L^{sd}(Z^\prime)$ is forced to be small by 
$\varepsilon_K$ and $\Delta M_K$ constraints so that $\Delta_L^{\nu\bar\nu}(Z')$ 
can be chosen to be $\ord(1)$.
\subsection{A Toy Model}
There is an interesting aspect of the possible contribution of a 
$Z^\prime$ to the $\Delta I=1/2$ rule in the case in which the suppression factor $\delta$ does not vanish. One can relate the physics responsible for the 
missing piece in ${\rm Re} A_0$ to the one in $\epe$, $\varepsilon_K$, $\Delta M_K$ and rare decays $\kpn$ and $\klpn$ and consequently obtain correlations 
between the related observables.

In order to illustrate this we consider a model for the $\Delta_R^{sd}(Z^\prime)$
coupling: 
\be\label{tuning}
\frac{\Delta_R^{sd}(Z^\prime)}{\Delta_L^{sd}(Z^\prime)}=
-\frac{1}{2} R_Q(1+h R^2_Q), \qquad  R_Q\equiv\frac{\langle \hat Q_1^\text{VLL}((M_{Z^\prime})\rangle}{\langle \hat Q_1^\text{LR}((M_{Z^\prime})\rangle}\approx -0.01
\ee
where $h=\ord(1)$. This implies
\be
\delta=\frac{1}{2} R_Q(1-4 h)^{1/2} +\ord(R_Q^2)
\ee
which shows that by a proper choice of the parameter $h$ one can suppress NP 
contributions to $\Delta M_K$ to the level that it agrees with experiment.

In this model we find 
\be\label{epsilonZprime}
\varepsilon_K(Z^\prime)=-\frac{\kappa_\eps e^{i\varphi_\eps}}{\sqrt{2}(\Delta M_K)_\text{exp}}\frac{({\rm Re}\Delta^{sd}_L)({\rm Im}\Delta^{sd}_L)}{ M_{Z^\prime}^2}
\langle \hat Q_1^\text{VLL}((M_{Z^\prime})\rangle\delta^2\equiv\tilde\varepsilon_K(Z^\prime)e^{i\varphi_\eps},
\ee

\be\label{Delta MKprime}
\Delta M_K(Z^\prime)=\frac{({\rm Re}\Delta^{sd}_L)^2}{M_{Z^\prime}^2}\langle \hat Q_1^\text{VLL}((M_{Z^\prime})\rangle\delta^2,
\ee
where $\varphi_\eps = (43.51\pm0.05)^\circ$ and $\kappa_\eps=0.94\pm0.02$ \cite{Buras:2008nn,Buras:2010pza} takes into account 
that $\varphi_\eps\ne \tfrac{\pi}{4}$ and includes long distance effects in $\IM( \Gamma_{12})$ and $\IM (M_{12})$.
The shift in the function $X(K)$ is in view of (\ref{tuning}) given 
by 
\be\label{XLK1}
\Delta X(K)=\left[\frac{\Delta_L^{\nu\bar\nu}(Z')}{g^2_{\rm SM}M_{Z'}^2}\right]
              \frac{\left[\Delta_L^{sd}(Z')\right]}{\lambda_t}.
\ee

While the $\delta$ is at this stage not fixed, it will be required to 
be non-vanishing in case SM predictions for $\varepsilon_K$ and 
$\Delta M_K$ will disagree with data  once the parametric and hadronic 
uncertainties will be reduced.  Moreover independently of $\delta$, 
as long as it is non-vanishing these formulae together with 
(\ref{Basicrelation}) imply correlations
\be\label{Brelation2}
\tilde\varepsilon_K(Z^\prime)=
-\frac{\kappa_\eps}{\sqrt{2} r_{\Delta M} }\left[\frac{{\IM \Delta_L^{sd}(Z^\prime)}}{{\RE \Delta_L^{sd}(Z^\prime)}}\right], \qquad r_{\Delta M}=\left[\frac{(\Delta M_K)_\text{exp}}{\Delta M_K(Z^\prime)}\right]\,,
\ee

\be\label{Basicrelation1}
\left(\frac{\varepsilon'}{\varepsilon}\right)_{Z^\prime}= \frac{3.5}{\kappa_\eps}\, \tilde\varepsilon_K(Z^\prime) \left[\frac{P\%}{20\%}\right] r_{\Delta M}\,.
\ee

Already without a detail numerical analysis we note the following 
general properties of this model:
\begin{itemize}
\item
$\Delta M_K(Z^\prime)$ is strictly positive.
\item
As $P$ is also positive $\epe$ and $\varepsilon_K$ are correlated with each other. Therefore this scenario can only work if the SM predictions for both observables are either below or above the data.
\item 
The ratio of NP contributions to $\epe$ and $\varepsilon_K$ depends only on  
the product of $P$ and  $r_{\Delta M}$.
\item
For $P=20\pm 10$, NP contribution to $\epe$ is predicted to be by an 
order of magnitude larger than in $\varepsilon_K$.  This tells us that 
in order for $Z^\prime$ contribution to be relevant for the $\Delta I=1/2$ 
rule and simultaneously be consistent with the data on $\epe$, its 
contribution to $\varepsilon_K$ must be small implying that the SM 
value for $\varepsilon_K$ must be close to the data.
\end{itemize}

The correlations in (\ref{Brelation2}) and (\ref{Basicrelation1}) together 
with the condition (\ref{condition2}) allow to test this NP scenario in a 
straight forward manner as follows:

\paragraph{Step 1}

We will set $r_{\Delta M}=4$, implying that $Z^\prime$ contributes $25\%$ 
of the measured value of $\Delta M_K$. In view of a large uncertainty in $\eta_{cc}$ and consequently in $(\Delta M_K)_\text{SM}$  this value is plausible 
and used here only to illustrate the general structure of what is 
going on. In this manner (\ref{Basicrelation1})  gives us the 
relation between NP contributions to $\varepsilon_K$ and $\epe$. Note that 
this relation does not involve $\bsi$ and only $P$. But the SM contribution 
to $\epe$ involves explicitly $\bsi$. Therefore the correlation of the resulting total $\epe$ and $\varepsilon_K$ will depend on the values of $P$ and $\bsi$ 
as well as CKM parameters. Note that to obtain these results it was not necessary to specify the value of $\Delta_L^{sd}(Z^\prime)$.  But already this step will tell us 
which combination of $P$ and $\bsi$ are simultaneously consistent with 
data on $\epe$ and $\varepsilon_K$.

\paragraph{Step 2}

In order to find $\Delta_L^{sd}(Z^\prime)$ and to test whether the results of Step 1 
are consistent with the LHC data, we use condition (\ref{condition2}). 
As we will see below LHC implies an upper bound on $\Delta_R^{qq}(Z^\prime)$ 
as a function of $M_{Z^\prime}$. For fixed $M_{Z^\prime}$ setting 
 $\Delta_R^{qq}(Z^\prime)$ at a value consistent with this bound allows to determine  the minimal value of ${\rm Re} \Delta_L^{sd}(Z^\prime)$ as a function of $P$ and  $\bsi$. 
 Combining finally these results in Section~\ref{LHCZprime}  with the bound on ${\rm Re} \Delta_L^{sd}(Z^\prime)$  from the LHC we will 
finally be able 
to find out what are the maximal values of $P$ consistent with all available 
constraints and this will also restric the values of $\bsi$.

 Having  ${\rm Re} \Delta_L^{sd}(Z^\prime)$ as a function of of 
$P$, $\bsi$ and $\Delta_R^{qq}(Z^\prime)$,
 we can next use the relation  (\ref{Brelation2}) to calculate ${\rm Im} \Delta_L^{sd}(Z^\prime)$ as a function of 
$\tilde\varepsilon_K(Z^\prime)$.  We will then find that only a certain 
range of the values of ${\rm Im} \Delta_L^{sd}(Z^\prime)$ is consistent with 
the data on $\varepsilon_K$ and $\epe$ and this range depends on 
$P$, $\bsi$ and $\Delta_R^{qq}(Z^\prime)$.

\paragraph{Step 3}

With this information on the allowed values of the coupling $\Delta_L^{sd}(Z^\prime)$  we can find  correlation between the branching ratios for $\kpn$ and $\klpn$ and the correlation between these two branching ratios and $\epe$. To this end  $\Delta_L^{\nu\nu}(Z^\prime)$  has to be suitably chosen.

\subsection{Scaling Laws in the  Toy Model}\label{scalings}
 While the outcome of this procedure  depends on the assumed value of 
$r_{\Delta M}$, the relations (\ref{Brelation2}) and (\ref{Basicrelation1})
 allow to find out what happens for different values of $r_{\Delta M}$. To this 
end let us note the following facts.

The correlation between NP contributions to $\epe$ and $\varepsilon_K$ in 
(\ref{Basicrelation1}) depends only on the product of 
$P$ and $r_{\Delta M}$. But one should remember that 
the full results for $\epe$ and $\varepsilon_K$ that include also SM 
contributions depend on the scenario $a)-f)$ for CKM parameters considered in 
Section~\ref{sec:4}
and on $\bsi$, present explicitly in the SM contribution.
In a given CKM scenario there is a specific room left for NP contribution to 
$\varepsilon_K$ which restricts the allowed range for $\tilde\varepsilon_K$, 
which dependently on scenario considered could be negative or positive. Thus 
dependently on $P$, $\bsi$ and the CKM scenario $a)-f)$, one can adjust $r_{\Delta M}$ to satisfy simultaneously the data on $\epe$ and $\varepsilon_K$. But 
as $r_{\Delta M}$ is predicted in the model considered to be positive and 
long distance contributions, at least within the large $N$ approach  \cite{Buras:2014maa}, although small, are also predicted to be positive, $r_{\Delta M}$ cannot be too small.

Once the agreement on  $\epe$ and $\varepsilon_K$ is achieved it is crucial 
to verify whether the selected values of $P$ and $\bsi$ are consistent 
with the LHC bounds on the couplings ${\rm Re}\Delta_L^{sd}(Z^\prime)$ and 
$\Delta_R^{q q}(Z^\prime)$ which are related to  $P$ and $\bsi$ through the 
relation (\ref{condition2}). The numerical factor $-3.4$ in this equation 
valid for $Z^\prime$  is as seen in 
(\ref{condition4}) modified to $-2.4$ in the case of $G^\prime$. Otherwise the correlations 
between $\epe$, $\varepsilon_K$ and $r_{\Delta M}$ given above 
are valid also for $G^\prime$,
 although the bounds on ${\rm Re}\Delta_L^{sd}(G^\prime)$ and 
$\Delta_R^{q q}(G^\prime)$ from the LHC differ from $Z^\prime$ case as 
we will see in  Section~\ref{GprimeNum}.

\begin{figure}[!tb]
 \centering
\includegraphics[width = 0.45\textwidth]{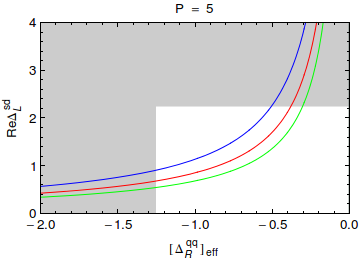}
\includegraphics[width = 0.45\textwidth]{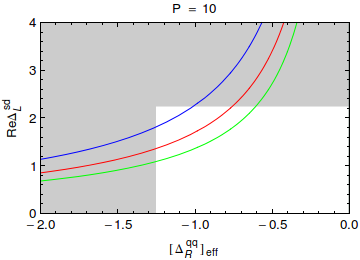}
\includegraphics[width = 0.45\textwidth]{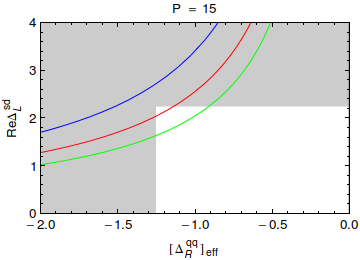}
\includegraphics[width = 0.45\textwidth]{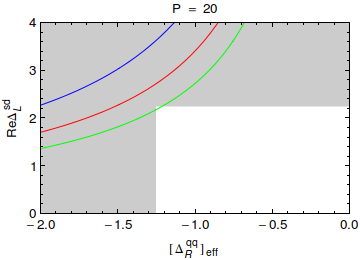}
\caption{\it  ${\rm Re} \Delta_L^{sd}(Z^\prime)$ versus $|[\Delta_R^{q q}(Z^\prime)]_\text{eff}|$ for $P = 5,~10,~15,~20$ and $\bsi = 0.75$ 
(blue), $1.00$ (red) and $1.25$ (green). The gray area is basically excluded by the LHC. See Section~\ref{LHCZprime}.
}\label{fig:ReDLvsDR}~\\[-2mm]\hrule
\end{figure}

In order to be prepared for the improvement of the LHC bounds in question 
we define
\be\label{Deltaeff}
[\Delta_R^{q q}(Z^\prime)]_\text{eff}=\Delta_R^{q q}(Z^\prime)\left[\frac{3\tev}{M_{Z^\prime}}\right]^2.
\ee
In four panels in 
Fig.~\ref{fig:ReDLvsDR}, corresponding to four values of $P$ indicated in each of them, we plot 
$|[\Delta_R^{q q}(Z^\prime)]_\text{eff}|$ as a function of ${\rm Re} \Delta_L^{sd}(Z^\prime)$ for different values of $\bsi$. For $M_{G^\prime}=M_{Z^\prime}$ the corresponding plot for
$G^\prime$ can be obtained from Fig.~\ref{fig:ReDLvsDR} by either rescaling 
upwards all values of $P$ by a factor of $1.4$ or scaling down either 
$|[\Delta_R^{q q}(Z^\prime)]_\text{eff}|$ or ${\rm Re} \Delta_L^{sd}(Z^\prime)$ 
by the same factor.  We will show such a  plot in Section~\ref{GprimeNum}.

 As we will discuss in  Section~\ref{LHCZprime}
 the values in  the gray area corresponding 
to  $|[\Delta_R^{q q}(Z^\prime)]_\text{eff}|\ge 1.25$ and 
$|\Delta_L^{sd}(Z^\prime)|\ge 2.3$  are basically ruled out by 
the LHC\footnote{ As mentioned in Section~\ref{LHCZprime} the complete 
exclusion of the grey area would require more intensive study of points 
corresponding to larger values of $\Delta_R(Z^\prime)$ and $M_{Z^\prime}< 3\tev$.}. We also note that
 while for $P=5$ and $P=10$ and $\bsi\ge 1.0$, the required
values of  ${\rm Re} \Delta_L^{sd}(Z^\prime)$ are in the ballpark of unity, 
for  $P=20$ they are generally larger than two implying for 
 ${\rm Re} \Delta_L^{sd}(Z^\prime)=2.3$
\be
\alpha_L=\frac{[{\rm Re} \Delta_L^{sd}(Z^\prime)]^2}{4\pi}=0.42~.
\ee
As $\alpha_L$ is not small let us remark that  in
in the case of a $U(1)$ gauge symmetry  for even larger values of $\alpha_L$  it is difficult to avoid a Landau pole at higher scales. 
However, if only the coupling 
$\Delta_L^{sd}(Z^\prime)$ is large,  a simple renormalization group 
analysis shows that these scales are much larger than the LHC 
scales. Moreover, if $Z^\prime$ is associated 
with a  non-abelian gauge symmetry that is asymptotically free 
${\rm Re} \Delta_L^{sd}(Z^\prime)$  could be even higher allowing to 
reach values of $P$ as high as $25-30$. We will see in Section~\ref{GprimeNum} 
that this is in fact the case for $G^\prime$.

 In this context a rough estimate of the perturbativity upper bound 
on $\Delta_L^{sd}(Z^\prime)$ can be made by considering the loop expansion 
parameter\footnote{A.J.B would like to thank Bogdan Dobrescu,  Maikel de Vries  and Andreas Weiler 
for discussions on this issue.}
\be\label{loop}
L=N\frac{[\Delta_L^{sd}(Z^\prime)]^2}{16\pi^2}
\ee
where $N=3$ is the number of colours. 
For $\Delta_L^{sd}(Z^\prime)=2.5,~3.0,~3.5$ one has $L=0.12,~0.17,~0.23$, respectively,  implying that using 
$\Delta_L^{sd}(Z^\prime)$ as large as $2.3$ can certainly be defended.

\subsection{Strategy}
This discussion and an independent numerical analysis using the general formulae 
presented above leads us to the conclusion
that for the goals of the present paper it is sufficient to consider only the 
following two scenarios for $Z^\prime$ couplings that satisfy the hierarchy 
(\ref{couplingsh}):

\paragraph{Scenario A}

This scenario is represented by our toy model constructed above.  It 
provides significant contribution to the $\Delta I=1/2$ rule 
without violating constraints from $\Delta F=2$ processes. Here 
in addition to  $\Delta_{L}^{sd}(Z^\prime)$ and  $\Delta_R^{qq}(Z^\prime)$ of $\ord(1)$
 also a small $\Delta_{R}^{sd}(Z^\prime)$ satisfying (\ref{tuning}) is required. Undoubtedly this scenario is fine-tuned but cannot be 
excluded at present. Moreover, it implies certain correlations between 
various observables and it is interesting to investigate them numerically. 
The three steps procedure outlined above allows to study transparently this 
scenario.

\paragraph{Scenario B}

Among flavour violating couplings only   $\Delta_{L}^{sd}(Z^\prime)$ is non-vanishing or  at all relevant.
In this case only SM operator contributes to $\varepsilon_K$ and $\Delta M_K$ 
and we deal with scenario LHS for flavour violating couplings not allowing 
for the necessary shift in  $\RE A_0$ due to $\Delta M_K$ constraint but 
still providing interesting results for $\epe$. Indeed only QCD penguin
 operator $Q_6$ contributes as in  Scenario A to the NP part in 
$K_L\to\pi\pi$ in an important manner. But  $\RE A_0^{\rm NP}$  in this 
scenario is very small and there is no relevant 
correlation between $\Delta I=1/2$ rule and remaining observables. The 
novel part of our analysis in this scenario  relative to our previous 
papers is the analysis of $\epe$ and 
of its correlation with $\kpn$ and $\klpn$.

\section{Numerical Analysis}\label{sec:4}
\subsection{Preliminaries}
In order to proceed we have to describe how we treat parametric and hadronic 
uncertainties in the SM contributions as this will determine  the room 
left for NP contributions in the observables discussed by us.

First in order to simplify the numerical analysis we will set all parameters in 
Table~\ref{tab:input}, except for $\vub$ and $\vcb$, at their central values.
 Concerning the latter two we will investigate six scenarios for them in 
order to stress the importance of their determination in the context of 
the search for NP through various observables. In order to bound the 
parameters of the model and to take hadronic and 
parametric uncertainties into account we will first only require that in Scenario B the results for $\Delta M_K$ and $\varepsilon_K$  
including NP contributions satisfy 
\be\label{CK}
0.75\le \frac{\Delta M_K}{(\Delta M_K)_{\rm SM}}\le 1.25,\qquad
2.0\times 10^{-3}\le |\varepsilon_K|\le 2.5 \times 10^{-3}.
\ee
However, it will be interesting to see what happens when the allowed range for
$\varepsilon_K$ is reduced to $3\sigma$ range around its experimental value.
In Scenario A 
which is easier numerically we will see more explicitly what happens 
to $\Delta M_K$ and $\varepsilon_K$  and the latter $3\sigma$ range will be more 
relevant than the  use of (\ref{CK}).

We will set $M_{Z^\prime}=3~\tev$ as our nominal value. This is an appropriate 
value for being consistent with ATLAS and CMS experiments 
although as we will discuss below such a mass puts an 
upper bound on   $\Delta_R^{qq}(Z')$. The scaling laws in  \cite{Buras:2013dea}
and our discussion in Section~\ref{scalings}
 allow us to translate our results to other values of $M_{Z^\prime}$. In particular 
when  $\Delta_L^{sd}(Z')$ is bounded by $\Delta S=2$ observables, NP 
effects in $\Delta F=1$ decrease with increasing  $M_{Z^\prime}$. Therefore 
in order that NP plays a role in the $\Delta I=1/2$ rule and the 
involved couplings are in perturbative regime,  $M_{Z^\prime}$ should be smaller 
than $5~\tev$ and consequently in the reach of the upgraded LHC.

Concerning the values of $\Delta_L^{sd}(Z')$ the numerical analyses in 
Scenarios A and B differ in the following manner from each other:
\begin{itemize}
\item
In Scenario A, in which ${\rm Re}A_0$ plays an important role, we will 
use the three step procedure outlined in the previous section. In 
this manner we will find that   $\Delta_L^{sd}(Z^\prime)\ge 1$ in order for $Z^\prime$ 
to play any role in the $\Delta I=1/2$ rule.
\item
In Scenario B, we can proceed as in our previous papers by using the 
parametrization 
\be\label{Zprimecouplings}
\Delta_L^{sd}(Z')=-\tilde s_{12} e^{-i\delta_{12}},
\ee
and searching for the allowed oases in the space
$(\tilde s_{12},\delta_{12})$ that satisfy  the constraints in (\ref{CK}) 
or  the stronger $3\sigma$ constraint for $\varepsilon_K$.
 In this scenario  $\Delta_L^{sd}(Z^\prime)$ will turn out to be very small.
 We will not show the results for these oases as they can be found 
in \cite{Buras:2012jb}. 
\end{itemize}

Having determined $\Delta_L^{sd}(Z')$  we can proceed to calculate the 
$\Delta F=1$ observables and study correlations between them. 
Here 
additional uncertainties will come from $\bsi$ which is hidden in the 
condition  (\ref{condition2}) so that it does not appear explicitly in 
NP contributions but affects the SM contribution to $\epe$. Also $Z^\prime$
coupling to neutrinos has to be fixed.

\begin{table}[!tb]
\center{\begin{tabular}{|l|l|}
\hline
$G_F = 1.16637(1)\times 10^{-5}\gev^{-2}$\hfill\cite{Beringer:1900zz}	&  $M_W = 80.385(15) \gev$\hfill\cite{Beringer:1900zz}  \\
$\sin^2\theta_W = 0.23116(13)$\hfill\cite{Beringer:1900zz} & 
$\alpha(M_Z) = 1/127.9$\hfill\cite{Beringer:1900zz} \\				
$\alpha_s(M_Z)= 0.1185(6) $\hfill\cite{Beringer:1900zz}	&  $m_K= 497.614(24)\mev$
	\hfill\cite{Nakamura:2010zzi} \\
$m_u(2\gev)=(2.1\pm0.1)\mev $ 	\hfill\cite{Aoki:2013ldr}						&  $m_\pi=135.0\mev$ \\
$m_d(2\gev)=(4.68\pm0.16)\mev$	\hfill\cite{Aoki:2013ldr}							&  $F_\pi = 129.8\mev$ \\
$m_s(2\gev)=(93.8\pm2.4) \mev$	\hfill\cite{Aoki:2013ldr}				&   $F_K = 156.1(11)\mev$\hfill\cite{Laiho:2009eu}	
             \\
$m_c(m_c) = (1.279\pm 0.013) \gev$ \hfill\cite{Chetyrkin:2009fv}					&          
$|V_{us}|=0.2252(9)$\hfill\cite{Amhis:2012bh}\\
$m_b(m_b)=4.19^{+0.18}_{-0.06}\gev$\hfill\cite{Beringer:1900zz} 			& 
$|V^\text{incl.}_{ub}|=(4.41\pm0.31)\times10^{-3}$\hfill\cite{Beringer:1900zz}\\
$m_t(m_t) = 163(1)\gev$\hfill\cite{Laiho:2009eu,Allison:2008xk} &  
$|V^\text{excl.}_{ub}|=(3.23\pm0.31)\times10^{-3}$\hfill\cite{Beringer:1900zz}\\	
$\eta_{cc}=1.87(76)$\hfill\cite{Brod:2011ty} 	&  $|V_{cb}|=(40.9\pm1.1)\times
10^{-3}$\hfill\cite{Beringer:1900zz} \\			
$\eta_{tt}=0.5765(65)$\hfill\cite{Buras:1990fn} & $\hat B_K= 0.75$   \\	
$\eta_{ct}= 0.496(47)$\hfill\cite{Brod:2010mj} & $\kappa_\epsilon=0.94(2)$\hfill\cite{Buras:2008nn,Buras:2010pza}             \\		
		
\hline
\end{tabular}  }
\caption {\textit{Values of the experimental and theoretical
    quantities used as input parameters.}}
\label{tab:input}~\\[-2mm]\hrule
\end{table}

Finally uncertainties due  to the values of the CKM elements 
$\vcb$ and $\vub$ have to be considered. These uncertainties are at first sight 
absent in $Z^\prime$ contributions 
but affect the SM predictions for $\varepsilon_K$ and $\epe$ and consequently 
indirectly also  $Z^\prime$ contributions through the size of allowed range 
for $\Delta^{sd}_{L}(Z^\prime)$ in both scenarios A and B. Indeed 
 $\epe$ and $\klpn$ depend in the SM on ${\IM \lambda_t}$, while $\varepsilon_K$ and $\kpn$ on both ${\IM \lambda_t}$ and ${\RE \lambda_t}$. 
Now within the 
accuracy of better than $0.5\%$
\be
{\IM \lambda_t}=\vub\vcb\sin\gamma, \qquad {\RE \lambda_t}=-{\IM \lambda_t}\cot(\beta-\beta_s)
\ee
with $\gamma$ and $\beta$ being the known angles of the unitarity triangle and 
$-\beta_s\approx 1^\circ$ is the phase of $V_{ts}$ after the minus sign has been 
factored out.
Consequently, within the SM not only $\epe$ and $\varepsilon_K$ but also the branching 
ratios for $\kpn$ and $\klpn$ will depend sensitively on the chosen values for $\vcb$ and $\vub$.

One should 
recall that the typical values for $\vub$ and $\vcb$ extracted from 
{\it inclusive} decays are (see \cite{Ricciardi:2012pf,Gambino:2013rza} and 
refs therein)\footnote{We prefer to quote for the central value of $\vcb$ the most recent value from \cite{Gambino:2013rza} than the one given in Table~\ref{tab:input}.}
\be\label{inclusive}
\vub=4.1\times 10^{-3}, \qquad  \vcb=42.0\times 10^{-3}
\ee
while the typical values extracted from {\it exclusive} decays read \cite{Ricciardi:2013cda,Bailey:2014tva}
\be\label{exclusive}
\vub=3.2\times 10^{-3}, \qquad  \vcb=39.0\times 10^{-3}.
\ee
As the determinations of $\vub$ and $\vcb$ are independent of each other it will
be instructive to consider the following 
scenarios for these elements:
\begin{align}
 a)&\qquad \vub = 3.2\times 10^{-3}\qquad \vcb = 39.0\times 10^{-3}\qquad ({\rm purple)} \label{sa}\\
b)& \qquad \vub = 3.2\times 10^{-3}\qquad \vcb = 42.0\times 10^{-3}\qquad ({\rm cyan)}\\
c)& \qquad \vub = 4.1\times 10^{-3}\qquad \vcb = 39.0\times 10^{-3}\qquad ({\rm magenta)}\\
d)& \qquad \vub = 4.1\times 10^{-3}\qquad \vcb = 42.0\times 10^{-3}\qquad ({\rm yellow)}\\
e)& \qquad \vub = 3.7\times 10^{-3}\qquad \vcb = 40.5\times 10^{-3}\qquad ({\rm green)}\\
f)&\qquad \vub = 3.9 \times 10^{-3}\qquad \vcb = 42.0\times 10^{-3} \qquad ({\rm blue)} \label{sf}
\end{align}
 where we also included two additional scenarios, one for averaged 
values of $\vub$ and $\vcb$ and the last one ($f)$) particularly suited 
for the analysis of Scenario A.  We also give the colour coding for these 
scenarios used in the plots.

Concerning the parameter $\hat B_K$ which enters the evaluation of
$\varepsilon_K$ the world average from 
lattice QCD is $\hat B_K=0.766\pm 0.010$ \cite{Aoki:2013ldr}, very close 
to the strictly large $N$ limit value $\hat B_K=0.75$. On the other hand 
the recent calculation within the dual approach 
to QCD gives $\hat B_K=0.73\pm 0.02$ \cite{Buras:2014maa}. Moreover, 
the analysis in \cite{Gerard:2010jt} indicates that in the absence 
of significant $1/N^2$ corrections to the leading large $N$ value one should have 
$\hat B_K\le 0.75$. It is an interesting question whether this result will be confirmed by future lattice calculations which have a better control over the uncertainties than it is 
possible within the approach in \cite{Buras:2014maa,Gerard:2010jt}.
For the time being it 
is a very good approximation to set simply $\hat B_K=0.75$.
Indeed compared to the present uncertainties from $\vcb$ and $\vub$ 
in $\varepsilon_K$ proceeding in this manner is fully justified.

Concerning the value of $\gamma$ we will just set $\gamma=68^\circ$. This 
is close to central values from recent determinations \cite{LHCb-CONF-2013-006,Fleischer:2010ib,Aaij:2013zfa} and varying $\gamma$ 
simultaneously with 
$\vcb$ and $\vub$ would not improve our analysis.

As seen in Table~\ref{tab:CKM} the six scenarios for CKM parameters
 imply rather different values of ${\IM \lambda_t}$ and ${\RE \lambda_t}$ 
and consequently different values for various 
observables considered by us. This is seen in this table where we give SM values for 
$\varepsilon_K$, $\Delta M_K$, $\Delta M_s$, $\Delta M_d$, $S_{\psi K_S}$, 
$\epe$, $\mathcal{B}(\klpn)$ and $\mathcal{B}(\kpn)$
together with their experimental values. To this end we have used the 
central values of the remaining parameters, relevant for $B_{s,d}^0$ systems 
collected in \cite{Buras:2013ooa}.  For completeness we give also 
the values for $\overline{\mathcal{B}}(B_s\to \mu^+\mu^-)$ and 
${\mathcal{B}}(B_d\to \mu^+\mu^-)$.

We would like to warn the reader that the SM values for various observables 
in Table~\ref{tab:CKM} 
have been obtained directly by using CKM parameters from tree-level decays 
and consequently differ from SM results obtained usually from Unitarity Triangle fits that include constraints from processes in principle 
affected by NP.
 
 We note that for a given choice of $\vub$, $\vcb$ and $\gamma$ the 
SM predictions can differ sizably from the data but these departures 
are different for different scenarios: 
\begin{itemize}
\item
Only in scenario $a)$ does $S_{\psi K_S}^{\rm SM}$ agree fully with the data.  On the other hand in the remaining scenarios  $Z^\prime$ 
contributions 
to $B_d^0-\bar B_d^0$ are required to bring the theory to agree with the data. 
But then also $\Delta M_s$ and $\Delta M_d$ have to receive new contributions, 
even in the case of scenario $a)$. 
As in the models considered here $Z^\prime$ flavour violating couplings involving $b$-quarks are not fixed, this can certainly be achieved. 
We refer to 
\cite{Buras:2012jb,Buras:2013qja} for details. 
\item
On the other hand $\varepsilon_K$ is definitely below the experimental value in 
scenario $a)$ but roughly consistent with experiment in other scenarios leaving still some room for NP contributions. 
{ In particular in scenarios $d)$ and $f)$ it is close to its experimental value.}
\item
$\Delta M_K$ is as expected the same in all scenarios and roughly $10\%$ below 
its experimental value. But we should remember that the large uncertainty 
in $\eta_{cc}$ corresponds  to $\pm 40\%$ uncertainty in $\Delta M_K$ {
and still sizable NP contributions are allowed.}
\item
The dependence of $\mathcal{B}(\klpn)$ on scenario considered is 
large but moderate in the case of $\mathcal{B}(\kpn)$.
\item
 We emphasize strong dependence on $\vcb$ and consequently on $\vts$ of the branching 
ratios  $\overline{\mathcal{B}}(B_s\to \mu^+\mu^-)$ and 
${\mathcal{B}}(B_d\to \mu^+\mu^-)$. For exclusive 
values of $\vcb$ both branching ratios are significantly lower than 
the official SM values \cite{Bobeth:2013uxa} obtained  using $\vcb=42.4\times  10^{-3}$.
\end{itemize}

In Scenario B, where the constraint from $\Delta I=1/2$ is absent we will 
have more freedom in adjusting NP parameters to improve in each of the 
scenarios $a)-f)$ the agreement of the theory with data but within Scenario A 
we will find that only for certain scenarios of CKM parameters it will be possible to fit the data.

\begin{figure}[!tb]
 \centering
\includegraphics[width = 0.45\textwidth]{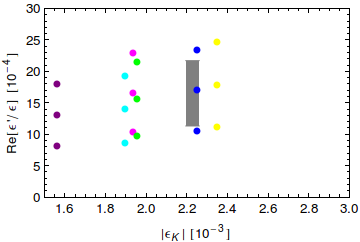}
\includegraphics[width = 0.45\textwidth]{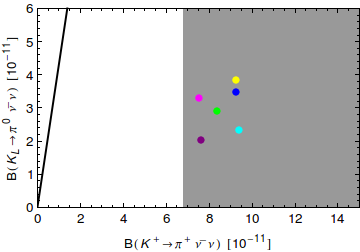}
\includegraphics[width = 0.45\textwidth]{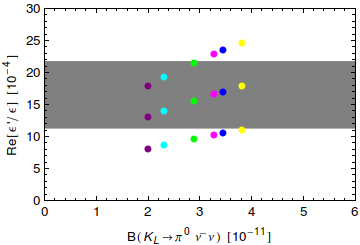}
\includegraphics[width = 0.45\textwidth]{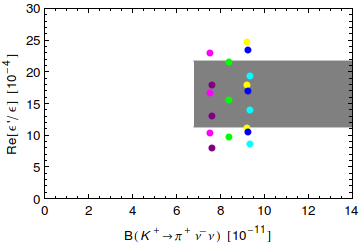}
\caption{\it  SM central values for $\epe$, $\varepsilon_K$, $\mathcal{B}(\klpn)$ and 
$\mathcal{B}(\klpn)$   for scenarios a) (purple), b) (cyan), c) (magenta), d) (yellow),  e) (green) 
and f) (blue) and different values of 
$B_6^{(1/2)} = 0.75,~1.00,~1.25$ corresponding to the increasing value of $\epe$  for fixed colour.   Gray region: 2$\sigma$
experimental range of $\epe$ and $3\sigma$ for $\varepsilon_K$. }\label{SMResults}~\\[-2mm]\hrule
\end{figure}

In Fig.~\ref{SMResults} we summarize those results of Table~\ref{tab:CKM} that      will help us in following our numerical analysis in 
various NP scenarios  presented by us. In particular we observe in the lower left panel strong 
correlation between $\epe$ and $\mathcal{B}(\klpn)$.  Fig.~\ref{SMResults} 
shows graphically how important  the determination of $\vub$, $\vcb$ and 
$\bsi$ in the indirect search for NP is. Let us hope that at the end of this 
decade there will be only a single point representing the SM in each of these 
four panels.

\begin{table}[!tb]
{\renewcommand{\arraystretch}{1.3}
\begin{center}
\begin{tabular}{|c||c|c|c|c|c|c|c|}
\hline
&$a)$& $b)$&$ c)$ &$ d)$& $e)$& $f)$& Data\\
\hline
\hline
$\IM\lambda_t\, [10^{-4}]$&$ 1.16$&$ 1.25$ & $1.48$ & $1.60$ & $1.39$& $1.52$ & $-$\\
\hline
$\RE\lambda_t\, [10^{-4}]$& $-2.90$&$-3.40$ &$-2.76$ & $-3.25$ & $-3.07$&$-3.29$ &$-$\\
\hline
 $S^{\rm SM}_{\psi K_S}$ & $0.664$&$ 0.622$ &$ 0.808$ & $0.765$ & $0.726$&$ 0.736$ &$ 0.679(20)$ \\
\hline
$\Delta M_s\,[\text{ps}^{-1}]$ & $15.92$ & $18.44$ & $15.99$& $18.51$ & $17.19$ &$ 18.49 $&$17.69(8)$\\
\hline
$\Delta M_d\,[\text{ps}^{-1}]$ & $ 0.47$ & $0.54$ & $0.47$ & $0.54$ & $0.50$& $ 0.54$&
$0.510(4)$\\
\hline
$ \Delta M_K\,[10^{-3}\text{ps}^{-1}]$ & $4.70$ & $4.72 $ & $ 4.70 $ & $4.71$ & $4.71$& $ 4.72$ &$5.293(9)$\\
\hline
$|\varepsilon_K|\,[10^{-3}]$ & $1.56$ & $1.89$ & $1.93$ & $2.35$ & $1.96$& $ 2.25 $ &$2.228(11)$\\
\hline
$\epe\,[10^{-4}](B_6^{(1/2)} = 0.75)$&$ 8.0$&$ 8.6$ & $10.2$ & $11.0$ & $9.6$&$ 10.5 $&$16.5\pm 2.6$\\
\hline
$\epe\,[10^{-4}](B_6^{(1/2)} = 1.00)$& $12.9$&$13.9$ &$16.5$ & $17.8$ & $15.5$&$ 16.9$&
$16.5\pm 2.6$\\
\hline
$\epe\,[10^{-4}](B_6^{(1/2)} = 1.25)$& $17.8$&$19.2$ &$22.8$ & $24.6$ & $21.4$&$ 23.4 $ &
$16.5\pm 2.6$ \\
\hline
$\mathcal{B}(K_L\to \pi^0\nu\bar\nu)\,[10^{-11}]$ & $2.01$ & $2.33$ & $3.29$ & $3.82$ & $2.89$ &$ 3.45$ &$\le 2.6\cdot 10^{-8}$ \\
\hline
$\mathcal{B}(K^+\to \pi^+\nu\bar\nu)\,[10^{-11}]$ & $ 7.65 $ & $9.40$ & $7.54$& $ 9.25$& $8.40$ & $ 9.28$ &$17.3^{+11.5}_{-10.5}$ \\
\hline
$\overline{\mathcal{B}}(B_s\to \mu^+\mu^-)\,[10^{-9}]$ & $3.00$ & $3.47$ & $3.01$ & $3.48$ & $3.23$ &$ 3.48$ &$ 2.9\pm0.7$ \\
\hline
$\mathcal{B}(B_d\to \mu^+\mu^-)\,[10^{-10}]$ & $ 0.94 $ & $1.09$ & $0.94$& $ 1.09$& $1.01$ & $ 1.09$ &$ 3.6^{+1.6}_{-1.4}$ \\
\hline
\end{tabular}
\end{center}}
\caption{\it Values of $\IM\lambda_t$, $\RE\lambda_t$ and of several observables within the SM for
various scenarios of CKM elements as discussed in the text.
\label{tab:CKM}}~\\[-2mm]\hrule
\end{table}

\subsection{LHC Constraints}\label{LHCZprime}
Finally, we should remember that  $Z^\prime$ couplings to quarks can be bounded 
by collider data as obtained from LEP-II and the LHC. In the case of LEP-II 
all the bounds  can be satisfied  in our models by using sufficiently small 
leptonic couplings. However, in the case of $\Delta_R^{qq}$ and $\Delta_L^{sd}$
 we have to 
 check whether the values $\Delta_R^{qq}(Z^\prime)=\ord(1)$ and 
$\Delta_L^{sd}(Z^\prime)=\ord(1)$ necessary for 
a significant $Z^\prime$ contribution to ${\rm Re} A_0$ are allowed by the 
ATLAS and CMS outcome of the search for narrow resonances using dijet mass 
spectrum in proton-proton collisions  and by the  effective operator 
bounds. 

\begin{figure}[ht]
	\begin{center}
		\includegraphics[scale=0.46]{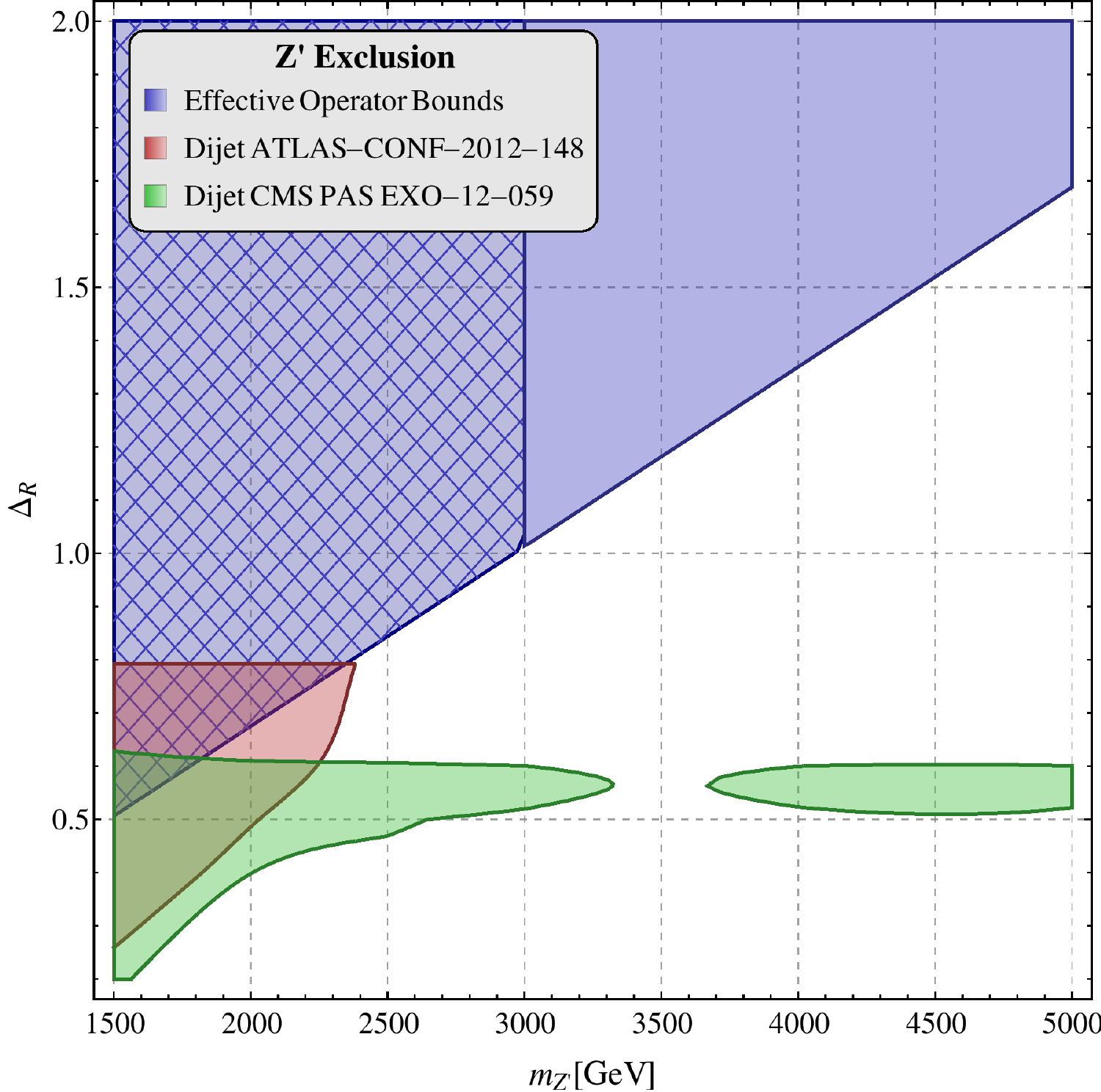}
\caption{\it Exclusion limits for the $Z'$  in the mass-coupling plane, from various searches at the LHC as found in \cite{Vries}. The blue 
region is excluded by effective operator limits studied by ATLAS \cite{ATLAS:2012pu} and CMS\cite{Chatrchyan:2012bf}. The dashed surface represents the 
region where the effective theory is not applicable, and the bounds here should be interpreted as a rough estimate. The red and green 
contours are excluded by dijet resonance searches by ATLAS \cite{ATLAS:2012qjz} and CMS \cite{CMS:kxa}. See additional comments in 
the text.} 
		\label{fig:zprimeexclusion}
	\end{center}
\end{figure}

Bounds of this sort can be found in \cite{Harris:2011bh,Domenech:2012ai,Redi:2013eaa,Chatrchyan:2013qha,Dobrescu:2013cmh} 
but the $Z^\prime$ models considered there have SM couplings or as in the 
case of \cite{Dobrescu:2013cmh} all diagonal 
couplings, both left-handed and right-handed, are flavour universal which is not the case of our models in which the hierarchy 
(\ref{couplingsh}) is assumed.

For this reason a dedicated analysis of our toy model has been performed \cite{Vries}\footnote{The details of this analysis will be presented elsewhere.}
using the most recent results from ATLAS  and CMS. The result of this study 
is presented in Fig.~\ref{fig:zprimeexclusion} and can be briefly summarized as follows:
\begin{itemize}
\item
The most up to date dijet searches from ATLAS \cite{ATLAS:2012qjz} and CMS \cite{CMS:kxa} allow to put an  upper bound on $|\Delta_R^{qq}(Z^\prime)|$ but only
for $|\Delta_R^{qq}(Z^\prime)|\le 0.8$. As seen in Fig.~\ref{fig:zprimeexclusion} this  maximal value is 
only allowed for $M_{Z^\prime}\ge 2.4\tev$.
\item
A second source of exclusion limits for $Z'$ boson couplings comes from effective operator limits, in this case from four-quark operators studied 
by both ATLAS \cite{ATLAS:2012pu} and CMS \cite{Chatrchyan:2012bf}.
As seen in Fig.~\ref{fig:zprimeexclusion} the upper bound on $|\Delta_R^{qq}(Z^\prime)|$
can be summarized by 
\be\label{DRLHCbound}
|\Delta_R^{qq}(Z^\prime)|\le 1.0\times \left[ \frac{M_{Z^\prime}}{3\tev}\right].
\ee
\end{itemize}

The following additional comments should be made in connection with results in 
Fig.~\ref{fig:zprimeexclusion}:
\begin{itemize}
\item
The dijet limits are only effective if the width of the $Z'$ or $G'$ is below $15\%$ for ATLAS and $10\%$ for CMS.
\item
The lack of exclusion limits for CMS around $M_{Z'}=3.5$ TeV are the result of a fluctuation in the data and therefore their exclusion limits.
\item
It is important to note that the limits from effective operator constraints 
should  not to be trusted when the center of mass energy of the experiment is bigger than the mass of the particle which is integrated out.  For this analysis the effective center of mass energy is $3\tev$.
\end{itemize}

While dijets constraints would still allow for $[\Delta_R^{q q}(Z^\prime)]_\text{eff}=1.25$ (see (\ref{Deltaeff})) we will use for it $1.0$ 
so that 
our nominal values will be
\be\label{LHCqqR}
\Delta_R^{qq}(Z^\prime)=-1.0, \qquad M_{Z^\prime}=3\tev
\ee
that is consistent with the bound in (\ref{DRLHCbound}).
As seen in (\ref{condition2}) the couplings $\Delta_R^{qq}(Z^\prime)$ and 
$\Delta_L^{sd}(Z^\prime)$ must have opposite signs in order to satisfy the
$\Delta I=1/2$ constraint. On the basis of the present LHC data it is not 
possible to decide which of the two possible sign choices for these couplings 
is favoured by the collider data but this could be in principle possible in 
the future.
The minus in $\Delta_R^{qq}(Z^\prime)$ is chosen here only to keep the coupling 
$\Delta_L^{sd}(Z^\prime)$ positive definite but presently the same results 
would be obtained with the other choice for signs of these two couplings.

As far as $\Delta_L^{sd}(Z^\prime)$ is concerned 
the derivation of corresponding bounds
is more difficult, since the experimental collaborations do not provide constraints for flavoured four quark interactions. However, there have been efforts to obtain these from the current data \cite{Domenech:2012ai,Davidson:2013fxa}.
In particular the analysis of the $\Delta S=2$ operator 
 in \cite{Davidson:2013fxa} turns out to be useful. With its help one finds
 the upper bound \cite{Vries}
\be\label{DLLHCbound}
|\Delta_L^{sd}(Z^\prime)|\le 2.3 \left[\frac{M_{Z^\prime}}{3\tev}\right].
\ee

Now, as seen in Fig.~\ref{fig:ReDLvsDR} with 
(\ref{LHCqqR}) the values $P=20-30$ require   ${\rm Re}\Delta_L^{sd}(Z^\prime)\approx 3-4$ dependently on the value of $\bsi$.  This would 
still 
be consistent with rough perturbativity bound  ${\rm Re}\Delta_L^{sd}(Z^\prime)\le 4$ discussed by us in Section~\ref{scalings}. However, 
the LHC bound 
in (\ref{DLLHCbound}), seems to exclude this possibility, although
 a dedicated analysis of this bound including simultaneously left-handed 
and right-handed couplings would be required to put this bound on a 
firm footing. We hope to return to such an analysis in the future. 
For the time being we conclude that the maximal values of $P$ possible 
in this NP scenario are in the ballpark of $16$, that is roughly of 
the size of SM QCD penguin contribution.

Indeed, combining the bounds on the couplings of $Z^\prime$ and its mass 
and using the relation (\ref{condition2}) we arrive at the upper bound
\be\label{PboundZp}
P\le 16\left[\frac{\bsi}{1.0}\right], \qquad (Z^\prime)~.
\ee
 This result is also seen in Fig.~\ref{fig:ReDLvsDR}.
 In principle for $\bsi$ significantly larger than unity one could 
increase the value of $P$ above 20 but as we will see soon this is not 
allowed when simultaneously the correlation between $\epe$ and $\varepsilon_K$ 
is taken into account.

 At this point it should be emphasized that the dashed surface in Fig.~\ref{fig:zprimeexclusion} has in fact not been completely excluded by 
ATLAS and CMS 
analyses and as an example $\Delta_R^{qq}(Z^\prime)=-1.5$ and 
$M_{Z^\prime}=2.5\tev$, allowing $P$ as high as $30$, is still a valid point.
While it is likely that a dedicated analysis of this model by ATLAS and CMS 
in this range of parameters would exclude the dashed surface completely, 
such an analysis has still to be done.

\subsection{Results}
\boldmath
\subsubsection{SM Results for $\epe$}
\unboldmath
We begin our presentation by discussing briefly the SM prediction for
$\epe$ given in Table~\ref{tab:CKM}
for different scenarios for CKM couplings and three values of $\bsi$. We observe 
that for $\bsi=1.00$, except for scenario $a)$, the SM is in good agreement 
with the data but in view of the experimental error NP at the level of 
$\pm 20\%$ can still contribute. In the past when $\bei=1.0$ was used $\epe$ 
for $\bsi=1.0$ was below the data, but with the lattice result $\bei=0.65\pm 0.05$ \cite{Blum:2012uk}
it looks like $\bsi\approx 1.0$ is the favourite value within the SM. 
Except for scenario $a)$ and $\bsi=1.25$ for which SM gives values consistent 
with experiment, for other two values of $\bsi$ we get either visibly lower or 
 visibly higher values of $\epe$ than measured and some NP is required to 
fit the data.
\subsection{Scenario A}

\begin{figure}[!tb]
 \centering
\includegraphics[width = 0.45\textwidth]{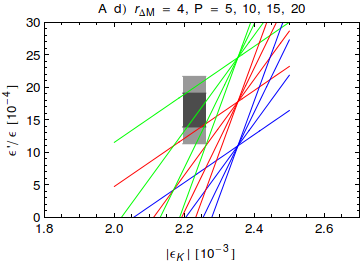}
\includegraphics[width = 0.45\textwidth]{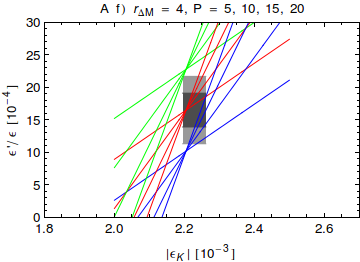}
\caption{\it $\epe$ versus $\varepsilon_K$ for scenario for scenario $d)$ and $f)$ for $r_{\Delta M} = 4$. Light(Dark) gray region:
experimental $2\sigma$($1\sigma$) range of $\epe$ and  $3\sigma$ range $2.195\times 
10^{-3}\leq 
|\varepsilon_K|\leq 2.261\times 10^{-3}$. Blue, red and green stands for 
$B_6^{(1/2)} = 0.75,\,1.00,\,1.25$, respectively and for $P$ we use $5,\,10,\,15,\,20$ (the steeper the line, the larger $P$).  
}\label{fig:epsvsepstoy}~\\[-2mm]\hrule
\end{figure}

The question then arises whether simultaneous agreement with the data for  ${\RE} A_0$, $\varepsilon_K$ and $\epe$ can be obtained in the 
toy $Z^\prime$ model introduced by us. 

We use the three step procedure suited for this scenario that we outlined 
in the previous section. Investigating all six scenarios $a)-f)$  for $(\vcb,\vub)$ we have found that only in scenarios $d)$ and $f)$ it 
is possible to obtain satisfactory agreement with the
data on $\epe$ and $\varepsilon_K$ for significant values of $P$. Indeed 
due to relation (\ref{Basicrelation1}) NP in $\varepsilon_K$ must be small 
in order to keep $\epe$ under control. As seen in Fig.~\ref{SMResults} this 
is only the case in these two CKM scenarios. Yet, as seen 
in Fig.~\ref{fig:epsvsepstoy}, even $d)$ and $f)$ scenarios can be 
distinguished 
 by the correlation between $\epe$ and $\varepsilon_K$   
demonstrating again how 
important it is to determine precisely $\vcb$ and $\vub$.

While, as seen in (\ref{Basicrelation1}), the correlation between NP contributions to   $\epe$ and $\varepsilon_K$ 
depends at fixed $r_{\Delta M}$ only on $P$, in the case of SM contributions it depends
explicitly on $\bsi$. Therefore we show in Fig.~\ref{fig:epsvsepstoy}   the lines  for  
$\bsi=0.75,~1.00,~1.25$ using the colour coding 
\be\label{bsicoding}
 \bsi= 0.75 ~\text{(blue)},\quad \bsi= 1.0~\text{(red)},\quad \bsi=1.25~\text{(green)}.
\ee
The three lines carrying the same colour correspond to four
values of $P=5,~10,~15,~20$. With increasing $P$ the lines become steeper. 
The dark(light) gray region corresponds to the $1(2)\sigma$ experimental
 range for 
$\epe$ and $3\sigma$ range for $\varepsilon_K$.

\begin{figure}[!tb]
 \centering
\includegraphics[width = 0.45\textwidth]{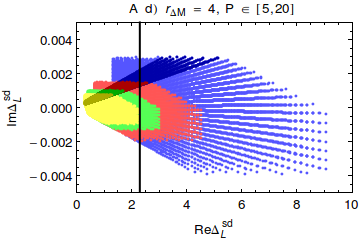}
\includegraphics[width = 0.45\textwidth]{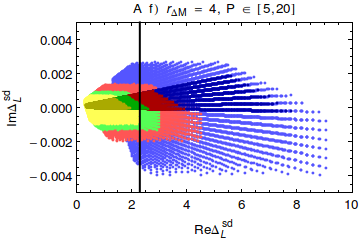}
\caption{\it Here we show the allowed values of $\re\Delta_L^{sd}$ and $\im\Delta_L^{sd}$ in scenario A $d)$ and $f)$ for $\Delta_R^{qq} = 
-0.5$ 
(blue), $-1$ (red), $-1.5$ (green) and $-2$ (yellow). We varied 
$P\in[5,\,20]$ and $B_6^{(1/2)}\in[0.75,1.25]$  and took only those $(\bsi,P)$ combinations that fulfill the constraints on $\epe$ 
($2\sigma$) and 
$\varepsilon_K$ (darker colours $3\sigma$ and lighter colours $2.0\cdot 10^{-3}\leq |\varepsilon_K|\leq 2.5\cdot 10^{-3}$). The vertical 
black 
line indicates the LHC bound in (\ref{DLLHCbound}). 
}\label{fig:oasestoyd}~\\[-2mm]\hrule
\end{figure}

Beginning with scenario $d)$ We observe that only
the following combinations of $P$ and $\bsi$ are consistent with this range:

\begin{itemize}
\item
For $\bsi=1.25$ only  $P=5,~10,~15$ are allowed when $1\sigma$ range for 
$\epe$ is considered. At $2\sigma$ also $P=20$ is allowed. Larger values 
of $P$ are only possible for $\bsi>1.25$. We conclude therefore that for 
$\bsi=1.25$ we find the upper bound $P\le 20$.
\item
For $\bsi=1.00$  the corresponding upper bound amounts to $P\le 10$.
\item
For $\bsi=0.75$ even for $P=5$ one cannot obtain simultaneous agreement with 
the data on $\epe$ and $\varepsilon_K$.
\end{itemize}

A rather different pattern is found for scenario $f)$:

\begin{itemize}
\item
For $\bsi=1.25$ the values  $P=5,~10,~15,~20$ are  not allowed even at $2\sigma$ range for $\epe$ but decreasing slightly $\bsi$ would 
allow 
values $P\ge 20$. 
\item
 On the other hand, in the case of  $\bsi=1.00$ 
 there is basically no restriction on $P$ from this correlation 
simply because in this scenario NP contributions to $\epsilon_K$ are 
small (see  Fig.~\ref{SMResults}). In fact in this case values of $P$ as high as $30$ would be allowed. { While such values are not 
possible in the 
case of $Z^\prime$ due to LHC constraint in (\ref{PboundZp}) we will see 
that they are allowed in the case of $G^\prime$.}
\item
Similar situation is found for $\bsi=0.75$ although here at $1\sigma$ for 
$\epe$ one finds the bound $P\ge 10$.
\end{itemize}

We conclude therefore that in view of the fact that NP effects in $\epe$ 
in our toy model are by an order of magnitude larger than in $\varepsilon_K$, 
 scenario $f)$ is particularly suited for allowing large values of $P$  as it
 avoids strong constraints from $\epe$ and $\varepsilon_K$.
In scenario $d)$ independently of the LHC we find $P<20$. { While in 
the case 
of $Z^\prime$ model at hand this virtue of scenario $f)$ cannot be fully 
used because of the LHC constraint (\ref{PboundZp}) we will see in the 
next section that it plays a role in the case of $G^\prime$ model.}
These findings are interesting as they 
imply that only for the inclusive determinations of $\vub$ and $\vcb$ $Z^\prime$ has a chance to contribute in a significant manner to the 
$\Delta I=1/2$ rule. 
 This assumes the absence of other mechanisms at work which  otherwise could help in this case 
if the exclusive determinations of these CKM parameters would turn out to be 
true.

In Fig.~\ref{fig:oasestoyd} we show with darker colours the allowed values of $\re\Delta_L^{sd}$ and $\im\Delta_L^{sd}$ in scenario A 
 for CKM values $d)$ and 
$f)$ that correspond to the values of $P$ and $\bsi$ selected by the light gray region in Fig.~\ref{fig:epsvsepstoy}. In lighter 
colours we show the allowed values of $\re\Delta_L^{sd}$ and $\im\Delta_L^{sd}$ using~(\ref{CK}) as constraint for $\varepsilon_K$. As 
 for $M_{Z^\prime}=3\tev$ only 
values $|\Delta_R^{qq}|\le 1.0$ are allowed by the LHC bound in (\ref{DRLHCbound}), the {\it green} and {\it yellow} ranges are ruled out 
but we show 
them anyway as this demonstrates the power of the LHC in constraining our 
model. Among the remaining areas the {\it red} one is favoured as it 
corresponds to smaller values of $\re\Delta_L^{sd}$  for a given $P$ and this is the reason why $\Delta_R^{qq}=-1.0$ has been chosen as 
nominal value for this 
coupling. This feature is not clearly seen in this figure where we varied $P$ but this is evident from plots in Fig.~\ref{fig:ReDLvsDR}.  
The vertical black 
line shows the LHC bound in (\ref{DLLHCbound}). Only values on the left of this 
line are allowed.

We have investigated the correlation between $\mathcal{B}(\klpn)$ and $\mathcal{B}(\kpn)$ for scenarios $d)$ and $f)$ finding the following 
pattern 
that follows from the fact that in Scenario A,  as can be seen in Fig.~\ref{fig:oasestoyd},  ${\rm Re} \Delta_L^{sd}(Z^\prime)=\ord(1)$.
 In view of this, the neutrino coupling  $\Delta_L^{\nu\nu}(Z^\prime)$ must be 
 sufficiently small in order to be consistent with the data on $\mathcal{B}(\kpn)$. But as seen in  Fig.~\ref{fig:oasestoyd} ${\rm Im} 
\Delta_L^{sd}(Z^\prime)$ is required to be small in order to satisfy the data on $\epe$ and $\varepsilon_K$. The smallness of both  
$\Delta_L^{\nu\nu}(Z^\prime)$ and ${\rm Im} \Delta_L^{sd}(Z^\prime)$ implies in this scenario negligible NP contributions to  
$\mathcal{B}(\klpn)$. Thus the main message from this exercise is  that $\mathcal{B}(\klpn)$ remains SM-like, while  $\mathcal{B}(\kpn)$ can 
be modified 
 but this modification depends on the size of the unknown coupling  $\Delta_L^{\nu\nu}(Z^\prime)$ and changing its sign one can obtain both 
suppression or 
enhancement  of $\mathcal{B}(\kpn)$ relative to the SM value. For 
 $\Delta_L^{\nu\nu}(Z^\prime)$ in the ballpark of $5\times 10^{-4}$ significant 
enhancements or suppressions can be obtained. In view of this simple 
pattern and low predictive power we refrain from showing any plots.

Yet, the requirement of strongly suppressed leptonic couplings implies 
that unless  $\Delta_{L,R}^{sb}(Z^\prime)$ and $\Delta_{L,R}^{db}(Z^\prime)$ are 
sizable, in Scenario A NP contributions to rare $B_{s,d}$ decays with 
neutrinos and charged leptons in the final state are predicted to be small. 
On the other hand these effects could be sufficiently large  in $\Delta B=2$ 
processes to cure SM problems in scenarios $d$ and $f$ seen in Table~\ref{tab:CKM}.

While for a fixed value of  $\Delta_L^{\nu\nu}(Z^\prime)$ there exist 
correlations between $\epe$ and $\mathcal{B}(\kpn)$ such correlations 
are more interesting in the case of Scenario B which we will discuss 
next.

\subsection{Scenario B}

\begin{figure}[!tb]
 \centering
\includegraphics[width = 0.5\textwidth]{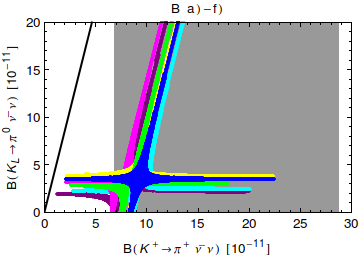}
\caption{\it $\mathcal{B}(\klpn)$ versus $\mathcal{B}(\kpn)$ for scenario a) (purple), b) (cyan), c) (magenta), d) (yellow),  e) (green) 
and f) (blue). Gray region:
experimental range of $\mathcal{B}(\kpn)$.  The black line corresponds to the Grossman-Nir bound.}\label{fig:pKLvsKpA}~\\[-2mm]\hrule
\end{figure}

Here we proceed as in \cite{Buras:2012jb} except that we use scenarios 
$a)-f)$ for  $(\vcb,\vub)$ and also present results for $\epe$. To this end 
 we use colour coding for  these scenarios in (\ref{sa})-(\ref{sf}) and the 
one for $\bsi$ in 
 (\ref{bsicoding})  and set
\be\label{setting}
\Delta_R^{qq}(Z^\prime)=0.5,\,1.0, \qquad \Delta_L^{\nu\nu}(Z^\prime)=0.5 \,
\ee
with darker(lighter) colours representing $\Delta_R^{qq}(Z^\prime)=1.0(0.5)$. 
These values of $\Delta_R^{qq}(Z^\prime)$  satisfy LHC bounds. The 
neutrino coupling can be chosen as in our previous papers because
the coupling  $\Delta_L^{sd}(Z^\prime)$ will be bounded by $\Delta M_K$ and 
$\varepsilon_K$ to be very small and this choice is useful 
as it allows to see the impact of $\epe$ constraint on our results for 
rare decays $\kpn$ and $\klpn$ obtained in \cite{Buras:2012jb} without this 
constraint.

\begin{figure}[!tb]
 \centering
\includegraphics[width = 0.45\textwidth]{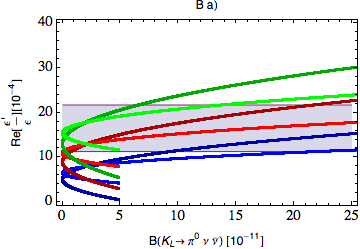}
\includegraphics[width = 0.45\textwidth]{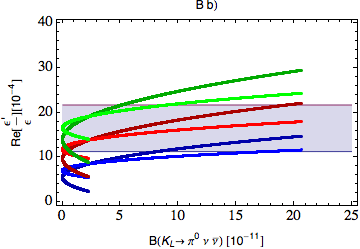}
\includegraphics[width = 0.45\textwidth]{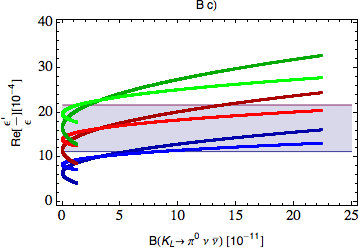}
\includegraphics[width = 0.45\textwidth]{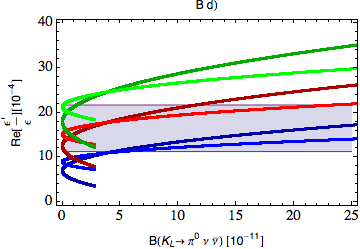}
\includegraphics[width = 0.45\textwidth]{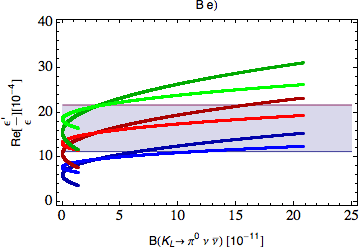}
\includegraphics[width = 0.45\textwidth]{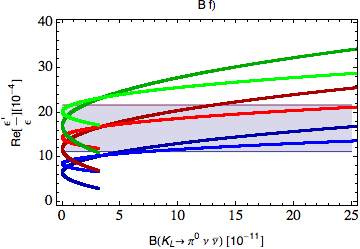}
\caption{\it  $\epe$ versus $\mathcal{B}(\klpn)$ for scenario $a)-f)$ and different values of $B_6^{(1/2)} = 0.75$ (blue), $B_6^{(1/2)} = 
1.00$ (red), $B_6^{(1/2)} = 1.25$ (green) and $\Delta_R^{qq}(Z^\prime)=1.0(0.5)$ for darker(lighter) colours. Gray region: 2$\sigma$
experimental range of $\epe$.  }\label{fig:epsvsBr0}~\\[-2mm]\hrule
\end{figure}

\begin{figure}[!tb]
 \centering
\includegraphics[width = 0.45\textwidth]{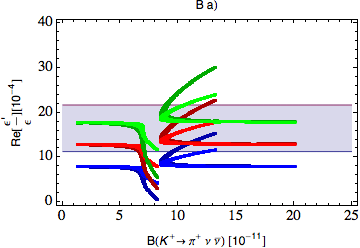}
\includegraphics[width = 0.45\textwidth]{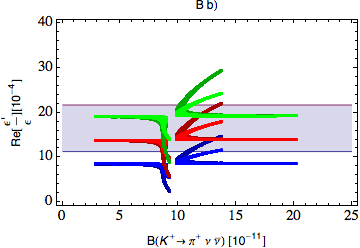}
\includegraphics[width = 0.45\textwidth]{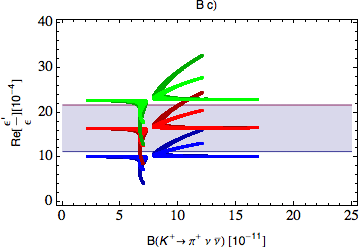}
\includegraphics[width = 0.45\textwidth]{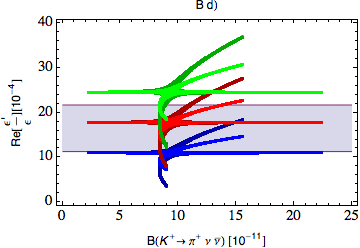}
\includegraphics[width = 0.45\textwidth]{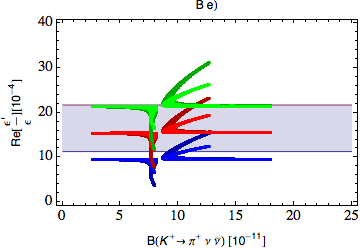}
\includegraphics[width = 0.45\textwidth]{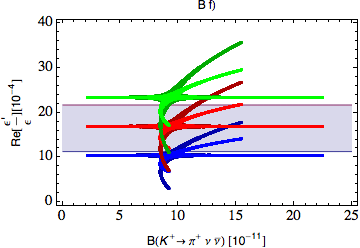}
\caption{\it  $\epe$ versus $\mathcal{B}(\kpn)$ for scenario $a)-f)$ and different values of $B_6^{(1/2)} = 0.75$ (blue), $B_6^{(1/2)} = 
1.00$ (red), $B_6^{(1/2)} = 1.25$ (green) and $\Delta_R^{qq}(Z^\prime)=1.0(0.5)$ for darker(lighter) colours. Gray region: 2$\sigma$
experimental range of $\epe$.  }\label{fig:epsvsBrpiu}~\\[-2mm]\hrule
\end{figure}

We find that due to the absence of the constraint from the $\Delta I=1/2$ 
rule in all six scenarios for  $(\vcb,\vub)$ agreement with the data on 
$\varepsilon_K$ and $\epe$ can be obtained. 
In Fig.~\ref{fig:pKLvsKpA} we show the correlation between $\mathcal{B}(\klpn)$ and $\mathcal{B}(\kpn)$ for the six  scenarios $a)-f)$ for 
$(\vcb,\vub)$. In Figs.~\ref{fig:epsvsBr0} and~\ref{fig:epsvsBrpiu} we show correlations of $\epe$ with  $\mathcal{B}(\klpn)$ and 
$\mathcal{B}(\kpn)$, 
respectively.

We make the following observations:
\begin{itemize}
\item
The plot in  Fig.~\ref{fig:pKLvsKpA} is
familiar from other NP scenarios.
$\mathcal{B}(\klpn)$ can be strongly enhanced on one of the  branches and 
then $\mathcal{B}(\kpn)$ is also enhanced. But  $\mathcal{B}(\kpn)$ can 
also be enhanced without modifying  $\mathcal{B}(\klpn)$. The last feature 
is not possible within the SM and any model with  minimal flavour 
violation in which these two branching ratios are strongly correlated.
\item
As seen in  Fig.~\ref{fig:epsvsBr0} except for the smallest values of $\mathcal{B}(\klpn)$, where this branching 
ratio is below the  SM predictions, in each scenario there is a strong 
correlation between $\epe$ and this branching ratio so that for fixed 
$\bsi$ the increase of $\epe$ uniquely implies the increase of  $\mathcal{B}(\klpn)$. In this case as seen in Fig.~\ref{fig:pKLvsKpA} also 
$\mathcal{B}(\kpn)$ increases so
that we have actually a triple correlation.
\item
We note that even a small increase of $\epe$ for fixed values of $\bsi$ 
implies a strong increase of  $\mathcal{B}(\klpn)$. But this hierarchy 
applies only for $\Delta_R^{qq}(Z^\prime)$ and $\Delta_L^{\nu\nu}(Z^\prime)$ 
being of the same order as assumed in  (\ref{setting}). Introducing 
a hierarchy in these couplings would change the effects in favour of $\epe$ 
or  $\mathcal{B}(\klpn)$ relative to the results presented by us. In the case of $Z$ boson with FCNCs analyzed in Section~\ref{sec:5b}, 
where all diagonal 
couplings are fixed, definite results for this correlation will be obtained.
\item
Values of $\bsi=1.25$ are disfavored for scenarios $c)-f)$ unless 
$\mathcal{B}(\klpn)$ is suppressed with respect to the SM value.
\item
For  $\bsi=1.0$ the branching ratio $\mathcal{B}(\klpn)$ can reach 
values as high as $10^{-10}$ but in view of the experimental error in 
$\epe$ this is not required by $\epe$.
\item
For $\bsi=0.75$ SM prediction for $\epe$ is in all scenarios $a)-f)$  visibly below 
the data and curing this problem with $Z^\prime$ exchange enhances 
 $\mathcal{B}(\klpn)$ typically above $1.5\times 10^{-10}$. 
\item
The main message from these plots is that values of  $\mathcal{B}(\klpn)$ 
as large as several $10^{-10}$ are not possible when $\epe$ constraint is 
taken into account unless the coupling $\Delta_R^{qq}(Z^\prime)$ is chosen 
to be much smaller than assumed by us.
\item
The correlation between $\epe$ and  $\mathcal{B}(\kpn)$ is more involved 
as here also real part of $\Delta_L^{sd}(Z^\prime)$ plays a role. In 
particular we observe that $\mathcal{B}(\kpn)$  can increase without 
affecting $\epe$ at all. But then it is bounded from above by $K_L\to\mu^+\mu^-$  although this bound depends on the value of the 
$Z^\prime$ 
axial vector coupling to muons which is not specified here.  If this coupling  equals $\Delta_L^{\nu\nu}(Z^\prime)$  then as seen in 
Fig.~10 
in \cite{Buras:2012jb} 
values 
of $\mathcal{B}(\kpn)$ above $15\cdot 10^{-11}$ are excluded. 
\end{itemize}

We emphasize that the correlation between $\epe$ and the branching ratio 
$\mathcal{B}(\klpn)$  shown in 
Figs.~\ref{fig:epsvsBr0} and~\ref{fig:epsvsBrpiu}
differs markedly from many other NP scenarios, 
in particular LHT \cite{Blanke:2007wr} and SM with four generations \cite{Buras:2010pi}, where 
$\epe$ was modified by electroweak penguin contributions. There, the increase 
of $\mathcal{B}(\klpn)$ implied the decrease of $\epe$ and only the 
values of $\bsi$ significantly larger than unity allowed large enhancements 
of $\mathcal{B}(\klpn)$. However, the correlations in 
Figs.~\ref{fig:epsvsBr0} and~\ref{fig:epsvsBrpiu} are valid for 
 the assumed  $\Delta_R^{qq}(Z^\prime)$. For the opposite sign of $\Delta_R^{qq}(Z^\prime)$
the values of $\epe$ are flipped along the horizontal ``central'' line without 
the change in the branching ratios which do not depend on this coupling. 
 Similar flipping  the sign of   $\Delta_L^{\nu\nu}(Z^\prime)$ would change 
the correlation between $\epe$ and  $\mathcal{B}(\klpn)$ into anticorrelation.
\boldmath
\subsection{The Primed Scenarios and the $\Delta I=1/2$ Rule}\label{PRIMED}
\unboldmath
Clearly the solution for the missing piece in ${\rm Re} A_0$ can also be obtained by choosing $\Delta_R^{sd}(Z^\prime)$ and 
$\Delta_L^{qq}(Z^\prime)$ to 
be $\ord(1)$ instead of  $\Delta_L^{sd}(Z^\prime)$ and $\Delta_R^{qq}(Z^\prime)$, 
respectively. Interchanging $L$ and $R$ in the hierarchies (\ref{couplingsh}) 
would then lead from the point of view of low energy flavour violating processes to the same conclusions which can be understood as follows.

In this primed scenario the operator $Q_6^\prime$ replaces $Q_6$ and 
as the matrix element $\langle Q_6^\prime\rangle_0$ differs by sign from 
$\langle Q_6\rangle_0$, the $\Delta I=1/2$ rule requires the product
 $\Delta_R^{sd}(Z^\prime)\times \Delta_L^{qq}(Z^\prime)$ to be positive. Choosing 
then positive $\Delta_L^{qq}(Z^\prime)$ instead of a  negative $\Delta_R^{qq}(Z^\prime)$ in Scenario A 
our results for $\epe$ and  ${\rm Re} A_0$ remain unchanged as also the $\Delta S=2$ analysis remains unchanged. 
Similarly our analysis of $\kpn$ and $\klpn$ is not modified as these decays 
are insensitive to $\gamma_5$. The only change takes place in $K_L\to\mu^+\mu^-$  where for a fixed muon coupling NP contribution has 
opposite sign to 
the scenarios considered by us. But this change can be compensated by a flip of 
the sign of the muon coupling which without a concrete model is not fixed.

On the other hand the difference between primed and unprimed scenarios 
could possibly 
be present in other processes, like the ones studied at the LHC, in which 
the constraints on the couplings could depend on whether the bounds 
on a negative product $\Delta_L^{sd}(Z^\prime)\times \Delta_R^{qq}(Z^\prime)$ or 
a positive product $\Delta_R^{sd}(Z^\prime)\times \Delta_L^{qq}(Z^\prime)$ are 
more favourable for the $\Delta I=1/2$ rule. However, presently, as discussed above, only separate bounds on the couplings involved and not their products are 
available. Whether the future bounds on these products will improve the situation of the  $\Delta I=1/2$ rule remains to be seen.

\boldmath
\section{Coloured Neutral Gauge Bosons $G^\prime$}\label{sec:5}
\unboldmath
\subsection{Modified Initial Conditions}
In various NP scenarios neutral gauge bosons with colour ($G^\prime$) are present. One 
of the prominent examples of this type are Kaluza-Klein gluons in Randal-Sundrum scenarios that belong to the adjoint representation of the 
colour $SU(3)_c$. In what 
follows we will assume that these gauge bosons carry a common mass $M_{G^\prime}$ and being in the octet representation of $SU(3)_c$
couple to fermions in the same manner as gluons do. However, we will allow 
for different values of their left-handed and right-handed couplings. Therefore  up to the colour matrix $t^a$, the couplings to quarks will 
be again parametrized by:
\be\label{couplingsA}
\Delta_L^{s d}(G^\prime),\qquad \Delta_R^{s d}(G^\prime), \qquad \Delta_L^{q q}(G^\prime),\qquad \Delta_R^{qq}(G^\prime)
\ee
and the hierarchy in (\ref{couplingsh}) will be imposed.

Calculating then the tree-diagrams with $G^\prime$ gauge boson exchanges and expressing  the result in terms of the operators encountered in 
previous sections we 
find that the initial conditions at $\mu=M_{G^\prime}$ are modified.

The new initial conditions for the operators entering $K\to\pi\pi$ read now at 
LO as follows
\begin{align}
\begin{split}
C_3(M_{G^\prime})
& = \left[-\frac{1}{6}\right]\frac{\Delta_L^{s d}(G^\prime)\Delta_L^{q q}(G^\prime)}{4 M^2_{G^\prime}}, \qquad 
C_3^\prime(M_{G^\prime})
 = \left[-\frac{1}{6}\right]\frac{\Delta_R^{s d}(G^\prime)\Delta_R^{q q}(G^\prime)}{4 M^2_{G^\prime}}
 \,,\end{split}\label{C3A}\\
\begin{split}
C_4(M_{G^\prime})
& = \left[\frac{1}{2}\right]\frac{\Delta_L^{s d}(G^\prime)\Delta_L^{q q}(G^\prime)}{4 M^2_{G^\prime}}, \qquad 
C_4^\prime(M_{G^\prime})
 = \left[\frac{1}{2}\right]\frac{\Delta_R^{s d}(G^\prime)\Delta_R^{q q}(G^\prime)}{4 M^2_{G^\prime}}
 \,,\end{split}\label{C4A}\\
\begin{split}
C_5(M_{G^\prime})
& = \left[-\frac{1}{6}\right]\frac{\Delta_L^{s d}(G^\prime)\Delta_R^{q q}(G^\prime)}{4 M^2_{G^\prime}}, \qquad 
C_5^\prime(M_{G^\prime})
 = \left[-\frac{1}{6}\right]\frac{\Delta_R^{s d}(G^\prime)\Delta_L^{q q}(G^\prime)}{4 M^2_{G^\prime}}
 \,,\end{split}\label{C5A}\\
\begin{split}
C_6(M_{G^\prime})
& = \left[\frac{1}{2}\right]\frac{\Delta_L^{s d}(G^\prime)\Delta_R^{q q}(G^\prime)}{4 M^2_{G^\prime}}, \qquad 
C_6^\prime(M_{G^\prime})
 = \left[\frac{1}{2}\right]\frac{\Delta_R^{s d}(G^\prime)\Delta_L^{q q}(G^\prime)}{4 M^2_{G^\prime}}
 \,.\end{split}\label{C6A}
\end{align}
Again due to the hierarchy in (\ref{couplingsh}) the contributions 
of primed operators can be neglected. Moreover, due the non-vanishing 
value of $C_6(M_{G^\prime})$ the dominance of the operator $Q_6$  
is this time even more pronounced than in the case of a colourless $Z^\prime$. 
Indeed we find now

\begin{equation}\label{C5C6c} 
\left[\begin{array}{c} C_5(m_c) \\
                         C_6(m_c)
    \end{array}\right]
 = \left[\begin{array}{cc} 0.86 & 0.19\\
                          1.13 & 3.60
    \end{array}\right]  \left[\begin{array}{c} -1/6 \\
                         1/2
    \end{array}\right] \frac{\Delta_L^{s d}(G^\prime)\Delta_R^{q q}(G^\prime)}{4 M^2_{G^\prime}}.
  \end{equation}
Consequently 
\be
 C_5(m_c)= -0.05  \frac{\Delta_L^{s d}(G^\prime)\Delta_R^{q q}(G^\prime)}{4 M^2_{G^\prime}}\qquad   C_6(m_c)= 1.61\frac{\Delta_L^{s 
d}(G^\prime)\Delta_R^{q q}(G^\prime)}{4 M^2_{G^\prime}}.
\ee

Also the initial conditions for $\Delta S=2$ transition change:
\begin{align}\label{equ:WilsonZA}
\begin{split}
C_1^\text{VLL}(M_{G^\prime})=\left[\frac{1}{3}\right]\frac{(\Delta_L^{sd}(G^\prime)^2}{2M_{G^\prime}^2},  & \qquad     
C_1^\text{VRR}(M_{G^\prime}) =\left[\frac{1}{3}\right]\frac{(\Delta_R^{sd}(G^\prime))^2}{2M_{G^\prime}^2}
\,,\end{split}\\
\begin{split}
 C_1^\text{LR}(M_{G^\prime}) =\left[-\frac{1}{6}\right]\frac{\Delta_L^{sd}(G^\prime)\Delta_R^{sd}(G^\prime)}{ M_{G^\prime}^2}, & \qquad 
C_2^\text{LR}(M_{G^\prime})= \left[-1\right]\frac{\Delta_L^{sd}(G^\prime)\Delta_R^{sd}(G^\prime)}{ M_{G^\prime}^2}\,.
\end{split}
 \end{align}
The NLO QCD corrections to tree-level coloured gauge boson exchanges at $\mu=M_{G^\prime}$ to $\Delta S=2$ are not known. They are expected 
to be small due to small 
QCD coupling at this high scale and serve mainly to remove certain renormalization scheme and matching scale uncertainties. More important 
is 
the  RG evolution from low energy scales to $\mu=M_{G^\prime}$ necessary 
to evaluate $\langle Q_1^\text{VLL}(M_{G^\prime})\rangle$ and 
 $\langle Q_{1,2}^\text{LR}(M_{G^\prime})\rangle$. Here we include NLO QCD 
corrections using the technology in \cite{Buras:2001ra}. Again 
$Q_{1}^\text{VLL}$
remains the only operator in 
scenario B while   $Q_{1,2}^\text{LR}$ contributing in scenario A  help in 
solving the problem with $\Delta M_K$.

\boldmath
\subsection{${\rm Re}A_0$ and ${\rm Im}A_0$}
\unboldmath

Proceeding as in the case of a colourless $Z^\prime$ we find
\be\label{RENPFINALC}
{\RE} A_0^{\rm NP}={{\RE} \Delta_L^{sd}}(G^\prime) K^c_6(M_{G^\prime}) \left[0.7\times 10^{-8}\gev\right], 
\ee
\be\label{IMNPFINALC}
{\IM} A_0^{\rm NP}={{\IM}\Delta_L^{sd}}(G^\prime) K^c_6(M_{G^\prime}) \left[0.7\times 10^{-8}\gev\right],
\ee
where we have defined $\mu$-independent factor
\be
K_6(M_{G^\prime})=-r^c_6(\mu) \Delta_R^{qq}(G^\prime)\, \left[ \frac{3\tev}{M_{G^\prime}}\right]^2 \, \left[ \frac{114\mev}{m_s(\mu) + 
m_d(\mu)}\right]^2 \,B_6^{(1/2)}\, 
\ee
with the renormalization group factor $r^c_6(\mu)$ defined by
\be
C_6(\mu) = \left[\frac{1}{2}\right]
\frac{\Delta_L^{s d}(G^\prime)\Delta_R^{q q}(G^\prime)}{4 M^2_{G^\prime}} r^c_6(\mu).
\ee
Even if formulae (\ref{RENPFINALC}) and (\ref{IMNPFINALC}) involve an explicit 
factor $0.7$ instead of $1.4$ in the case of the colourless case, this decrease  is overcompensated by the value of $r^c_6$ which 
for $\mu=1.3\gev$ is found to be  $r^c_6=3.23$, that is by roughly a factor of 
three larger than $r_6$ in the colourless case.

Demanding now that $P\%$ of the experimental value of ${\rm Re}A_0$ in 
(\ref{N1}) comes from $G^\prime$ contribution, we arrive at the condition:
\be\label{condition1c}
{\RE \Delta_L^{sd}}(G^\prime) K^c_6(M_{G^\prime}) = 7.8\, \left[\frac{P\%}{20\%}\right].
\ee
Consequently the couplings ${\rm Re} \Delta_L^{sd}(G^\prime)$ and $\Delta_R^{q q}(G^\prime))$ 
must have opposite signs and  must satisfy 
\be\label{condition4}
{\rm Re} \Delta_L^{sd}(G^\prime)\Delta_R^{q q}(G^\prime)\left[ \frac{3\tev}{M_{Z^\prime}}\right]^2\bsi= -2.4\, 
\left[\frac{P\%}{20\%}\right].
\ee

In view of the fact that  $r^c_6$ is larger than $r_6$ by a factor of 2.9, ${\RE \Delta_L^{sd}}$ can be by a factor of $1.4$ smaller than in 
the colourless case in order to reproduce the data on  ${\rm Re}A_0$. 

We also find
\be
{\IM} A_0^{\rm NP}=\frac{\IM \Delta_L^{sd}}{\RE \Delta_L^{sd}} \left[\frac{P\%}{20\%}\right]\left[5.4\times 10^{-8}\gev\right]\,.
\ee

\boldmath
\subsection{$\Delta M_K$ Constraint}
\unboldmath
Beginning with LHS scenario B we find that
due to the modified initial conditions $\Delta S(K)$ is by the 
colour factor $1/3$ suppressed relative to the colourless case
\be\label{Zprime1VLLc}
\Delta S(K)=0.8\,
\left[\frac{\Delta_L^{sd}(G^\prime)}{\lambda_t}\right]^2
\left[\frac{3\tev}{M_{G^\prime}}\right]^2.
\ee
Consequently  allowing conservatively that NP contribution is at most as large as the short distance SM contribution to $\Delta M_K$ we find 
the bound 
on a real $\Delta^{sd}_L(G^\prime)$
\be\label{DeltaMKboundc}
|\Delta^{sd}_L(G^\prime)|\le 0.007\, \left[\frac{M_{G^\prime}}{3\tev}\right].
\ee
This softer bound is still in conflict with (\ref{condition1c})  
and we conclude that also in this case the LHS scenario does not provide a significant 
 NP contribution to  ${\rm Re} A_0$ when $\Delta M_K$ constraint 
is taken into account.  On the other hand in this scenario 
there are no NP contributions to $\kpn$ and $\klpn$ because of the  vanishing $G^\prime\nu\bar\nu$ coupling.  This fact offers of 
course an important test of this 
scenario.

In scenario A for couplings assuming first for simplicity
that the 
couplings $\Delta^{sd}_{L,R}(G^\prime)$ are real, we find
\be\label{DMZnewc}
 \Delta M_K(G^\prime)   =
 \frac{(\Delta_L^{sd}(G^\prime))^2}{3M_{G^\prime}^2} \langle Q_1^\text{VLL}(M_{G^\prime})\rangle 
\left[1+\left(\frac{\Delta_R^{sd}(G^\prime)}{\Delta_L^{sd}(G^\prime)}\right)^2+6\left(\frac{\Delta_R^{sd}(G^\prime)}{\Delta_L^{sd}(G^\prime)
}\right) 
\frac{\langle Q^\text{LR}(M_{G^\prime})\rangle_c}{\langle Q_1^\text{VLL}(M_{G^\prime})\rangle}\right],
 \ee
with $\langle Q_1^\text{VLL}(M_{G^\prime})\rangle$  as before but
\be
\langle Q^\text{LR}(M_{G^\prime})\rangle_c\equiv -\frac{1}{6}\langle Q_1^\text{LR}(M_{G^\prime})\rangle-\langle 
Q_2^\text{LR}(M_{G^\prime})\rangle \approx  -143\, \langle Q_1^\text{VLL}(M_{G^\prime})\rangle.
\ee
We indicate with the subscript ''c'' that the initial conditions for 
Wilson coefficients are modified relative to the case of a colourless 
$Z^\prime$. Hadronic matrix elements remain of course unchanged except that 
in view of the absence of NLO QCD corrections at the high matching scale 
no {\it hats} are present.

Denoting then the analog of suppression factor $\delta$ by $\delta_c$ we 
find that the required suppression of $\Delta M_K$ is given by 
\be\label{deltac}
\delta_c=0.002 \left[\frac{r^c_6(m_c)}{3.23}\right] \Delta_R^{qq}(G^\prime)\, \left[ 
\frac{3\tev}{M_{G^\prime}}\right]B_6^{(1/2)}\left[\frac{20\%}{P\%}\right]\, 
\ee
and in our toy model is given by
\be
\delta_c=\left[1+\left(\frac{\Delta_R^{sd}(G^\prime)}{\Delta_L^{sd}(G^\prime)}\right)^2+6\left(\frac{\Delta_R^{sd}(G^\prime)}{\Delta_L^{sd}
(G^\prime)}\right) 
\frac{\langle Q^\text{LR}(M_{G^\prime})\rangle_c}{\langle Q_1^\text{VLL}(M_{G^\prime})\rangle}\right]^{1/2}\,.
\ee
Consequently also in this case the problem with $\Delta M_K$ can be solved 
by suitably adjusting the coupling $\Delta_R^{sd}(G^\prime)$. 

The expression for 
$\Delta_R^{sd}(G^\prime)$  in our toy model  
now reads 
\be\label{tuningc}
\frac{\Delta_R^{sd}(G^\prime)}{\Delta_L^{sd}(G^\prime)}=
-\frac{1}{6} R^c_Q(1+h (R^c_Q)^2), \qquad  R^c_Q\equiv\frac{\langle Q_1^\text{VLL}((M_{G^\prime})\rangle}{\langle  
Q_1^\text{LR}((M_{G^\prime})\rangle_c}\approx -0.7\times 10^{-2}
\ee
and consequently 
\be
\delta_c=\frac{1}{6} R^c_Q(1-36 h)^{1/2} +\ord((R^c_Q)^2)
\ee
which shows that by a proper choice of the parameter $h$ one can suppress NP 
contributions to $\Delta M_K$ to the level that it agrees with experiment.

We find then
\be\label{epscolour}
\varepsilon_K(G^\prime)=-\frac{\kappa_\eps e^{i\varphi_\eps}}{\sqrt{2}(\Delta M_K)_\text{exp}}\frac{({\rm Re}\Delta^{sd}_L(G^\prime))({\rm 
Im}\Delta^{sd}_L(G^\prime))}{3\, M_{G^\prime}^2}
\langle Q_1^\text{VLL}(M_{G^\prime})\rangle \delta_c^2\equiv\tilde\varepsilon_K(G^\prime)e^{i\varphi_\eps},\, ,
\ee

\be\label{DMKcolour}
\Delta M_K(G^\prime)=\frac{({\rm Re}\Delta^{sd}_L(G^\prime))^2}{3\, M_{G^\prime}^2}
\langle Q_1^\text{VLL}(M_{G^\prime})\rangle  \delta_c^2\,.
\ee

Consequently we find the correlations

\be\label{Brelation2c}
\tilde\varepsilon_K(G^\prime)=
-\frac{\kappa_\eps}{\sqrt{2} r_{\Delta M} }\left[\frac{{\IM \Delta_L^{sd}(G^\prime)}}{{\RE \Delta_L^{sd}(G^\prime)}}\right], \qquad 
r_{\Delta M}=\left[\frac{(\Delta M_K)_\text{exp}}{\Delta M_K(G^\prime)}\right]\,,
\ee

\be\label{Basicrelation1c}
\left(\frac{\varepsilon'}{\varepsilon}\right)_{G^\prime}= \frac{3.5}{\kappa_\eps}\, \tilde\varepsilon_K(G^\prime) 
\left[\frac{P\%}{20\%}\right] r_{\Delta M}\,.
\ee

We note that these correlations  are exactly the same as in the colourless case and we can use the three step procedure used in the latter 
case. 
But there 
 are the following differences which will change the numerical analysis:
\begin{itemize}
\item
The relation (\ref{condition4}) differs from the one in (\ref{condition2}) 
so that a smaller value of
 the product $|{\rm Re} \Delta_L^{sd}(G^\prime)\Delta_R^{q q}(G^\prime)|$ than of 
$|{\rm Re} \Delta_L^{sd}(Z^\prime)\Delta_R^{q q}(Z^\prime)|$ is required to 
obtain a given value of $P$.
\item
But the LHC constraints on $\Delta_R^{q q}(G^\prime)$, $\Delta_L^{sd}(G^\prime)$ and $M_{G^\prime}$ differ from the 
ones on $\Delta_R^{q q}(Z^\prime)$, $\Delta_L^{sd}(Z^\prime)$ and $M_{Z^\prime}$ and therefore in order 
to find out whether $G^\prime$ or $Z^\prime$ contributes more to ${\rm Re}A_0$ 
these constraints have to be taken into account.  See below.
\item
NP contributions to $\kpn$ and $\klpn$ vanish.
\end{itemize}

\begin{figure}[ht]
	\begin{center}
		\includegraphics[scale=0.46]{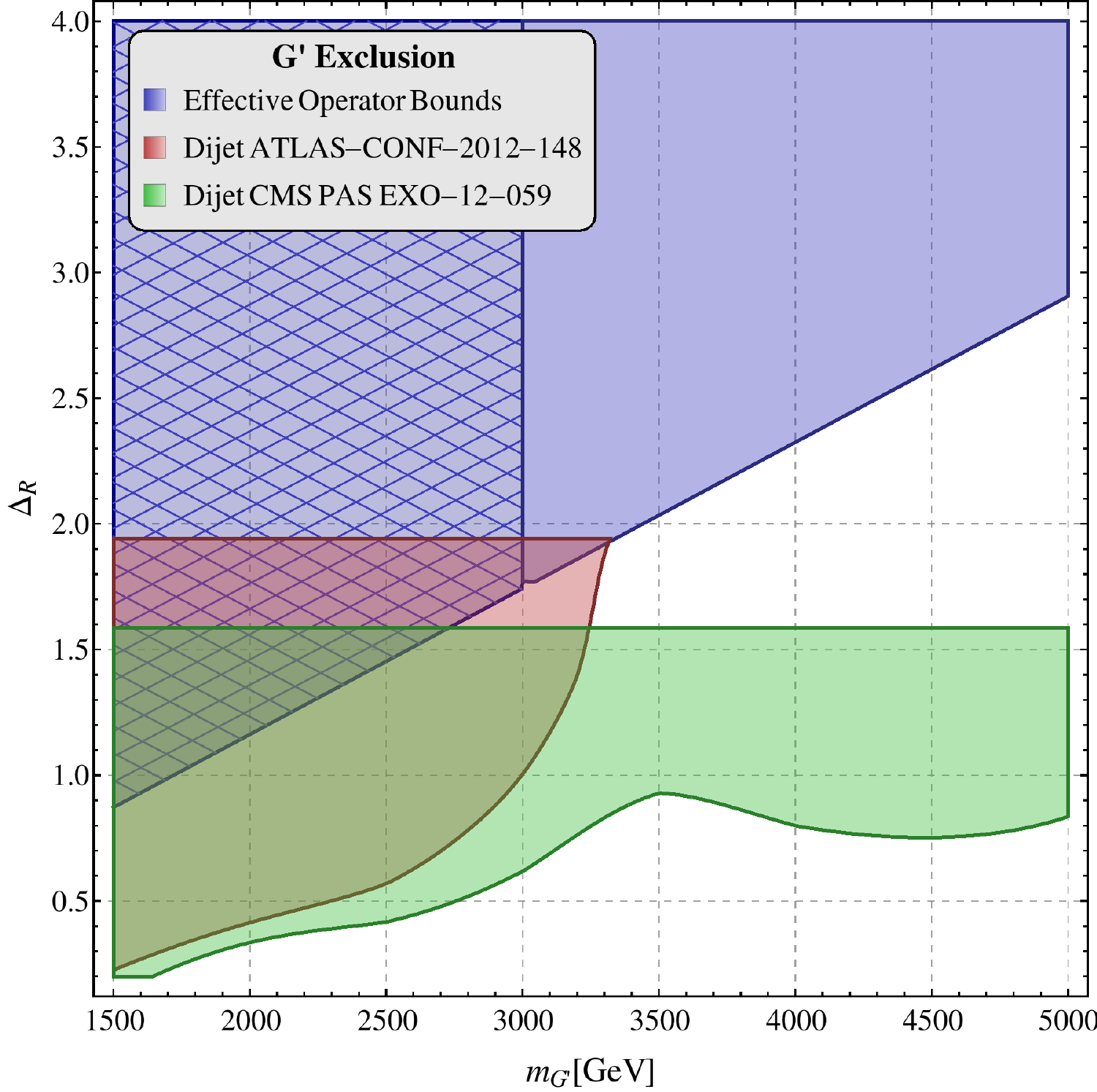}
		\caption{\it Exclusion limits for the $G'$ in the mass-coupling plane, from various searches at the LHC as found in \cite{Vries}. The blue region is 
excluded by effective operator bounds provided by ATLAS \cite{ATLAS:2012pu} and CMS\cite{Chatrchyan:2012bf}. The dashed surface represents the region 
where the effective theory is not applicable, and the bounds here should be interpreted as a rough estimate. The red and green contours are 
excluded by dijet resonance searches by ATLAS \cite{ATLAS:2012qjz} and CMS \cite{CMS:kxa}. See for additional comments in the text.
} 
		\label{fig:gprimeexclusion}
	\end{center}
\end{figure}

\subsection{Numerical Results}\label{GprimeNum}
\subsubsection{Scenario A}
In the case of Scenario A, we just follow the steps performed for $Z^\prime$
 but as correlation between $\epe$ and $\varepsilon_K$ is the 
same we just indicate for which values of $\bsi$ and $P$  this correlation 
is consistent with the data on $\epe$ and $\varepsilon_K$ and the LHC 
constraints on the relevant couplings.

Concerning the LHC constraints a dedicated analysis of our toy $G^\prime$ 
model has been performed in \cite{Vries} with the results given in 
Fig.~\ref{fig:gprimeexclusion}. Additional comments made in 
connection with the bounds on $Z^\prime$ couplings in Fig.~\ref{fig:zprimeexclusion} also apply here. In particular the complete exclusion 
of the dashed 
surface would require a new ATLAS and CMS study in the context of our simple 
model.

These results can be summarized as follows
\begin{itemize}
\item
From dijets constraints the upper bounds can only be obtained for 
$|\Delta_R^{qq}(G^\prime)|\le 1.9$ and at this value  only $M_{Z^\prime}\ge 3.3\tev$ is allowed.
\item
The effective operator bounds can be summarized by 
\be\label{DRLHCboundG}
|\Delta_R^{qq}(G^\prime)|\le 2.0\times \left[ \frac{M_{Z^\prime}}{3.5\tev}\right].
\ee
We note that the bound in this case is weaker than in the case of 
$Z^\prime$ which is partly the result of colour factors that 
suppress NP contributions.
\item
We are not aware of any  LHC bound on the $\Delta S=2$ operator in this 
case but we expect on the basis of the last finding that this bound 
is also weaker than the one on $\Delta_L^{sd}(Z^\prime)$ in (\ref{DLLHCbound}).
However, in the absence of any dedicated analysis we assume that the 
bound on $\Delta_L^{sd}(G^\prime)$ is as strong as the latter bound. A 
simple rescaling then gives
\be\label{DLLHCboundG}
|\Delta_L^{sd}(G^\prime)|\le 2.6 \left[\frac{M_{Z^\prime}}{3.5\tev}\right].
\ee
\end{itemize}

Even if a dedicated analysis of the latter bound would be necessary 
to put our analysis of LHC constraints on firm footing we 
conclude for the time being that $G^\prime$ copes much better with 
the missing piece in ${\rm Re}A_0$ than $Z^\prime$ and consequently 
can provide significantly larger contribution than 
the SM QCD penguin contribution. This is not only the result of the
weaker LHC bound on $\Delta_R^{qq}$ but also of different renormalization 
group effects as seen in (\ref{condition4}).

Putting all the factors together we conclude that $P$ as high as 
$30-35$ is still possible at present and this is sufficient to 
reproduce the $\Delta I=1/2$ rule within $5-10\%$. Indeed taking all 
these bounds into account and using (\ref{condition4}) we arrive 
at the bound
\be\label{PboundGp}
P\le 32\left[\frac{\bsi}{1.0}\right], \qquad (G^\prime)~.
\ee

\begin{figure}[!tb]
 \centering
\includegraphics[width = 0.45\textwidth]{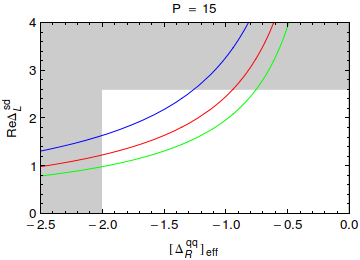}
\includegraphics[width = 0.45\textwidth]{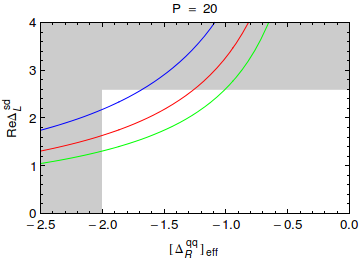}
\includegraphics[width = 0.45\textwidth]{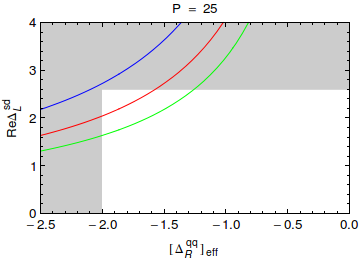}
\includegraphics[width = 0.45\textwidth]{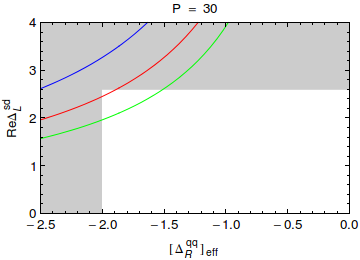}
\caption{\it  ${\rm Re} \Delta_L^{sd}(G^\prime)$ versus $|[\Delta_R^{q q}(G^\prime)]_\text{eff}|$ for $P = 15,~20,~25,~30$ and $\bsi = 
0.75$ 
(blue), $1.00$ (red) and $1.25$ (green). The gray area is basically excluded by the LHC. See additional comments in the text.
}\label{fig:ReDLvsDRc}~\\[-2mm]\hrule
\end{figure}

 In Fig.~\ref{fig:ReDLvsDRc}  we show the results for $G^\prime$ corresponding to 
Fig.~\ref{fig:ReDLvsDR}. As now the values of $P$ 
can be larger we  show  the results for $P=15,~20,~25,~30$. 
With the definition 
\be\label{DeltaeffG}
[\Delta_R^{q q}(G^\prime)]_\text{eff}=\Delta_R^{q q}(G^\prime)\left[\frac{3.5\tev}{M_{Z^\prime}}\right]^2
\ee
the values in  gray area correspond to
to  $|[\Delta_R^{q q}(G^\prime)]_\text{eff}|\ge 2.00$ and 
${\rm Re}\Delta_L^{sd}(G^\prime)\ge 2.6$. Even if these values are already ruled out by 
the LHC it is evident that $G^\prime$ can provide significantly larger values 
of $P$ than $Z^\prime$. We do not show the plot corresponding to Fig.~\ref{fig:epsvsepstoy} as this correlation is also valid in the case of 
$G^\prime$ except 
that now also larger values of $P$, like 25-30, are allowed that correspond 
to steeper lines than $P=20$ in Fig.~\ref{fig:epsvsepstoy}.

\subsubsection{Scenario B}
In the case of Scenario B in the absence of $\Delta I=1/2$ constraint 
and  NP contributions to $\kpn$ and $\klpn$  we can only illustrate how 
going from $Z^\prime$ to $G^\prime$ scenario modifies the allowed oases for 
 $\Delta_L^{sd}$ when the $\epe$, $\varepsilon_K$ and $\Delta M_K$ constraints 
are imposed. To this end  we set\footnote{ The case of $\Delta_R^{qq}(G^\prime)=1.0$ and $M_{G^\prime}=3.0\tev$ is ruled out by dijet data from CMS and direct 
comparison with $Z^\prime$ for these parameters is not possible.}
\be\label{DELTAR}
\Delta_R^{qq}(G^\prime)=\Delta_R^{qq}(Z^\prime)= 0.5, \qquad M_{G^\prime}=M_{Z^\prime}=3.0\tev
\ee
and use in the $G^\prime$  case  the formula 
(\ref{eprimeZprime}) with ${\IM} A_0^{\rm NP}$ given 
in (\ref{IMNPFINALC}). For  the corresponding  contributions to $\varepsilon_K$ and $\Delta M_K$ we use the shift 
in the function $S$ given this time in (\ref{Zprime1VLLc}).

In order to understand better the results below it should be noted that for the same values of the couplings 
$\Delta_R^{qq}$ and $\Delta_L^{sd}$ the contribution of $G^\prime$ to 
$\epe$ is by a factor of $1.4$ larger than  the $Z^\prime$ contribution. 
In the case of $\Delta M_K$ and $\varepsilon_K$ it is opposite: 
$G^\prime$ contribution is by a factor of $3$ smaller than in the $Z^\prime$ 
case.

\begin{figure}[!tb]
\begin{center}
\includegraphics[width=0.45\textwidth] {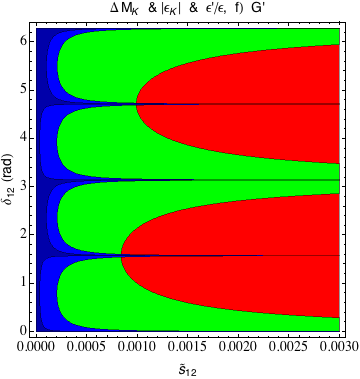}
\includegraphics[width=0.45\textwidth] {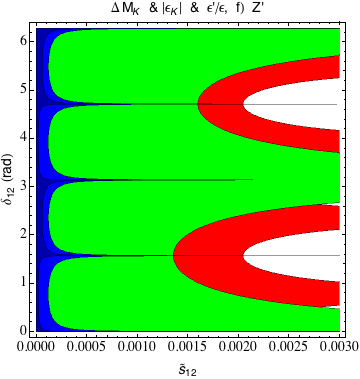}
\includegraphics[width=0.45\textwidth] {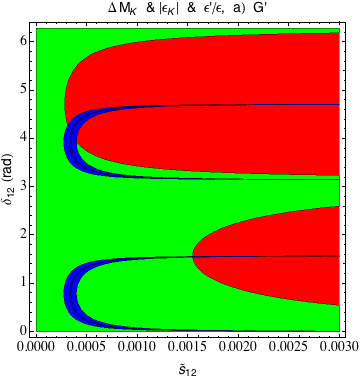}
\includegraphics[width=0.45\textwidth] {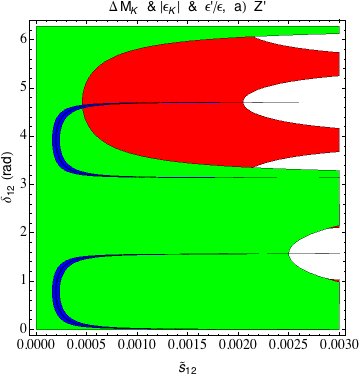}
\caption{\it  Ranges for $\Delta M_K$ (red region) and $\varepsilon_K$ (blue region)  satisfying the bounds in Eq.~(\ref{CK}) (lighter 
blue) and within its $3\sigma$ experimental range (darker blue) and $\epe$ 
(green region) within its $2\sigma$ range $[11.3,\,21.7]\cdot 10^{-4}$ for $B_6^{(1/2)} =1$ and $\Delta_R^{qq} = 0.5$ (green) for CKM scenario $f)$  (top) and  $a)$ (down) and $G'$ (left) and $Z'$ (right).
}\label{fig:oasesKc}~\\[-2mm]\hrule
\end{center}
\end{figure}

In Fig.~\ref{fig:oasesKc} we compare the oases obtained in this manner for $G^\prime$ 
with those obtained for $Z^\prime$ for $\bsi=1.00$ and the scenarios $f)$ and 
$a)$  for $(\vcb,\vub)$.  To this end we have used $2\sigma$ constraint 
for $\epe$ with (\ref{DELTAR}) shown in {\it green.}
For $\varepsilon_K$ we impose either
softer constraint (lighter blue region) in (\ref{CK}) or a tighter $3\sigma$ experimental range (darker blue). 

 We observe the following features:
\begin{itemize}
\item
In all plots the $3\sigma$ constraint from $\varepsilon_K$ (dark blue) determines the 
allowed oasis simply because the present experimental error on $\epe$ is 
unfortunately significant.
\item
The bound on $\Delta_L^{sd}$ from $\varepsilon_K$ is stronger in the case of 
$Z^\prime$. On the other hand the corresponding bound from $\epe$ is stronger 
in the case of $G^\prime$. Both properties follow from the different numerical 
factors in $\epe$ and $\varepsilon_K$ summarized above.
\item
In scenario $f)$, the coupling   $\Delta_L^{sd}$ can vanish as SM value for 
$\varepsilon_K$ is very close to the data. This is not the case in scenario 
$a)$ in which the SM value is well below the data and NP is required to 
enhance $\varepsilon_K$.
\item
In spite of weak constraint from $\epe$, also $\epe$ in scenario $a)$ has 
to be enhanced. This helps to distinguish between two oases that follow from 
$\varepsilon_K$ favouring the one with smaller $\delta_{12}$ in which 
$\epe$ is enhanced over its SM value. But the large experimental error 
on $\epe$ does not allow to exclude the second oasis in which $\epe$ is 
suppressed unless $1\sigma$ constraint on $\epe$ is used.
\end{itemize}

In presenting these results we have set $\bsi=1.0$. Choosing different 
values would change the role of $\epe$ but we do not show these results 
as it is straightforward to deduce the pattern of NP effects for these 
different values of $\bsi$. Similar comment applies to other CKM scenarios.

\boldmath
\section{The Case of $Z$ Boson with FCNCs}\label{sec:5b}
\unboldmath

\subsection{Preliminaries}
We will next discuss the scenario of $Z$ with FCNC couplings in order to demonstrate  that the missing piece in ${\rm Re}A_0$ cannot come 
from this 
corner as this would imply total destruction of the SM agreement with the data on 
${\rm Re}A_2$. Still interesting results for $\epe$ and its correlation 
with the branching ratios for $\kpn$ and $\klpn$ can be found. They are more 
specific than in the $Z^\prime$ case due to the knowledge of all flavour diagonal couplings of $Z$ and of its mass.

Indeed the only freedom in the kaon system in this NP scenario are the complex 
couplings $\Delta^{sd}_{L,R}(Z)$. Its detailed phenomenology  
including  $\Delta S=2$ transitions and rare kaon decays has been presented by us in \cite{Buras:2012jb}. This section generalizes that 
analysis to $K\to\pi\pi$ 
decays, in particular $\epe$ constraint will eliminate some portion of the large enhancements found by us for the branching ratios of rare 
$K$ decays.

In order to understand better our results for $\kpn$ and $\klpn$ in the 
presence of simultaneous constraints from $\epe$ and $K_L\to\mu^+\mu^-$ in 
addition to $\Delta S=2$ constraints let us recall that $\epe$ puts 
constraints only on imaginary parts of NP contributions while $K_L\to\mu^+\mu^-$  only on the real ones. As demonstrated already  in 
\cite{Buras:2012jb} the 
impact of the latter constraint on  $\kpn$ and $\klpn$   depends strongly 
on the scenario for the $Z$ flavour violating couplings: LHS, RHS, LRS, ALRS 
and to lesser extent on the CKM scenarios considered. Moreover, it 
has different impact on $\kpn$ and $\klpn$ as the latter  decay is 
only sensitive to imaginary parts in NP contributions.
Let summarize briefly these findings adding right away brief comments on  
 $\epe$: 
\begin{itemize}
\item
In the LHS scenario the branching ratio for $K_L\to\mu^+\mu^-$ is strongly
enhanced relatively to its SM value and this limits possible enhancement 
of $\mathcal{B}(\kpn)$.
But $\kpn$ receives also NP contribution from imaginary parts 
so that its branching ratio is strongly correlated with the one for $\klpn$ on 
the branch on which both branching can be significantly modified. As we will 
see below the imposition of the $\epe$ contraint will eliminate some part 
of these modifications but this will depend on $\bsi$ and scenarios for 
CKM parameters considered.
\item
In RHS scenario the $K_L\to\mu^+\mu^-$  constraint has a different 
impact on $\kpn$. Indeed, as $K_L\to\mu^+\mu^-$ is sensitive to axial-vector couplings there is a sign flip in NP contributions to the 
relevant decay amplitude  while there 
is no sign flip in the case of $\kpn$. Consequently  the impact of 
$K_L\to\mu^+\mu^-$ on $\kpn$  is now much weaker on the branch where there is 
no NP contribution to $\klpn$ but on the branch where $\kpn$ and $\klpn$
 are strongly correlated we will find the impact of $\epe$ 
constraint.
\item
In the LRS scenario there are no NP contributions to $K_L\to\mu^+\mu^-$ 
so that, as already found in Fig. 30 of \cite{Buras:2012jb} very large 
NP effects in $\kpn$ and $\klpn$ without $\epe$ constraint can be found. 
$\epe$  will again constrain both decays on the branch where these 
decays are strongly correlated but leaving the other branch unaffected.

\item
In the ALRS scenario NP contributions to $\kpn$ and $\klpn$ vanish. 
$\epe$ receives NP contributions  but they are 
unaffected by the ones in $K_L\to\mu^+\mu^-$. In this scenario then $\epe$ 
is not correlated with rare $K$ decays and the only question we can ask is
 how NP physics contributions to $\epe$ are correlated with the ones present 
in  $\varepsilon_K$.
\end{itemize}

\boldmath
\subsection{ ${\rm Re}A_0$ and  ${\rm Re}A_2$ }
\unboldmath
It is straight forward to calculate the values  of the Wilson coefficients  
entering NP part of the $K\to\pi\pi$ Hamiltonian. 
The non-vanishing Wilson coefficients at $\mu=M_{Z}$ are then given at the 
LO as follows
\begin{align}
\begin{split}
C_3(M_{Z})
& = -\left[\frac{g}{6 c_W}\right]\frac{\Delta_L^{s d}(Z)}{4 M^2_Z}, \qquad 
C_5^\prime(M_Z)= -\left[\frac{g}{6 c_W}\right]\frac{\Delta_R^{s d}(Z)}{4 M^2_Z}
 \,,\end{split}\label{C3Z}\\
\begin{split}
C_7(M_Z)
& = -\left[\frac{4 g s_W^2}{6 c_W}\right]\frac{\Delta_L^{s d}(Z)}{4 M^2_Z}, 
\qquad
C_9^\prime(M_Z) =
 -\left[\frac{4 g s_W^2}{6 c_W}\right]\frac{\Delta_R^{s d}(Z)}{4 M^2_Z}\,
\,,\end{split}\label{C7Z}\\
\begin{split}
C_9(M_Z)
& = \left[\frac{4 g c_W^2}{6 c_W}\right]\frac{\Delta_L^{s d}(Z)}{4 M^2_Z}, 
\qquad
C_7^\prime(M_Z) = \left[\frac{4 g c_W^2}{6 c_W}\right]\frac{\Delta_R^{s d}(Z)}{4 M^2_Z}
\,.\end{split}\label{C9Z}
\end{align}
We have used the known flavour conserving couplings of $Z$ to quarks which 
are collected in the same notation in the appendix in \cite{Buras:2013dea}.
  The $SU(2)_L$ gauge coupling constant $g(M_Z)=0.652$.
We note that the values of the coefficients in front of $\Delta_{L,R}$ are 
in the case of $C_9$ and $C_7^\prime$ by a factor of three larger than for 
the remaining coefficients. 

We will first discuss the LHS scenario so that $\Delta_R^{s d}(Z)=0$. Similar 
to $Z^\prime$ scenarios only left-right operators are relevant at low energy scales but this time it is the electroweak penguin operator 
$Q_8$ that dominates 
the scene. Concentrating then on the operators $Q_7$ and $Q_8$, the relevant 
one-loop anomalous dimension matrix in the $(Q_7,Q_8)$ basis is very similar 
to the one in  (\ref{reduced})
\begin{equation}
\hat \gamma^{(0)}_s = 
\left(
\begin{array}{cc}
2 & -6  \\ \svs
0 & -16
   \end{array}
\right).
\label{reducedZ}
\end{equation}

Performing the renormalization group evolution from $M_Z$ to $m_c=1.3\gev$ 
we find
\be\label{LOC7C8}
 C_7(m_c)= 0.87\, C_7(M_Z)\qquad   C_8(m_c)= 0.76\,  C_7(M_Z).
\ee
Due to the large element $(1,2)$ in the matrix (\ref{reducedZ}) and 
the large anomalous dimension of the $Q_8$ operator represented by the $(2,2)$ 
element in (\ref{reducedZ}), the two coefficients are comparable in size.
But  the matrix elements $\langle Q_7\rangle_{0,2}$ are colour suppressed 
which is not the case of  $\langle Q_8\rangle_{0,2}$ and within a good 
approximation we can neglect the contributions of $Q_7$. In summary, it 
is sufficient to keep only $Q_8$ contributions in the decay amplitudes in this scenario for flavour violating $Z$ couplings.

We find then
\be\label{RENPZ}
{\RE} A_0^{\rm NP}={\RE} C_8(m_c)\langle Q_8(m_c)\rangle_0,\qquad
{\RE} A_2^{\rm NP}= {\RE} C_8(m_c)\langle Q_8(m_c)\rangle_2.
\ee

Now the relevant hadronic matrix elements of $Q_8$ operator are given 
as follows
\be
\frac{\langle Q_8(m_c)\rangle_2}{\langle Q_6(m_c)\rangle_0}\approx
-\frac{R_8}{R_6}\frac{F_\pi}{2\sqrt{2}(F_K-F_\pi)}=-1.74 \,\frac{B_8^{(3/2)}}{B_6^{(1/2)}}\,,
\ee
\be\label{Zrule}
\frac{{\RE} A_2^{\rm NP}}{{\RE} A_0^{\rm NP}}=\frac{\langle Q_8(m_c)\rangle_2}{\langle Q_8(m_c)\rangle_0}\approx 
\frac{F_\pi}{\sqrt{2}F_K}\frac{B_8^{(3/2)}}{B_8^{(1/2)}}=0.59 
\frac{B_8^{(3/2)}}{B_8^{(1/2)}},
\ee
with $B_8^{(3/2)}=B_8^{(1/2)}=1$ in the large $N$ limit but otherwise expected to 
be $\ord(1)$ as confirmed in the case of  $B_8^{(3/2)}$ by lattice QCD 
\cite{Blum:2012uk}.

It is evident from (\ref{Zrule}) that the explanation of the missing piece 
in ${\RE} A_0$ with $Z$ exchange would totally destroy the agreement of the SM 
with the data on  ${\RE} A_2$. Rather we should investigate the constraint 
on ${\RE} \Delta_L^{sd}(Z)$ which would allow us to keep this agreement in 
the presence of $Z$ with FCNC couplings.

 Demanding then that at most $P\%$ of the experimental value of ${\rm Re}A_2$ 
in (\ref{N1}) comes from $Z$ contribution, we arrive at the condition
\be\label{condition3}
|{\rm Re} \Delta_L^{sd}(Z) K_8(Z)| \le 6.2\times 10^{-4}\, \left[\frac{P\%}{10\%}\right],
\ee
where 
\be
K_8(M_{Z})=- r_8(\mu) \,  \left[\frac{114\mev}{m_s(\mu) + m_d(\mu)}\right]^2 \,
\left[\frac{B_8^{(3/2)}}{0.65}\right]\, .
\ee
The renormalization group factor $r_8(m_c)=0.76$ is defined by
\be 
C_8(\mu)=r_8(\mu) C_7(M_Z)
\ee
with $C_7(M_Z)$ given in (\ref{C7Z}).

Consequently we arrive at the condition 
\be\label{boundLHS}
|{\rm Re} \Delta_L^{sd}(Z)|\frac{B_8^{(3/2)}}{0.65}\le 8.2\times 10^{-4}\left[\frac{P\%}{10\%}\right].
\ee

In fact this bound is weaker than the one following from $\Delta M_K$. Replacing  $M_{Z^\prime}$ by $M_Z$ the 
bound in (\ref{DeltaMKbound}) is now replaced by 
\be\label{DeltaMKboundZ}
|\Delta^{sd}_L(Z)|\le 1.2 \times 10^{-4}.
\ee
Consequently imposing the $\Delta M_K$ bound in the numerical analysis below 
we are confident that no relevant NP contribution to ${\RE} A_2$ is present.

\boldmath
\subsection{$\epe$, $\kpn$ and $\klpn$}
\unboldmath
We could as in the $Z^\prime$ case calculate separately NP contribution to $\epe$. However,  in the present case the initial conditions for 
Wilson coefficients are at the electroweak scale as in the SM and it is easier to modify the functions $X$, $Y$ and $Z$ entering the 
analytic formula (\ref{epeth}). We find then 
the shifts
\be\label{eprimeshifts}
\Delta X=\Delta Y =\Delta Z= c_W\frac{8\pi^2}{g^3}\frac{{\rm Im}\Delta_L^{sd}(Z)}{{\rm Im}\lambda_t}\, .
\ee 
In doing this we include in fact all operators whose Wilson coefficients 
are affected by NP but effectively only the operator $Q_8$ is really 
relevant. The final formula for $\epe$ in LHS scenario is then given by
\be\label{epsLHS}
\left(\frac{\varepsilon'}{\varepsilon}\right)_\text{LHS}=
\left(\frac{\varepsilon'}{\varepsilon}\right)_\text{SM}+
\left(\frac{\varepsilon'}{\varepsilon}\right)^L_{Z}
\ee
where the second term stands for the modification related to the 
shifts in (\ref{eprimeshifts}).

It should be emphasized that the shifts in (\ref{eprimeshifts}) should only be  used
 in the formula (\ref{epeth}) so that ${\rm Im}\lambda_t$ cancels the one 
present in the SM contribution. $\Delta X$ can also be used in the case of 
$\klpn$. However, in the case of $\kpn$, where also real parts matter 
one should use  the general formula
\be\label{kpnshifts}
\Delta X= c_W\frac{8\pi^2}{g^3}\frac{\Delta_L^{sd}(Z)}{\lambda_t}\, .
\ee 
or equivalently simply use the formulae for $\kpn$ and $\klpn$ in the 
LHS scenario in \cite{Buras:2012jb}.

\subsection{Numerical Analysis in the LHS Scenario}

\begin{figure}[!tb]
\begin{center}
\includegraphics[width=0.45\textwidth] {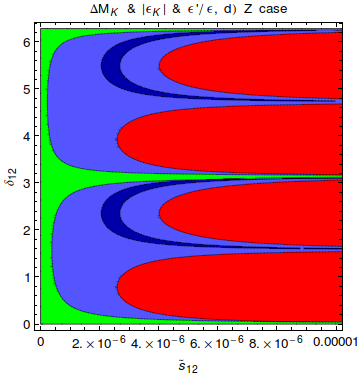}
\includegraphics[width=0.45\textwidth] {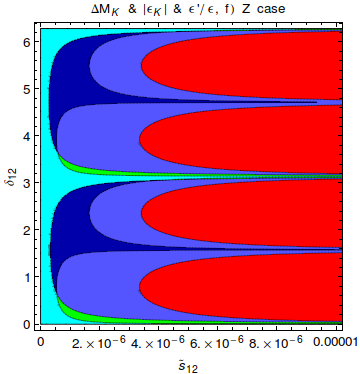}
\includegraphics[width=0.45\textwidth] {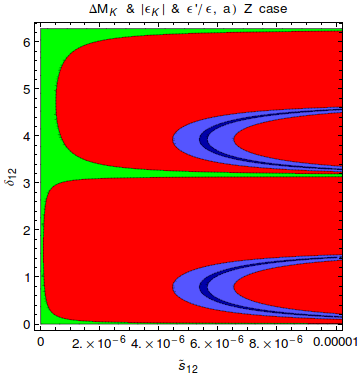}
\caption{\it  Ranges for $\Delta M_K$ (red region) and $\varepsilon_K$ (blue region)  satisfying the bounds in Eq.~(\ref{CK}) (lighter 
blue) and within its $3\sigma$ experimental range (darker blue) and $\epe$ 
(green region) within its $2\sigma$ range $[11.3,\,21.7]\cdot 10^{-4}$ for $B_6^{(1/2)} =1$ for CKM scenario $d)$ (top left), $f)$ 
(top right) and $a)$ (down). { The cyan region in case $f)$ corresponds to the overlap between the green and dark blue region.}
}\label{fig:oasesKLHSZ}~\\[-2mm]\hrule
\end{center}
\end{figure}

\begin{figure}[!tb]
\begin{center}

\includegraphics[width=0.45\textwidth] {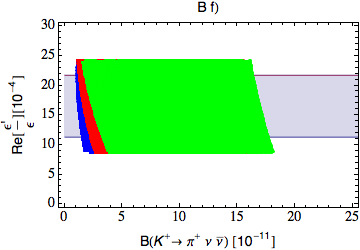}
\includegraphics[width=0.45\textwidth] {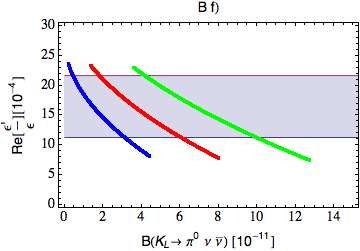}
\caption{\it $\epe$ versus $\mathcal{B}(\kpn)$ (left) and $\epe$ versus
 $\mathcal{B}(\klpn)$ (right) in LHS for scenario  $f)$ including the constraints from $\Delta M_K$, $\varepsilon_K$ from Eq.~(\ref{CK}), $\epe$ 
within 
its $3\sigma$ experimental range for $B_6^{(1/2)} = 0.75$ (blue)  $B_6^{(1/2)} = 1$ (red) and $B_6^{(1/2)} = 1.25$ (green) and 
$\mathcal{B}(K_L\to \mu^+\mu^-)\leq 2.5\cdot
10^{-9}$. Gray range: experimental $2\sigma$ range for $\epe$.
}\label{fig:epevsKpinu}~\\[-2mm]\hrule
\end{center}
\end{figure}

\begin{figure}[!tb]
\begin{center}
\includegraphics[width=0.45\textwidth] {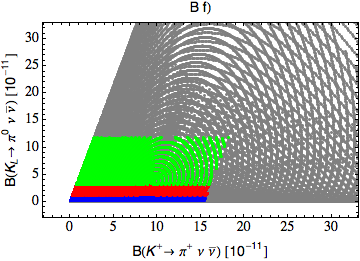}
\caption{\it  $\mathcal{B}(\klpn)$ versus
$\mathcal{B}(\kpn)$ in LHS for scenario  $f)$ including the constraints from $\Delta M_K$, $\varepsilon_K$ from Eq.~(\ref{CK}) (gray 
region) and 
$\epe$ within 
its $3\sigma$ experimental range for $B_6^{(1/2)} = 0.75$ (blue)  $B_6^{(1/2)} = 1$ (red) and $B_6^{(1/2)} = 1.25$ (green) and 
$\mathcal{B}(K_L\to \mu^+\mu^-)\leq 2.5\cdot
10^{-9}$. 
}\label{fig:KpinuZCKMd}~\\[-2mm]\hrule
\end{center}
\end{figure}

In \cite{Buras:2012jb} we have performed a detailed analysis of $\kpn$ and 
$\klpn$ decays in this NP scenario 
imposing the constraints listed above and from $K_L\to\mu^+\mu^-$ decay 
that is only relevant for $\kpn$. The 
present analysis generalizes that analysis in two respects:
\begin{itemize}
\item
We consider several scenarios $a)-f)$ for CKM parameters.
\item
We analyze the correlation between $\epe$ and the branching ratios for $\kpn$ 
and $\klpn$.
\end{itemize}

It is straight forward to convince oneself that unless ${\rm Im}\Delta_L^{sd}(Z)=
\ord(10^{-8})$ the shifts in (\ref{eprimeshifts}) imply modifications of 
$\epe$ that are not allowed by the data. In turn NP contributions to 
$\varepsilon_K$ are negligible and the model can only agree with data on 
$\varepsilon_K$ for which also the SM agrees with them. 
Similar to Scenario A in $Z^\prime$ case only 
scenarios $d)$ and $f)$ survive the $\epe$ constraint. This can be seen in the oases plots in Fig.~\ref{fig:oasesKLHSZ}. In scenario $d)$ 
shown there, 
and even more in scenario $f)$, 
there is an overlap region of the blue ($\varepsilon_K$) and green ($\epe$) range whereas in $a)$ and also in the other CKM scenarios 
there is none.   However, while in scenario $d)$ there is a clear overlap between the $2\sigma$ range of $\epe$ and the larger range of 
$\varepsilon_K$ in~Eq.~(\ref{CK}) (lighter blue), when using  the smaller experimental $3\sigma$ range of $\varepsilon_K$ (darker blue) 
the overlap is tiny. In contrast in scenario $f)$ the cyan region corresponds to the overlap of the darker blue and green region.
Therefore in Fig.~\ref{fig:epevsKpinu} we 
show the correlation of $\epe$ and branching ratios for $\kpn$ and 
$\klpn$ and in Fig.~\ref{fig:KpinuZCKMd} for the correlation between 
 $\kpn$ and  $\klpn$ only for the $f)$ scenario. However, we checked 
that in scenario $d)$ similar results are obtained and this is also the case 
of RHS, LRS and ALRS scenarios considered below. Therefore in the reminder of this section only results for scenario $f)$ will be shown.

 Comparing these results with those in the plots in Figs.~\ref{fig:pKLvsKpA}, \ref{fig:epsvsBr0} and 
 \ref{fig:epsvsBrpiu} for $Z^\prime$ we observe that they are more specific as 
the diagonal couplings of $Z$ and its mass are known and only selected 
 CKM scenarios are allowed.
While significant deviations from SM values for $\epe$,  $\mathcal{B}(\klpn)$, 
 and  $\mathcal{B}(\kpn)$  
are in principle possible, 
the bounds from $\epe$ and $K_L\to\mu^+\mu^-$ that are 
imposed in these plots do not allow  very large 
enhancements of both branching ratios. In particular the bound from 
$\epe$ does not allow large enhancements of $\mathcal{B}(\klpn)$
 that we found in \cite{Buras:2012jb}. 
This analysis shows again how important the $\epe$ constraint is. 
The correlation between $\mathcal{B}(\klpn)$ versus
$\mathcal{B}(\kpn)$ shown in Fig~\ref{fig:KpinuZCKMd} demonstrates 
in a spectacular manner the action of  $\epe$ and $K_L\to\mu^+\mu^-$ constraints. Without them the full gray region would still be allowed 
by 
$\Delta M_K$ and $\varepsilon_K$ constraints.

 The correlation in the right panel of  Fig.~\ref{fig:epevsKpinu} is similar to the one encountered in other NP scenarios 
in which NP in $\epe$ is dominated by electroweak penguins and 
the increase of  $\mathcal{B}(\klpn)$ implies automatically the suppression 
of $\epe$. Therefore only for $\bsi>  1.0$, where $\epe$ within the SM is 
above the data, large enhancements of  $\mathcal{B}(\klpn)$ are possible. 
For the same sign of the neutrino coupling in Scenario B for $Z^\prime$ and 
$\Delta_R^{qq}(Z^\prime)>0$ the correlation between $\epe$ and 
$\mathcal{B}(\klpn)$ is different, as seen in Fig.~\ref{fig:epsvsBr0}, 
because there the QCD penguin operator $Q_6$ instead of $Q_8$ encountered 
here is at work.

\subsection{The RHS Scenario}

\begin{figure}[!tb]
\begin{center}
\includegraphics[width=0.45\textwidth] {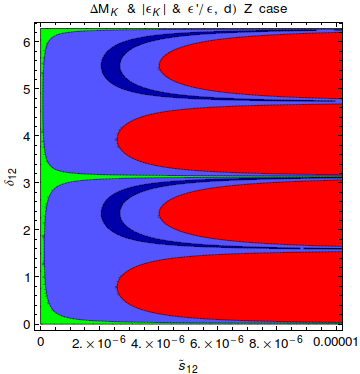}
\includegraphics[width=0.45\textwidth] {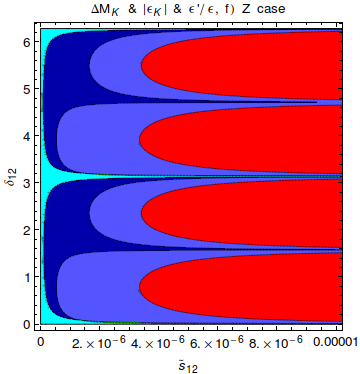}
\caption{\it As in Fig.~\ref{fig:oasesKLHSZ} but for RHS. 
}\label{fig:oasesKRHSZ}~\\[-2mm]\hrule
\end{center}
\end{figure}

\begin{figure}[!tb]
\begin{center}

\includegraphics[width=0.45\textwidth] {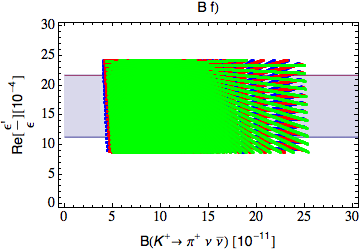}
\includegraphics[width=0.45\textwidth] {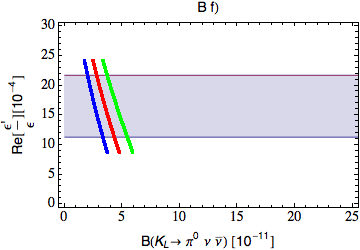}
\caption{\it As in Fig.~\ref{fig:epevsKpinu} but for RHS. 
}\label{fig:epevsKpinuRHS}~\\[-2mm]\hrule
\end{center}
\end{figure}

We discuss next the RHS scenario as here the pattern of NP effects differs from the LHS case. 
In this scenario NP in $K\to\pi\pi$ is  dominated by left-right primed operators.
 This time both $Q_6^\prime$ and $Q_8^\prime$ have to be considered although 
at the end only the latter operator will be important. 
Within a very good approximation we have
\be\label{RENPZ1}
 A_0^{\rm NP}=C^\prime_6(m_c)\langle Q^\prime_6(m_c)\rangle_0+C^\prime_8(m_c)\langle Q^\prime_8(m_c)\rangle_0,
\ee
\be\label{RENPZ2}
A_2^{\rm NP}= C^\prime_8(m_c)\langle Q^\prime_8(m_c)\rangle_2
\ee
where 
\be
C^\prime_6(m_c)=r_6^\prime(m_c) C^\prime_5(M_Z), \qquad C^\prime_8(m_c)=r_8^\prime(m_c) C^\prime_7(M_Z)
\ee
with
\be
r_6^\prime(m_c)\approx r_8^\prime(m_c)=r_8(m_c)=0.76\,.
\ee
Moreover, one has
\be
\langle Q^\prime_6(m_c)\rangle_0=-\langle Q_6(m_c)\rangle_0,\qquad
\langle Q^\prime_8(m_c)\rangle_{0,2}=-\langle Q_8(m_c)\rangle_{0,2}.
\ee

Proceeding as in the LHS scenario we again find that one cannot explain the 
missing piece in ${\RE} A_0$ with $Z$ exchange without totally destroying the agreement of the SM with the data on  ${\RE} A_2$. Due to the different 
initial conditions the upper bound in (\ref{boundLHS}) is replaced by a stronger bound
\be\label{boundRHS}
|{\rm Re} \Delta_R^{sd}(Z)|\left[\frac{B_8^{(3/2)}}{0.65}\right]\le 2.5\times 10^{-4}\left[\frac{P\%}{10\%}\right].
\ee
But in RHS scenario the bound on $|{\rm Re} \Delta_R^{sd}(Z)|$ from $\Delta M_K$ is the same 
as the one for $|{\rm Re} \Delta_L^{sd}(Z)|$ in LHS scenario and consequently 
no problem with  ${\RE} A_2$ arises after the bound from $\Delta M_K$ has been 
taken into account.

Taking first into account both $Q^\prime_6$ and $Q^\prime_8$ contributions to $\epe$ we 
have
\be\label{eprimeZ}
\left(\frac{\varepsilon'}{\varepsilon}\right)_{Z}=
-\frac{\omega_+}{|\varepsilon_K|\sqrt{2}}
\left[\frac{{\IM} A_0^{\rm NP}}{{\RE}A_0}(1-\Omega_{\rm eff})-\frac{{\IM} A_2^{\rm NP}}{{\RE}A_2}\right],
\ee
where ${\RE}A_0$ and ${\RE}A_2$ are to be taken from (\ref{N1}).

While both $Q^\prime_6$ and $Q^\prime_8$ contribute, the latter operator wins 
easily this competition because it is not only enhanced through the 
$\Delta I=1/2$ rule  relative to $Q_6^\prime$  contribution to $\epe$ but also because its Wilson coefficient is larger 
than the one of  $Q^\prime_6$. This is in contrast to the competition between 
$Q_6$ and $Q_8$ in the SM, where the much larger Wilson coefficient of $Q_6$ 
overcompensates the $\Delta I=1/2$ rule effect in question. Thus keeping 
only the $Q^\prime_8$ operator we find  within an 
excellent approximation 
\be\label{eprimeZfinal}
\left(\frac{\varepsilon'}{\varepsilon}\right)^R_{Z}=
\frac{\omega_+}{|\varepsilon_K|\sqrt{2}}\frac{{\IM} A_2^{\rm NP}}{{\RE}A_2}=
-5.3 \times 10^3 \,  \left[\frac{114\mev}{m_s(\mu) + m_d(\mu)}\right]^2 \,
\left[\frac{B_8^{(3/2)}}{0.65}\right]\,{\rm Im} \Delta_R^{sd}(Z)
\ee
implying that ${\rm Im} \Delta_R^{sd}(Z)$ must be $\ord(10^{-8})$ in order for 
$\epe$ to agree with experiment. Then similar to the LHS case just 
discussed NP contribution to $\varepsilon_K$ are negligible and consequently 
only scenarios $d)$ and $f)$ for CKM parameters survive the test.

The final formula for $\epe$ in RHS scenario is now given by
\be\label{epsRHS}
\left(\frac{\varepsilon'}{\varepsilon}\right)_\text{RHS}=
\left(\frac{\varepsilon'}{\varepsilon}\right)_\text{SM}+
\left(\frac{\varepsilon'}{\varepsilon}\right)^R_{Z}
\ee
where the second term is given in (\ref{eprimeZfinal}).

As far as $\kpn$ and $\klpn$ are concerned we can use the formulae in 
\cite{Buras:2012jb}. Equivalently in the case of RHS scenario  one can just 
make a shift in the function $X(K)$:
\be\label{XLKZ}
\Delta X(K)=\left[\frac{\Delta_L^{\nu\bar\nu}(Z)}{g^2_{\rm SM}M_{Z}^2}\right]
             \left[\frac{\Delta_R^{sd}(Z)}{\lambda_t}\right], \qquad 
\Delta_L^{\nu\bar\nu}(Z)=\frac{g}{2c_W}.
\ee

\begin{figure}[!tb]
\begin{center}
\includegraphics[width=0.45\textwidth] {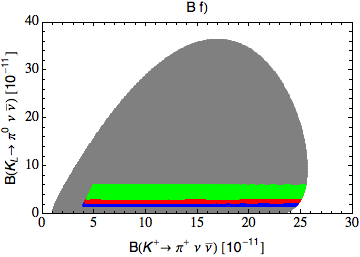}
\caption{\it  $\mathcal{B}(\klpn)$ versus
$\mathcal{B}(\kpn)$ for scenario  $f)$ as in Fig.~\ref{fig:KpinuZCKMd} but for RHS.
}\label{fig:KpinuZCKMdRHS}~\\[-2mm]\hrule
\end{center}
\end{figure}

Repeating the analysis performed in the LHS scenario for the RHS scenario 
we find the results in Figs.~\ref{fig:oasesKRHSZ}-\ref{fig:KpinuZCKMdRHS}. The main messages from these plots when compared with Figs.~\ref{fig:oasesKLHSZ}-\ref{fig:KpinuZCKMd} are as follows:
\begin{itemize}
\item
The constraint from $\epe$ is stronger not allowing as large enhancements 
of $\mathcal{B}(\klpn)$ as in the LHS case,
\item
The constraint from $K_L\to\mu^+\mu^-$ is weaker allowing larger enhancements 
of  $\mathcal{B}(\kpn)$.
\end{itemize}

These results are easy to understand. As already discussed in \cite{Buras:2012jb}  the outcome for the allowed values of $\Delta^{sd}_R(Z)$ following from 
$\Delta M_K$ and $\varepsilon_K$ is identical to the one for  $\Delta^{sd}_L(Z)$.  This is confirmed in Fig.~\ref{fig:oasesKRHSZ} which 
should be compared 
with  Fig.~\ref{fig:oasesKLHSZ}.
But the Wilson coefficient $C_8^\prime(m_c)$ is by a factor of three 
larger than $C_8(m_c)$ in the LHS case. The difference in sign of these 
two coefficients is compensated by the one of hadronic matrix elements 
so that simply the suppression of $\epe$ through NP  and the $\epe$ constraint in Fig.~\ref{fig:oasesKRHSZ}
is by a factor of three 
 stronger than in the LHS case in  Fig.~\ref{fig:oasesKLHSZ}. On the other hand for a given value of 
 $\Delta^{sd}_R(Z)$ the branching ratios $\mathcal{B}(\klpn)$ and $\mathcal{B}(\kpn)$
 are not modified. But the values of  ${\rm Im}\Delta^{sd}_R(Z)$ are now stronger bounded from above by $\epe$ than in the LHS case 
which implies stronger upper bound on  $\mathcal{B}(\klpn)$ as clearly 
seen in Fig.~\ref{fig:epevsKpinuRHS}.  While this also has an impact on $\mathcal{B}(\kpn)$  on 
the branch where the two branching ratios are strongly correlated, on 
the second branch where  ${\rm Re}\Delta^{sd}_R(Z)$ matters, the 
weaker constraint from $K_L\to\mu^+\mu^-$ allows for larger enhancements of $\mathcal{B}(\kpn)$ than in 
the LHS case. The difference in this pattern between LHS and RHS scenarios 
is best seen when comparing Fig.~\ref{fig:KpinuZCKMd} with Fig.~\ref{fig:KpinuZCKMdRHS}.

\subsection{The LRS and ALRS Scenarios} 
When both $\Delta_L^{sd}(Z)$ and $\Delta_R^{sd}(Z)$ are present
the general formula for $\epe$ is given as follows
\be\label{epsgeneral}
\left(\frac{\varepsilon'}{\varepsilon}\right)=
\left(\frac{\varepsilon'}{\varepsilon}\right)_\text{SM}+
\left(\frac{\varepsilon'}{\varepsilon}\right)^L_{Z}+
\left(\frac{\varepsilon'}{\varepsilon}\right)^R_{Z}
\ee
with the last two terms representing LHS and RHS contributions discussed 
above. Imposing relations between  $\Delta_L^{sd}(Z)$ and $\Delta_R^{sd}(Z)$, 
which characterize LRS and ALRS scenarios, one can calculate $\epe$ in 
these scenarios.

As far as rare decays are concerned in LRS scenario NP contributions to 
$K_L\to\mu^+\mu^-$ vanish which allows in principle 
for larger enhancement of $\mathcal{B}(\kpn)$ than it is possible in other scenarios. 
On the other hand for fixed values of $\Delta_L^{sd}(Z)=\Delta_R^{sd}(Z)$ 
the $\epe$ constraint is by a factor of four larger than in the LHS 
case because the operators $Q_8$ and $Q_8^\prime$ contribute to $\epe$ 
with the same 
sign. Therefore it is evident that NP effects in $\mathcal{B}(\klpn)$  
will be even smaller than in the RHS scenario.

\begin{figure}[!tb]
\begin{center}

\includegraphics[width=0.45\textwidth] {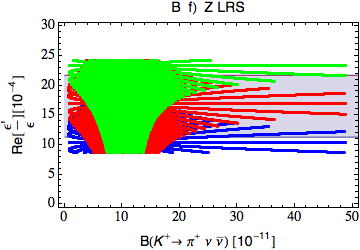}
\includegraphics[width=0.45\textwidth] {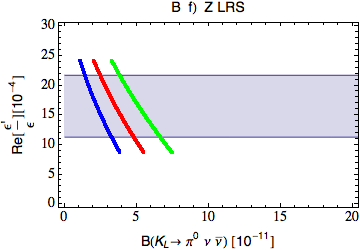}
\caption{\it As in Fig.~\ref{fig:epevsKpinu} but for LRS. 
}\label{fig:epevsKpinuLRS}~\\[-2mm]\hrule
\end{center}
\end{figure}

\begin{figure}[!tb]
\begin{center}
 \includegraphics[width=0.45\textwidth] {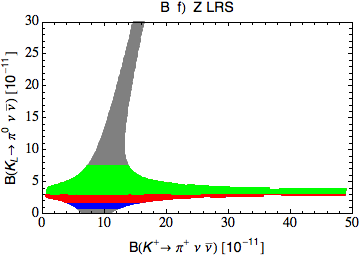}

\caption{\it  $\mathcal{B}(\klpn)$ versus
$\mathcal{B}(\kpn)$ for scenario $d)$ and $f)$ as in Fig.~\ref{fig:KpinuZCKMd} but for LRS.
}\label{fig:KpinuZCKMdLRS}~\\[-2mm]\hrule
\end{center}
\end{figure}

But now comes another effect which suppresses NP contributions in  $\mathcal{B}(\klpn)$   even further. Indeed 
one should recall that in the LRS  scenario the $\Delta S=2$ 
analysis is more involved than in LHS and RHS scenarios because of 
the presence of LR operators which as we have seen in Scenario A for 
the $Z^\prime$ play an essential role in allowing to satisfy constraints 
from $\Delta M_K$ and ${\rm Re} A_0$. But in the case at hand the 
constraints from  $\Delta M_K$ and $\varepsilon_K$ imply simply much smaller allowed values of  $\Delta_L^{sd}(Z)=\Delta_R^{sd}(Z)$ and in turn smaller NP effects  in the branching ratios $\mathcal{B}(\klpn)$ and $\mathcal{B}(\kpn)$. This is partially compensated by the fact that now for fixed  $\Delta_L^{sd}(Z)=\Delta_R^{sd}(Z)$  NP contributions to the amplitudes for $\klpn$ and $\kpn$ are 
enhanced by a factor of two and in the case of $\kpn$ by the absence of 
$K_L\to\mu^+\mu⁻$  constraint. The final result of this competition is shown in 
Figs.~\ref{fig:epevsKpinuLRS} and \ref{fig:KpinuZCKMdLRS}. In particular 
$\mathcal{B}(\kpn)$ can be very much enhanced. Comparison of Figs.~\ref{fig:KpinuZCKMd} (LHS), \ref{fig:KpinuZCKMdRHS} (RHS) and   \ref{fig:KpinuZCKMdLRS} (LRS)
 could one day allow us to distinguish between these three scenarios provided 
 deviations from the SM predictions will be sizable.

In the ALRS scenario NP contributions to $\kpn$ and $\klpn$ vanish but 
$\epe$ is modified. For the same values of $\Delta_R^{sd}(Z)=-\Delta_L^{sd}(Z)$ 
NP effect in $\epe$ is only by a factor of two larger than in LHS scenario 
because the contribution of $Q_8^\prime$ operator to $\epe$ is partially 
cancelled by the one of $Q_8$. Moreover as in the LRS scenario the values of 
the coupling $\Delta_R^{sd}(Z)=-\Delta_L^{sd}(Z)$ must be reduced in order 
to satisfy the  $\Delta M_K$ and $\varepsilon_K$ constraints. But 
on the whole the results do not look interesting and we refrain from showing 
any plots.

\begin{figure}[!tb]
\begin{center}
\includegraphics[width=0.4\textwidth] {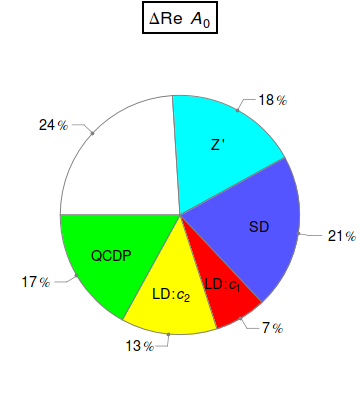}
\includegraphics[width=0.4\textwidth] {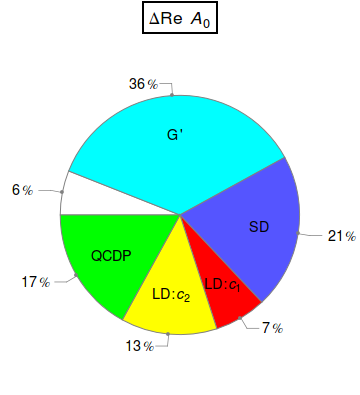}
\caption{\it Budgets of different enhancements of ${\rm Re}A_0$, denoted here 
by $\Delta{\rm Re}A_0$. $Z^\prime$ and $G^\prime$ denote the contributions 
calculated in the present paper. The remaining  coloured 
contributions come from the 
SM dynamics as calculated in \cite{Buras:2014maa}. The white region stands 
for the missing piece.
}\label{fig:piechart}~\\[-2mm]\hrule
\end{center}
\end{figure}

\section{Summary and Conclusions}\label{sec:6}
In the present paper we had two  main goals:
\begin{itemize}
\item
to investigate whether a subleading 
part of the $\Delta I=1/2$ rule,  at the level of $20-30\%$,  could be due to NP contributions originating 
in tree-level FCNC transitions mediated by a heavy colourless gauge boson 
$Z^\prime$ or an $SU(3)_c$ colour octet of gauge bosons $G^\prime$,
\item
to extent our previous analysis of tree level $Z^\prime$ and $Z$ FCNCs in 
\cite{Buras:2012jb} to the ratio $\epe$ and as a byproduct to update the 
SM analysis of this ratio. This was in particular motivated by the rather precise value
of $\bei$ obtained from QCD lattice calculations \cite{Blum:2012uk} that governs the electroweak penguin 
contributions to $\epe$.
\end{itemize}

As 
the experimental value for the smaller amplitude ${\rm Re}A_2$ 
has been successfully explained within the SM, both within dual 
representation of  QCD as a theory of weakly interacting mesons \cite{Buras:2014maa}
and by QCD lattice calculations \cite{Boyle:2012ys,Blum:2011pu,Blum:2011ng,Blum:2012uk}  we concentrated our analysis in 
the context of the first goal on the large amplitude 
${\rm Re}A_0$ which is by a factor of 22 larger than  ${\rm Re}A_2$ and its 
experimental value is not fully explained in these two approaches. In 
order to protect ${\rm Re}A_2$ from modifications we searched 
for NP that would have the property of the usual QCD penguins. They are capable 
  of shifting upwards ${\rm Re}A_0$ by an amount that  at scales $\ord(1\gev)$ is
roughly by a factor of three larger than  ${\rm Re}A_2$  without producing 
any relevant modification in the latter amplitude up to small isospin breaking effects.

However due to GIM mechanism the QCD penguin contribution within the SM 
is not large enough 
to allow within the dual approach to QCD to fully reproduce the experimental value of   ${\rm Re}A_0$ \cite{Buras:2014maa}.
Therefore 
we searched for a QCD-penguin like contribution that is not GIM suppressed. 
As we have demonstrated in the present paper, a neutral heavy gauge boson with 
FCNCs (with or without colour) and approximately flavour universal right-handed  diagonal couplings to 
quarks is capable of providing additional upward 
shift in ${\rm Re}A_0$ while  satisfying constraints from 
$\varepsilon_K$, $\Delta M_K$, $\epe$ and the LHC. Even if the 
structure of the relevant couplings must have a special hierarchy, summarized 
in (\ref{couplingsh}), (\ref{tuning}) and (\ref{tuningc}),
 we find this result interesting. Indeed our toy models for $Z^\prime$ and $G^\prime$  together with the dominant SM dynamics provide a better description of 
 the $\Delta I=1/2$ rule that it is presently possibly within the SM so that 
in these NP scenarios  we find that the values 
\be\label{N1aZprime}
R=\frac{{\rm Re}A_0}{{\rm Re}A_2}\approx 18~(Z^\prime),\qquad 
R=\frac{{\rm Re}A_0}{{\rm Re}A_2}\approx 21~(G^\prime)
\ee
can be obtained. This is fully compatible with the experimental value in (\ref{N1a}) even if in the case of $Z^\prime$ this ratio is visibly below the data. 
These results are summarized in Fig.~\ref{fig:piechart} where also the 
budget of different SM contributions calculated in \cite{Buras:2014maa} is 
shown.

We identified a {\it quartic} correlation between NP contributions to  ${\rm Re}A_0$, $\epe$, $\Delta M_K$ and $\varepsilon_K$ that offers means for more precise determination of the required properties of the 
neutral gauge bosons in question. Moreover, { in order to 
stay within perturbative regime for the couplings involved and explain the 
$\Delta I=1/2$ rule, $M_{Z^\prime}$ in 
Scenario A}
has to be at most few TeV so that these  simple extensions of the SM can be tested through the upgraded LHC and rare decays in the flavour precision era.

As our first goal, termed Scenario A, led to a fine-tuned scenario that 
could be ruled out one day, as a plan B, we have considered Scenario B 
for both tree-level heavy neutral gauge boson exchanges and $Z$ boson exchanges 
ignoring the $\Delta I=1/2$ rule constraint and concentrating on $\epe$ and its 
correlation with branching ratios for rare decays $\kpn$ and $\klpn$. { In 
this scenario  $M_{Z^\prime}$ can be well above the LHC range and its increase 
can be compensated by the increase of $Z^\prime$ couplings still fully within 
the perturbative regime.}

The most important findings of our paper are as follows: 
\begin{itemize}
\item
Within models containing only left-handed or only right-handed flavour-violating $Z^\prime$ or $G^\prime$ couplings to quarks it is impossible to generate any relevant 
contribution to  ${\rm Re}A_0$ without violating the constraint from 
$\Delta M_K$. The same applies to models with left-handed and right-handed couplings 
being equal or differing by sign.
\item
On the other hand $Z^\prime$ having in addition to $\Delta_L^{sd}(Z^\prime)=\ord(1)$, a small  right-handed coupling $\Delta_R^{sd}(Z^\prime)=\ord(10^{-3})$  and $M_{Z^\prime}$ in the reach of the LHC can improve the present status of $\Delta I=1/2$ rule, as summarized in (\ref{N1aZprime}), provided the diagonal coupling  $\Delta_R^{qq}(Z^\prime)=\ord(1)$. As demonstrated in \cite{Vries} and shown in 
Figs.~\ref{fig:zprimeexclusion} and \ref{fig:gprimeexclusion}
 such couplings are still allowed by the LHC data.
As seen in  (\ref{N1aZprime}) even larger values of $R$ can be obtained 
in $G^\prime$ scenario.
\item 
As far as $\epe$ is concerned, the interesting feature of this NP scenario is the absence of NP contributions to the electroweak penguin part of this ratio, 
a feature rather uncommon in many extensions of the SM. NP enters here 
only through QCD penguins and this implies interesting correlation between 
the new dynamics in $\epe$ and the $\Delta I=1/2$ rule. In particular, 
we have identified and interesting correlation 
between  NP contributions to  ${\rm Re}A_0$, $\epe$, $\varepsilon_K$ and 
$\Delta M_K$ which is shown in Fig.~\ref{fig:epsvsepstoy} for two sets of 
CKM parameters  which among the six considered by us are the only ones that
 allow simultaneous agreement for $\epe$ and $\varepsilon_K$ and significant 
contribution of $Z^\prime$ or $G^\prime$ to  ${\rm Re}A_0$. This means that
 only for the inclusive determinations of $\vub$ and $\vcb$ these 
heavy gauge bosons  have a chance to contribute in a significant manner to the $\Delta I=1/2$ rule. 
This assumes the absence of
 other mechanisms  at work which would help in this case 
if the exclusive determinations of these CKM parameters would turn out to be 
true.
\item
Interestingly, in Scenario A for $Z^\prime$  NP 
contributions to the  branching ratio for $\klpn$ are negligible when 
the experimental constraint for $\kpn$ is taken into account.
\item
As a byproduct we updated the values of $\epe$ in the SM stressing various uncertainties, originating in the values of $\vub$ and $\vcb$. In 
particular we have found that the best agreement of the SM with the 
data is obtained for $\bsi\approx 1.0$, that is close to the large $N$ limit 
of QCD.
\item
In the case of $Z^\prime$, in the context of scenario $B$, that is  ignoring the issue of 
the $\Delta I=1/2$ rule and concentrating on $Z^\prime$ with  
exclusively left-handed couplings, we have studied correlations between 
$\epe$ and the branching ratios for rare decays $\kpn$ and $\klpn$. In 
particular we have found that for $\bsi=0.75$ for which SM value of $\epe$ 
is much lower than the data, the cure of this problem through a $Z^\prime$ 
implies very enhanced values of $\mathcal{B}(\klpn)$. Simultaneously 
$\mathcal{B}(\kpn)$ is uniquely enhanced so that a triple correlation 
between these three observables exists. Figs.~\ref{fig:pKLvsKpA} and 
\ref{fig:epsvsBr0} show this in a transparent manner.
\item
We have also demonstrated that the SM $Z$ boson with FCNC couplings cannot 
provide the missing piece in ${\rm Re}A_0$ without violating the constraint 
from ${\rm Re}A_2$. Still the correlation between $\epe$, $\kpn$ and 
$\klpn$ can be used to test this NP scenario as demonstrated in Figs.~\ref{fig:epevsKpinu} and \ref{fig:KpinuZCKMd}. In particular very large enhancements 
of $\mathcal{B}(\klpn)$ found by us in \cite{Buras:2012jb} are excluded 
when the constraint from $\epe$ is taken into account: a property known 
from other studies.
\item
 We have also investigated various scenarios for flavour violating $Z$ 
couplings stressing different impact of $\epe$ and $K_L\to\mu^+\mu^-$ 
constraints
on rare branching ratios $\mathcal{B}(\kpn)$ and  $\mathcal{B}(\klpn)$.
In this context the comparison of Figs.~\ref{fig:KpinuZCKMd} (LHS), \ref{fig:KpinuZCKMdRHS} (RHS) and   \ref{fig:KpinuZCKMdLRS} (LRS)
 could one day allow us to distinguish between these three scenarios provided 
 deviations from the SM predictions will be sizable.
\end{itemize}

In summary a neutral $Z^\prime$ or $G^\prime$ with very special FCNC couplings summarized 
in (\ref{couplingsh}) and the mass in the reach of the LHC 
could be in principle responsible for 
the missing piece in ${\rm Re}A_0$. Whether  heavy gauge bosons with 
such properties exist should be answered by the LHC in this decade. 
 In particular a dedicated study of the dashed surface in Figs.~\ref{fig:zprimeexclusion} and \ref{fig:gprimeexclusion} in the context of 
our simple 
models would be very interesting as this would put the bounds used in our 
paper on firm footing. This applies also to the bounds on the coupling 
$\Delta_L^{sd}(G^\prime)$ and the fact that the bounds obtained in \cite{Vries} 
where derived under the condition that either $\Delta^{sd}_L$ or $\Delta^{qq}_R$  is vanishing. The presence of interferences between various contributions 
governed by these two couplings would not necessarily make 
the bounds on them stronger and could in fact soften them. Moreover in the former case the version of our models in which primed operator $Q_6^\prime$ is dominant could still provide the solution to the $\Delta I=1/2$ rule as discussed in 
Section~\ref{PRIMED}.

 If $Z^\prime$ or $G^\prime$ with such properties  do not 
exist, it is likely that the $\Delta I=1/2$ rule follows entirely from the 
SM dynamics.  Confirmation of this from lattice QCD would be in this case 
 important. 
 On the other hand any $Z^\prime$ with non-vanishing flavour 
violating couplings 
to quarks can have impact on $\epe$, $\kpn$ and $\klpn$ and the correlations 
between them. This also applies to scenario with 
flavour violating $Z$ couplings. In both cases the numerous plots presented by 
us should help in monitoring the exciting events to be expected at the LHC 
and in flavour physics in the second half of this decade.

\section*{Acknowledgements}
First of all we thank  Maikel de Vries for  providing the present bounds on the relevant couplings from the LHC and him and Andreas 
Weiler for illuminating discussions on the impact of LHC on our analysis.
Next
we would like to thank Matthias Jamin for updating the 
formula for $\epe$ within the SM. The discussions on LHC bounds with Bogdan Dobrescu, Robert Harris and Francois Richard are highly appreciated.
This research was done and financed in the context of the ERC Advanced Grant project ``FLAVOUR''(267104) and was partially
supported by the DFG cluster
of excellence ``Origin and Structure of the Universe''.

\bibliographystyle{JHEP}
\bibliography{allrefs}
\end{document}